\documentclass[11pt]{article}

\usepackage{amsmath,amssymb,latexsym}
\usepackage[T1]{fontenc}
\usepackage{epsfig}
\usepackage{fullpage}
\usepackage{graphicx}
\usepackage{makeidx}
\usepackage{slashed}
\usepackage{color}
\usepackage{xcolor}
\usepackage[normalem]{ulem}
\usepackage{subcaption}
\usepackage{feynmp}
\usepackage{cite}
\usepackage{hyperref}

\long\def\symbolfootnote[#1]#2{\begingroup%
\def\thefootnote{\fnsymbol{footnote}}\footnote[#1]{#2}\endgroup}

\def\as{\alpha_s}

\newcommand \slsh [1] {\not\!{#1}}
\newcommand{\defcol}{black}

\usepackage{soul}
\allowdisplaybreaks

\numberwithin{equation}{section}

\begin{document}

\begin{flushright}
Nikhef 2023-027
\end{flushright}

\vspace*{1.5cm}

\begin{center}
{\Large\sc Exponentiation of soft quark effects from the replica trick }\\[10ex] 
 { 
 Melissa van Beekveld$^{a,b}$,
 Leonardo Vernazza$^{c}$ and
 Chris D. White$^{d}$}\\[1cm]
 {\it 
$^a$ {Rudolf Peierls Centre for Theoretical Physics, Clarendon Laboratory, Parks Road, University of Oxford, Oxford OX1 3PU, UK}\\
$^b${Nikhef, Science Park 105, 1098 XG Amsterdam, NL}\\
$^c$ {INFN, Sezione di Torino, Via P. Giuria 1, I-10125 Torino, Italy}\\
$^d$ {Centre for Theoretical Physics, Department of Physics and Astronomy, Queen Mary
University of London, 327 Mile End Road, London E1 4NS, UK}\\}
\end{center}

\vspace{1.5cm}
\vspace{1cm}

\begin{abstract}
\noindent{
In this paper, we show that multiple 
maximally soft (anti-)quark and gluon emissions exponentiate 
at the level of either the amplitude or cross-section. 
We first show that such emissions can be captured by 
introducing new \emph{soft emission operators}, which 
serve to generalise the well-known Wilson 
lines describing emissions of maximally soft 
gluons. Next, we prove that vacuum expectation 
values of these operators exponentiate using 
the replica trick, a statistical-physics 
argument that has previously been used to 
demonstrate soft-gluon exponentiation properties 
in QCD. 
The obtained results are general, i.e.\ not tied to a particular scattering process. 
We illustrate our arguments by demonstrating the exponentiation of certain real and virtual corrections affecting subleading partonic channels in deep-inelastic scattering.
}
\end{abstract}

\vspace*{\fill}

\newpage
\reversemarginpar

\section{Introduction}
The understanding of contemporary collider physics experiments continues to rely on high-precision predictions in perturbative Quantum Chromodynamics (QCD). Predictions in QCD are typically done through a small-coupling expansion, resulting in a perturbative series. Next-to-leading order (NLO) accuracy has been achieved for all processes, whereas for particular processes we can reach as high as N$^3$LO~\cite{Baglio:2022wzu, Dreyer:2016oyx, Duhr:2021vwj, Anastasiou:2015vya, Duhr:2020seh}. The direct calculation of higher-order contributions is difficult in general, and often requires the development of new techniques to make obtaining results computationally feasible. This is further complicated by the fact that the coefficients of the perturbation expansion can diverge in certain kinematic regions. A well-known case of this is production of heavy (or off-shell) particles near threshold, for which perturbative cross-sections become enhanced due to the emission of soft and/or collinear radiation. In such cases, one may define a dimensionless {\it threshold variable} $\xi$, involving a ratio of kinematic invariants, and such that $\xi\rightarrow 0$ at threshold. The differential partonic cross-section turns out to involve terms of the form 
\begin{equation}
\frac{{\rm d}\hat{\sigma}}{{\rm d}\xi}\sim \sum_{n=0}\alpha_s^n\sum_{m=0}^{2n-1}
\left[c_{nm}^{(0)}\left(\frac{\log^m\xi}{\xi}\right)_+ + c_{nm}^{(1)}\log^m\xi+\ldots\right].
\label{threshold}
\end{equation}
That is, at each power of the coupling, one finds a series of logarithmic terms in $\xi$, where the first set (governed by the coefficients $c_{nm}^{(0)}$) involves the well-known {\it plus distribution}, thus ensuring that the differential cross-section is integrable, as required. These are known as {\it leading-power (LP)} threshold contributions, as they correspond to keeping the leading terms in a systematic expansion of the differential cross-section in powers of the threshold variable $\xi$, whose physical origin is the emission of radiation which is strictly soft and/or collinear.%
\footnote{We have neglected to show terms in Eq.~(\ref{threshold}) which involve a pure delta function $\delta(\xi)$, and which contribute at leading power. We will not be considering such contributions in this paper.} 
According to this nomenclature, the second set of terms in Eq.~(\ref{threshold}) comprises {\it next-to-leading power (NLP)} contributions, which are suppressed by a single power of $\xi$, and which potentially arise from {\it next-to-soft} and/or collinear emissions. Finally, the ellipsis in Eq.~(\ref{threshold}) denotes terms which are suppressed by further powers of $\xi$, and which are not formally singular as $\xi\rightarrow 0$.

The well-understood origin of LP terms allows us to sum them up to all orders in perturbation theory for some (classes of) observables, a procedure known as {\it resummation}. Different procedures exist, such as Feynman diagrammatic approaches~\cite{Parisi:1979xd,Curci:1979am,Sterman:1986aj,Catani:1989ne,Gatheral:1983cz,Frenkel:1984pz,Sterman:1981jc}, renormalisation group arguments~\cite{Forte:2002ni}, use of Wilson lines~\cite{Korchemsky:1992xv,Korchemsky:1993uz}, and soft-collinear effective theory (SCET)~\cite{Becher:2006nr,Schwartz:2007ib,Bauer:2008dt,Chiu:2009mg}. All of these have at heart that soft and collinear radiation factorises in a process-independent manner, making precise the quantum mechanical notion that radiation with a low (transverse) momentum has a large wavelength, and is thus insensitive to short-distance physics. One may then write down certain universal functions describing the threshold radiation, whose calculation at fixed orders in $\alpha_s$ leads to the resummation of entire towers of logarithms in perturbation theory. The most divergent logs at each order ($\sim \xi^{-1}\log^{2n-1}\xi$) are referred to as {\it leading logarithmic (LL)} terms, followed by {\it next-to-leading logarithmic (NLL)} and so on. The development of new resummation techniques forms a crucial part of the theory programme for current and forthcoming experiments.  

Until relatively recently, much less has been known about NLP threshold contributions, but there are by now well-established motivations for trying to include them in cross-section predictions. They may, for example, be numerically significant in certain scattering processes, such that they are comparable with missing higher-order contributions at either LP or fixed-order accuracy~\cite{Kramer:1996iq, Ball:2013bra,Bonvini:2014qga, Anastasiou:2015ema, Anastasiou:2016cez, vanBeekveld:2019cks,  vanBeekveld:2021hhv, Ajjath:2021lvg}. There is then a lengthy programme to be carried out of classifying the various NLP contributions, and either: (i) resumming them alongside LP contributions; (ii) using results for NLP terms at fixed-order to estimate missing higher-order corrections or improving fixed-order slicing schemes~\cite{Ebert:2018lzn,Boughezal:2018mvf,TorresBobadilla:2020ekr,Long:2023mvc,Catani:2022mfv,Ju:2023dfa,Abele:2021nyo}. However, it should also be emphasised that the characterisation of new types of contribution in QCD -- especially those that are potential windows to all-order structures in perturbation theory -- is a highly interesting field theoretical question in its own right. Recent years have seen a number of developments in the more formal theoretical physics literature, such as the relation of next-to-soft physics with asymptotic symmetries at past or future null infinity~\cite{Cachazo:2014fwa,Casali:2014xpa}, which be in turn be interpreted in terms of a conformal field theory living on the celestial sphere~\cite{Pasterski:2016qvg,Pasterski:2017kqt}. There has also been examination of the role that (next-to)-soft physics can play in (quantum) gravity in the high-energy limit~\cite{Akhoury:2011kq,Akhoury:2013yua,White:2011yy,Melville:2013qca,Luna:2016idw,Beneke:2021ilf,Beneke:2021umj,Beneke:2021aip,Beneke:2022pue,Beneke:2022ehj}
(see Ref.~\cite{Gross:1968in} for earlier work), which is indeed the exact limit probed by gravitational wave experiments such as LIGO. It is interesting that these very formal topics are directly relatable to cutting-edge phenomenology (see e.g. Refs.~\cite{White:2022wbr,Magnea:2021fvy,White:2014qia,Bonocore:2020xuj,Bonocore:2021qxh} for papers and reviews that make these links explicit), and the next few years are sure to see an interesting interplay between very different topics and ideas. 

Aside from the pioneering early works of Refs.~\cite{Low:1958sn,Burnett:1967km,DelDuca:1990gz}, studies of NLP effects in collider physics have utilised a number of techniques~\cite{Grunberg:2009yi,Soar:2009yh, Moch:2009hr,Moch:2009mu,deFlorian:2014vta,Presti:2014lqa, vanBijleveld:2023vck,vanBeekveld:2023gio,Agarwal:2023fdk,Buonocore:2023mne,Czakon:2023tld,Makarov:2023uet,Makarov:2023ttq,Engel:2023rxp,Bonocore:2015esa,Bonocore:2016awd,Gervais:2017yxv,Gervais:2017zky,Gervais:2017zdb,Laenen:2020nrt,DelDuca:2017twk,vanBeekveld:2019prq,Bonocore:2014wua,Bahjat-Abbas:2018hpv,Ebert:2018lzn,Boughezal:2018mvf,Boughezal:2019ggi,Bahjat-Abbas:2019fqa,Engel:2021ccn,Bonocore:2021cbv,Engel:2023ifn,Ajjath:2020ulr,Ajjath:2020sjk,Ajjath:2020lwb,Ahmed:2020caw,Ahmed:2020nci,Ajjath:2021lvg,AH:2022lpp,terHoeve:2023ehm,Kolodrubetz:2016uim,Moult:2016fqy,Feige:2017zci,Beneke:2017ztn,Beneke:2018rbh,Bhattacharya:2018vph,Beneke:2019kgv,Bodwin:2021epw,Moult:2019mog,Beneke:2019oqx,Liu:2019oav,Liu:2020tzd,Boughezal:2016zws,Moult:2017rpl,Chang:2017atu,Moult:2018jjd,Beneke:2018gvs,Ebert:2018gsn,Beneke:2019mua,Moult:2019uhz,Liu:2020ydl,Liu:2020eqe,Wang:2019mym,Beneke:2020ibj,Broggio:2021fnr,Broggio:2023pbu}, mirroring the situation at leading power. As these works make clear, the resummation of NLP contributions is starting to become possible, thus opening up a new chapter in the history of QCD perturbation theory. What makes this significantly more difficult than the study of LP contributions, however, is the fact that there are many different sources of NLP effect, each of which must be analysed in detail. Indeed, not all NLP terms in Eq.~(\ref{threshold}) arise from the emission of next-to-soft radiation. This has been emphasised in recent studies~\cite{Beneke:2020ibj,Beneke:2022obx,Beneke:2022zkz,Vernazza:2022wfl,vanBeekveld:2021mxn,vanBeekveld:2019lwy,vanBeekveld:2019prq}, which have examined the emission of soft (anti-)quarks in a range of collider processes including Higgs boson production, Drell-Yan (DY) production of a heavy vector boson, and deep inelastic scattering (DIS) (see also Ref.~\cite{Moult:2019uhz} for closely related work). Soft fermion emission is kinematically suppressed relative to gluon emission, such that the presence of soft (anti-)quarks in the final state is possible only from NLP onwards in the threshold expansion. However, there is a well-defined sense in which this soft fermion emission is simpler than other types of NLP contribution, such as those arising from next-to-soft radiation, next-to-soft operators for collinear gluon emission, or phase space effects. From a diagrammatic point of view, it is only the presence of fermion emission vertices -- evaluated in the leading soft approximation -- that makes such contributions differ from their soft gluonic counterparts. All other simplifications that are associated with soft gluon emission still apply, including factorisation of the phase space for all emitted soft partons. This suggests that it should be particularly simple to resum soft quark effects, at least at LL order, and it is the aim of this paper to present a general argument for this resummation.

The main ideas are as follows. We will first build on the results of Refs.~\cite{vanBeekveld:2019prq}, which defined {\it emission factors} describing the radiation of arbitrary partons from a given hard external leg of an amplitude.%
\footnote{Ref.~\cite{vanBeekveld:2019prq} used the term {\it emission operator} for the emission factors alluded to above. We here change this terminology to avoid confusion with the generalised soft-emission line operators to be discussed in what follows.} 
We will show that the emission of arbitrary numbers of soft partons can be expressed by combining these emission factors into an operator that generalises the well-known Wilson line governing the emission of soft gluons alone. Note that this is a different operator to the generalised Wilson line that has previously been considered in the study of next-to-soft physics\cite{Laenen:2008gt,White:2011yy,Bonocore:2020xuj}: the latter describes the emission of gluon radiation that is not strictly soft. By contrast, our generalised soft-emission line operator here encompasses purely soft partons, which may be gluons, or (anti-)quarks. It is thus matrix-valued in colour, spin and flavour space, which we will define shortly.

Similarly to the radiation of soft gluons, emission of multiple soft partons can be expressed in terms of vacuum expectation values of our generalised soft-emission line operators, which may in turn be written as a certain field theoretic path integral. We may then apply a novel argument known as the {\it replica trick}, which first arose in statistical physics (see e.g.~Ref.~\cite{Replica}), to show that soft-parton emissions of any type exponentiate at LL order, at either amplitude or cross-section level. Our use of this argument is very similar to an ongoing programme of work in QCD, aimed at calculating higher-loop soft gluon effects in multiparton scattering~\cite{Gardi:2010rn,Gardi:2013saa,Gardi:2011yz,Gardi:2011wa,Dukes:2013gea,Falcioni:2014pka,Gardi:2021gzz,Almelid:2017qju,Almelid:2015jia} (see Refs.~\cite{White:2015wha,Agarwal:2021ais} for reviews, and Refs.~\cite{Agarwal:2021him,Agarwal:2022wyk,Vladimirov:2015fea} for related work). To make the similarities clear we will review this below. A strength of the replica trick argument is that our results will be general, i.e.~not tied to any particular scattering process. Indeed, our analysis reveals that a large class of purely soft multiparton emissions exponentiates entirely. Given that fermion emissions are kinematically suppressed, this statement goes beyond NLP in the soft expansion, although the practical utility of this may be limited given that the associated terms in perturbation theory may overlap with those which have a different (next-to)soft origin. 

A convenient corollary of our results is that recent related conjectures emerge as a special case. In particular, Ref.~\cite{Beneke:2020ibj} considered Higgs-boson induced DIS, and focused on the kinematically subleading partonic channel in which a quark or anti-quark scatters in the initial state, rather than a gluon. This is the analogue of the gluon-initiated channel in conventional photon DIS, and the fact that it commences only at NLP is due to the emission of a fermion.%
\footnote{Similar calculations to those of Ref.~\cite{Beneke:2020ibj} exist for the cases of conventional DIS, and also for DY production~\cite{SCETunpublished}.}
Ref.~\cite{Beneke:2020ibj} showed that it is possible to resum LL NLP terms in this channel, provided one assumes that the leading singular terms in the virtual corrections to this process exponentiate, in the limit in which the emitted fermion is soft. A similar conjecture was made in the earlier Ref.~\cite{Moult:2019uhz}, which considered the process $e^+ e^-\rightarrow q\bar{q}g$, where the (anti-)quarks are collinear. If the quark and anti-quark carry momentum fractions $z$ and $\bar{z}=1-z$ of the total fermion momentum respectively, the authors noted that care is needed in examining the soft quark limit $z\rightarrow 0$: when including the virtual corrections, one must keep track of non-analytic dependence in $z$ as $z\rightarrow 0$, as this will affect the LL NLP terms after integration over the final state phase space. Both Refs.~\cite{Moult:2019uhz,Beneke:2020ibj} referred to terms in the soft quark limit of the fermion emission processes as {\it endpoint contributions}, and their conjectures amount to the statement that the one-loop virtual corrections to these endpoint contributions must exponentiate. As discussed in Refs.~\cite{Beneke:2020ibj,Vernazza:2022wfl}, the interpretation of endpoint contributions in SCET is delicate, and involves a refactorisation of SCET operators such as to manifestly extract the correct dependence on the soft momentum fraction. In the present paper, however, such contributions will emerge as a special case of our generalised soft-emission line operators. Exponentiation of endpoint contributions is then part of a more general exponentiation, involving arbitrary numbers of soft parton emissions. We hope that our results thus provide a useful complementary viewpoint to those of Refs.~\cite{Moult:2019uhz,Beneke:2020ibj}, whilst also providing an interesting further application of the replica trick. 

The structure of our paper is as follows. In Section~\ref{sec:replica}, we review the replica trick argument for the exponentiation of soft gluons, both at amplitude and cross-section level. In Section~\ref{sec:generalised}, we introduce our generalised soft-emission lines for the emission of soft (anti-)quarks and/or gluons, and explain how the replica trick argument generalises in this case. We then show two examples of how to apply this argument: in Section~\ref{sec:DISreal}, we show how the replica trick reproduces the exponentiation of real emission corrections in deep-inelastic scattering (DIS), that was conjectured in Ref.~\cite{Vogt:2010cv} and proven using much more cumbersome methods in Ref.~\cite{vanBeekveld:2021mxn}. Then, in Section~\ref{sec:DIS}, we show how a slightly modified version of the replica trick argument of Section~\ref{sec:generalised} can be used to confirm the expectations of Refs.~\cite{Moult:2019uhz,Beneke:2020ibj}, namely that the leading virtual corrections to the kinematically subleading partonic channels at NLO in DIS exponentiate. We discuss our results and conclude in Section~\ref{sec:discuss}, and certain technical details are collected in the appendix.

\section{The replica trick at leading power}
\label{sec:replica}

\begin{figure}[t]
\begin{center}
    \includegraphics[width=0.36\textwidth]{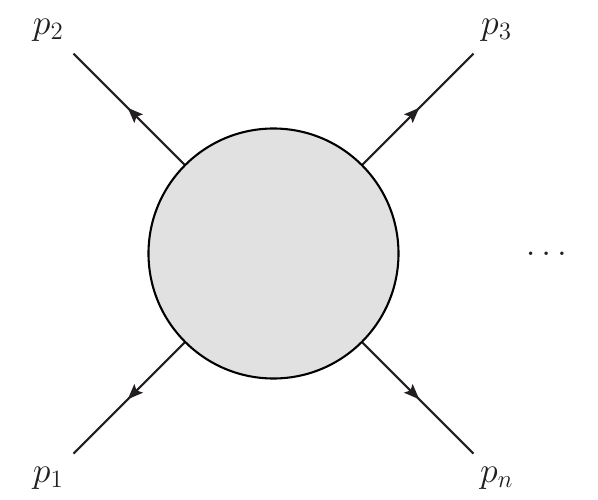}
    \caption{An $n$-point scattering amplitude, where all external momenta have been taken to be outgoing.}
    \label{fig:amplitude}
\end{center}
\end{figure}

In this section, we examine the replica trick argument for soft gluon exponentiation, first presented in Refs.~\cite{Laenen:2008gt,Gardi:2010rn}. Let us consider a scattering amplitude ${\cal A}_n$ with $n$ external partons and momenta labelled as shown in Fig.~\ref{fig:amplitude}.
As is well known, virtual QCD corrections to the amplitude lead to infrared (IR) singularities, associated with those kinematic regions in which the exchanged parton is {\it soft} (i.e.~it has vanishing 4-momentum), or {\it collinear} to any of the hard external legs. The correction has a universal form, meaning that for a general amplitude it may be factorised according to the following schematic formula~\cite{Dixon:2008gr}
\begin{equation}
  {\cal A}_n(\{p_i\})= \left[{\cal H}_n(\{p_i\},\{n_i\}) \otimes \,{\cal S}(\{\beta_i\})\right]
  \frac{\prod_{i=1}^n J(p_i,n_i)}{\prod_{i=1}^n {\cal J}(\beta_i,n_i)}\,.
  \label{Afac}
\end{equation}
Here ${\cal H}_n(\{p_i\},\{n_i\})$ is a process-dependent hard function, free of IR singularities. The soft function ${\cal S}(\{\beta_i\})$ depends only on the 4-velocities of the incoming and outgoing particles, and collects all singularities arising from soft radiation. This object is colour-connected to the hard function, as indicated by the symbol $\otimes$. The jet function $J(p_i,n_i)$ collects collinear singularities associated with emissions off the external leg with momentum $p_i$, and has a universal definition involving an auxiliary 4-vector $n_i$. Including both the soft and jet functions leads to a double-counting of radiation that is both soft and collinear, such that one has to remove each double-counted contribution. This is done though dividing out an eikonal jet function ${\cal J}(\beta_i,n_i)$ associated with leg $i$. The soft and (eikonal) jet functions all have universal definitions involving the appropriate partonic fields and Wilson line operators. Note that, given that the auxiliary vectors $\{n_i\}$ are arbitrary (apart from the condition $n_i\cdot p_i\neq 0$), dependence on them must cancel between the (eikonal) jet and hard functions, which in practice amounts to fixing the definition of the hard function in terms of the jet functions. 

Eq.~(\ref{Afac}) is known as the soft-collinear factorisation formula, and simplifies considerably if one only cares about the leading IR singularities of a given amplitude. In that case, all radiation must be soft (and collinear) so that one may remove the (eikonal) jet functions. Furthermore, higher-order contributions to the hard function can be neglected, such that this may be replaced by its first-order approximation. Note that this approximation does not have to correspond to the leading-order (LO) of a given inclusive scattering process, depending on whether or not we are considering a subleading partonic channel that only turns on at next-to-leading order (NLO) or higher. Thus, Eq.~(\ref{Afac}) reduces to 
\begin{equation}
  {\cal A}_n(\{p_i\})={\cal H}^{(0)}_n(\{p_i\}) \, \otimes \,{\cal S}(\{\beta_i\})\,,
  \label{Afac2}
\end{equation}
for some suitably defined first-order hard function ${\cal H}^{(0)}_n(\{p_i\})$ for an $n$-legged
hard-scattering processes. Classifying the leading IR singularities then amounts to calculating the soft function, which has the following operator definition as a vacuum expectation value~\cite{Dixon:2008gr}:
\begin{equation}
  {\cal S} \equiv  {\cal S}(\{\beta_i\})=\left\langle 0\left|
  \prod_{i=1}^n \Phi_i(A)\right|0\right\rangle
  \label{Sdef}
\end{equation}
where
\begin{equation}
  \Phi_i(A)={\cal P}\exp\left[ig_s\beta_i^\mu {\bf T}_i^a\int {\rm d}t_i
    A^a_\mu(t_i \beta_i)\right],
\label{Phidef}
\end{equation}
is a Wilson line operator along the classical straight-line contour $x_i^\mu = t_i \beta_i^\mu$ of external parton $i$, with $\beta_i^\mu$ the relevant 4-velocity. Furthermore, ${\bf T}^a_i$ is a colour generator in the appropriate representation of hard line $i$, $g_s$ is the strong coupling, and ${\cal P}$ denotes path-ordering of these colour generators along the integration contour. We can formally write the vacuum expectation value in Eq.~(\ref{Sdef}) as a path integral over the soft gauge field $A^a_\mu$ that couples to each external line:
\begin{equation}
{\cal S} =\int [{\cal D} A]\,
    \prod_{i=1}^n {\cal P}\exp\left( ig_s\beta_i^\mu{\bf T}_i^a
    \int {\rm d}t_i A_\mu^a(t_i\beta_i)\right)\, e^{iS[A]}\,,
  \label{pathint}
\end{equation}
where $S[A]$ is the action for the soft gauge field (note that this involves additional fields, which we have left implicit). Contracting the gluon fields with the action and carrying out the path integral (order-by-order) generates all possible Feynman diagrams in which the Wilson lines are connected by soft gluon graphs. 

It is well-known that one may write the soft function in an exponential form, where the logarithm contains certain sets of diagrams known as {\it webs}. The concept of a web was first introduced in QCD in the case of processes involving two coloured particles only~\cite{Gatheral:1983cz,Frenkel:1984pz,Sterman:1981jc}, but has more recently been generalised to arbitrary multiparton scattering processes~\cite{Gardi:2010rn} (see Refs.~\cite{White:2015wha,Agarwal:2021ais} for reviews). This property may be derived using the replica trick. To this end, we start with ${\cal S}$ defined as in Eq.~(\ref{pathint}), and consider its counterpart in a distinct, duplicated theory that contains $N$ identical copies, or {\it replicas}, of the original soft gauge field. Let us label these different fields by $\{A_\mu^{{\color{\defcol}{(i)}}a}\}$, where ${\color{\defcol}{(i)}}$ is the {\it replica index}. The soft gauge fields of different replicas are required to be non-interacting, so that the action for the replicated theory can simply be expressed as the sum of actions $S[A_\mu^{{\color{\defcol}{(i)}}}]$ for each gauge field.%
\footnote{Note that in diagrams where gluons couple via other fields off the Wilson lines (e.g.~to fermion bubbles), we must also replicate these other fields. We will not need to consider this complication explicitly here.}
The soft function in the replicated theory, denoted by $\color{\defcol}{{\cal S}_N}$, then takes the form
\begin{align}  
\label{SNdef}
  \color{\defcol}{{\cal S}_N}&= {\color{\defcol}{\prod_{i=1}^N}}\int [{\cal D}A^{{\color{\defcol}{(i)}}}] \,
    \prod_j^n\Phi_j(A^{{\color{\defcol}{(i)}}})\, e^{iS[A^{{\color{\defcol}{(i)}}}]} \\
  &={\cal S}^{N}.
\notag
\end{align}
In the first line, we now have a product of Wilson lines on each parton leg: one for each replica number. In the second line, we have recognised that the soft function in the replicated theory amounts to the original soft function raised to the power $N$, and the non-interacting nature of the replicas (leading to additivity of their actions) is crucial in this regard. A simple mathematical identity allows us to write
\begin{equation}
  {\cal S}^N=1+N\log{\cal S}+{\cal O}(N^2)\,,
  \label{Sexpand}
\end{equation}
from which we can immediately conclude that the logarithm of the soft function contains only those diagrams which in the replicated theory are ${\cal O}(N)$. This gives the following recipe for determining $\log{\cal S}$ (and thus the exponentiated soft function):
\begin{enumerate}
    \item Draw all possible Feynman diagrams in the replicated theory. These consist of Feynman diagrams that also occur in the original theory, but where each gluon emitted from a Wilson line now carries a replica index. 
    \item Find the dependence of each diagram on the number of replicas $N$.
    \item Take the ${\cal O}(N)$ part of each diagram, which gives the contribution of each diagram to the logarithm of the soft function.
\end{enumerate}
By taking the exponent of the logarithm of the soft function
one directly resums soft (and possibly collinear) eikonal emissions. 

\begin{figure}[t]
  \begin{center}
    \includegraphics[width=0.16\textwidth]{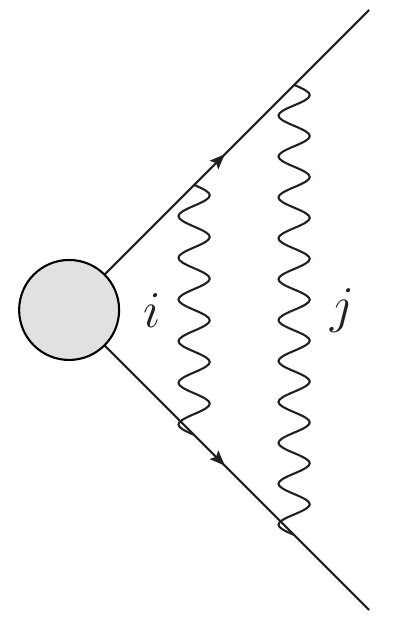}
    \caption{A diagram occurring in the replicated soft photon theory
      of Eq.~(\ref{Sabel}), for the case of two outgoing Wilson lines exchanging
      soft photons. There are two connected soft photon subdiagrams
      once the hard external lines are removed, consisting of
      individual soft photons, where $i$ and $j$ denote their replica
      indices.}
    \label{fig:QEDreplicas}
  \end{center}
\end{figure}

The simplest case to consider is that of QED. There are then no non-commuting colour matrices to worry about, and the soft function for the replicated theory reduces to
\begin{align}
  {\color{\defcol}{{\cal S}_N}} =\left[{\color{\defcol}{\prod_{i=1}^N}}\int {\cal D}A^{{\color{\defcol}{(i)}}}\right] \left[\prod_{k=1}^n
  \exp\left(ie\,{\color{\defcol}{\sum_{j=1}^N}} \int {\rm d} x_k^\mu A^{{\color{\defcol}{(j)}}}_\mu(x_k)\right)\right]
  \exp\left(i{\color{\defcol}{\sum_{l=1}^N}}S[A^{{\color{\defcol}{(l)}}}]\right).
  \label{Sabel}
\end{align}
This generates Feynman diagrams where each external line can emit photons, which may be joined by other fields off the Wilson lines. If we consider a diagram such as that shown in Fig.~\ref{fig:QEDreplicas}, which has 2 connected soft photon subdiagrams, there are $N$ choices for the first replica index $i$, and $N$ choices for the second replica index $j$. The diagram is thus ${\cal O}(N^2)$ and so does not contribute to the logarithm of the soft function. Likewise, a diagram with $m$ connected soft photon subdiagrams will be ${\cal O}(N^m)$, such that only diagrams with a single connected piece $m=1$ contribute to the logarithm of the soft function. This is the well-known exponentiation of connected soft photon diagrams that was first derived a long time ago using combinatoric methods~\cite{Yennie:1961ad}. 

Returning to a non-abelian context, we must face the fact that colour matrices associated with different replica fields do not commute with each other. To deal with this complication, Refs.~\cite{Gardi:2010rn, Laenen:2008gt} noted that one may rewrite the replicated soft function of Eq.~(\ref{SNdef}) in a form that instead makes clear the product of Wilson line operators associated with each individual hard leg $k$:
\begin{align}
  {\color{\defcol}{{\cal S}_N}}&= \int [{\cal D} A^{{\color{\defcol}{(1)}}}]\ldots [{\cal D}A^{{\color{\defcol}{(N)}}}]\,
  e^{i {\color{\defcol}{\sum_{j=1}^N}} S[A^{{\color{\defcol}{(j)}}}]} \, \prod_{k=1}^n 
  \left[\Phi_k(A^{{\color{\defcol}{(1)}}})\ldots \Phi_k(A^{{\color{\defcol}{(N)}}})\right].
  \label{SNdef2}
\end{align}
Here we have used the commuting nature of Wilson-line operators associated with different external legs, as they act on different partonic colour indices. Note that we are not allowed to reorder Wilson-line operators associated with the {\it same} leg, as they are non-commuting. There are then $N$ Wilson-line operators on each leg $k$, one for each replica gauge field, and such that the replica index increases as we go along the leg. Following Ref.~\cite{Gardi:2010rn, Laenen:2008gt}, we can implement this constraint in a convenient way by introducing a {\it replica-ordering operator} ${\cal R}$, such that the product of Wilson line operators on a given leg can be written 
\begin{equation}
    \prod_{{\color{\defcol}{j=1}}}^{\color{\defcol}{N}} {\cal P}\exp\left[ig_s\int {\rm d}x_k^\mu A_\mu^{{\color{\defcol}{(j)}}}
    \right]={\cal R}{\cal P}\exp\left[ig_s{\color{\defcol}{\sum_{j=1}^N}} \int {\rm d}x_k^\mu A_\mu^{{\color{\defcol}{(j)}}}(x_k)\right].
    \label{Wilsonprod}
\end{equation}
On the right-hand side, we now have a single path-ordered exponential, but where, in performing the Taylor expansion, ${\cal R}$ acts to reorder colour matrices where necessary, such that the replica index is increasing along the line. To clarify this, the explicit action of ${\cal R}$ on two colour generators ${\bf T}_k^{{\color{\defcol}{(i)}}}$ and ${\bf T}_k^{{\color{\defcol}{(j)}}}$ associated with replica indices $i$ and $j$ is defined as
\begin{equation}
    {\cal R}\left[{\bf T}_k^{{\color{\defcol}{(i)}}}{\bf T}_k^{{\color{\defcol}{(j)}}}\right]
    =\begin{cases} {\bf T}_k^{{\color{\defcol}{(i)}}}{\bf T}_k^{{\color{\defcol}{(j)}}},\quad i\leq j\\
    {\bf T}_k^{{\color{\defcol}{(j)}}}{\bf T}_k^{{\color{\defcol}{(i)}}},\quad i> j,
    \end{cases}
\end{equation}
so that ${\cal R}$ reorders the matrices only if the replica numbers are not increasing. This definition is straightforward to generalise to higher numbers of colour matrices. Armed with the replica-ordering operator ${\cal R}$, our replicated soft function of Eq.~(\ref{SNdef2}) now assumes the form 
\begin{align}
  {\color{\defcol}{{\cal S}_N}} &= \int [{\cal D} A^{{\color{\defcol}{(1)}}}]\ldots [{\cal D}A^{{\color{\defcol}{(N)}}}]\,
  e^{i {\color{\defcol}{\sum_{j=1}^N}} S[A^{{\color{\defcol}{(j)}}}]} \, \left[\prod_{k=1}^n 
  {\cal R}\,{\cal P}\, e^{ig_s\,{\color{\defcol}{\sum_{j=1}^N}} \int {\rm d}x_k^\mu A_\mu^{{\color{\defcol}{(j)}}}(x_k)}\right].
  \label{SNdef2b}
\end{align}
This provides a systematic prescription for ascertaining which diagrams contribute to the logarithm of the soft function. Contracting the fields within each replicated field theory leads to a set of Feynman diagrams. The colour factor of each such diagram will not be the usual colour factor of QCD perturbation theory, but will instead have colour matrices reordered by the operator ${\cal R}$. The contribution of each diagram in the replicated theory will depend upon the number of replicas $N$ in general. One may then take the ${\cal O}(N)$ part of each diagram as before.

\begin{table}[t]
    \begin{center}
        \begin{tabular}{c|c|c|c}
        Replica-ordering hierarchy $h$ & ${\cal R}[C(a)|h]$ & ${\cal R}[C(b)|h]$ & Multiplicity\\
        \hline
          $i=j$   & $C(a)$  & $C(b)$ & $N$ \\
          $i<j$ & $C(a)$ & $C(a)$ & $N(N-1)/2$ \\
           $i>j$ & $C(a)$ & $C(a)$ & $N(N-1)/2$         
        \end{tabular}
            \caption{Replica trick analysis of the diagrams in Fig.~\ref{fig:QCDreplicas}. For each diagram $D$ we give the various hierarchies $h$ of the ordering of the replica indices, together with the colour factor ${\cal R}[C(D)|h]$ one obtains by applying the replica-ordering operator ${\cal R}$. Finally, we give the number of ways of choosing replica numbers consistent with each hierarchy. }
            \label{tab:replica}
    \end{center}
\end{table}

\begin{figure}[t]
\begin{center}
    \includegraphics[width=0.50\textwidth]{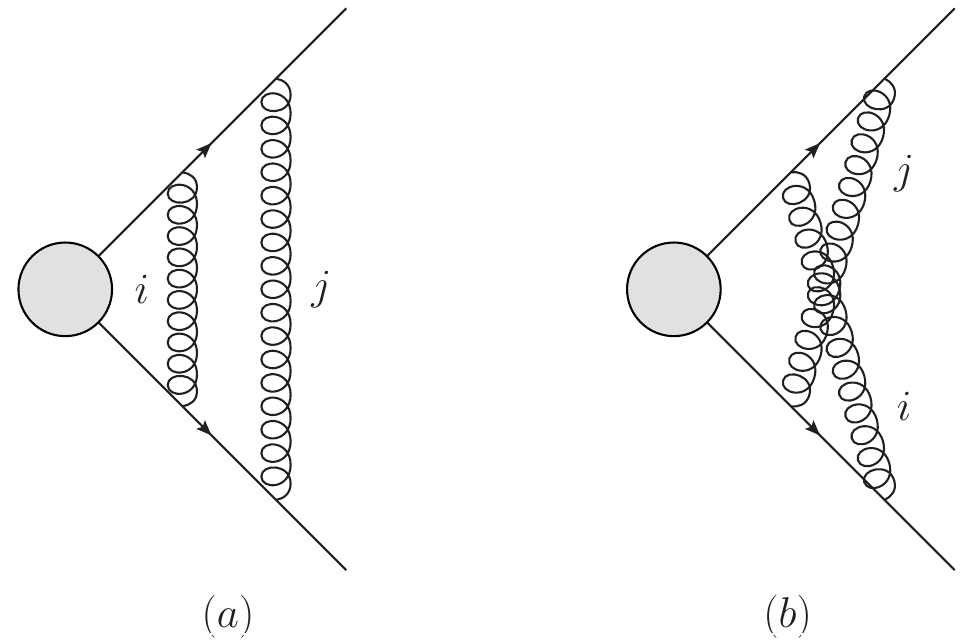}
    \caption{Two diagrams that potentially contribute to the logarithm of the QCD soft function, involving two outgoing Wilson lines. Replica indices $i$ and $j$ are labelled, as are the numbers of the Wilson lines.}
    \label{fig:QCDreplicas}
\end{center}
\end{figure}

The simplest non-trivial example of this procedure is the case of two separate gluon emissions between two Wilson lines, for which the two Feynman diagrams in the replicated theory are shown in Fig.~\ref{fig:QCDreplicas}. To calculate the colour factors, we must take into account all possible hierarchies of replica numbers, given that ${\cal R}$ potentially acts differently in each case. For example, the colour factor of diagram (a) is 
\begin{equation}
    {\cal R}\left[{\bf T}_{1}^{{\color{\defcol}{(i)}}}{\bf T}_{1}^{{\color{\defcol}{(j)}}}
    {\bf T}_{2}^{{\color{\defcol}{(i)}}}{\bf T}_{2}^{{\color{\defcol}{(j)}}}
    \right],
    \label{colfaca}
\end{equation}
with ${\bf T}_{k}$ a general colour operator acting on the external line with label $k=1,2$. 
If $i=j$ or $i<j$, the replica ordering operator will leave the colour factor intact. If $i>j$, however, it will reorder the colour matrices on each line, so that one has
\begin{equation}
    {\cal R}\left[{\bf T}_{1}^{{\color{\defcol}{(i)}}}{\bf T}_{1}^{{\color{\defcol}{(j)}}}
    {\bf T}_{2}^{{\color{\defcol}{(i)}}}{\bf T}_{2}^{{\color{\defcol}{(j)}}}
    \right]=\left[{\bf T}_{1}^{{\color{\defcol}{(j)}}}{\bf T}_{1}^{{\color{\defcol}{(i)}}}
    {\bf T}_{2}^{{\color{\defcol}{(j)}}}{\bf T}_{2}^{{\color{\defcol}{(i)}}}
    \right],\quad i>j\,.    
    \label{colfaca2}
\end{equation}
This has the same form as the original colour factor with the indices $i$ and $j$ interchanged, and so for all hierarchies of replica number, we obtain the colour factor of diagram (a). For a given hierarchy of replica orderings $h$ and diagram $D$, we denote the replica-ordered colour factor by ${\cal R}[C(D)|h]$. We may then summarise the above analysis as in table~\ref{tab:replica}, which gives the different hierarchies of the ordering of the two replica indices $i$ and $j$, together with the colour factor obtained by applying the replica ordering operator in each case. The table also details the findings for diagram (b). For this diagram, for some hierarchies, the action of ${\cal R}$ is to disentangle the crossed gluon pair, thus creating the colour factor of diagram (a). Finally, we give the multiplicity factor corresponding to the number of ways of choosing replica numbers consistent with the given hierarchy. The total contribution of a given diagram $D$ in the replicated theory will be given by its kinematic factor ${\cal K}(D)$, multiplied by the various colour factors weighted by the appropriate multiplicities. For diagram (a), for example, we have
\begin{align}
    {\cal K}(a)C_N(a)\,,\quad
    C_N(a)=NC(a)+2C(a)\left(\frac{N(N-1)}{2}\right)=N^2 C(a)\,,
\end{align}
where $C_N(D)$ denotes the effective colour factor for diagram $D$. According to the replica trick, we must take the ${\cal O}(N)$ part of this as the contribution to the logarithm of the soft function, and following Ref.~\cite{Gardi:2010rn} we may write this as
\begin{equation}
    {\cal K}(D)\tilde{C}(D),\quad \tilde{C}(D)\equiv C_N(D)\Big|_{{\cal O}(N)}.
    \label{tildeCdef}
\end{equation}
For the case of diagram (a), we thus find $\tilde{C}(a)=0$. In the case of diagram (b), we instead find 
\begin{align}
\tilde{C}(b)=C(b)-C(a)\,,
\end{align}
such that this diagram indeed contributes to the logarithm of the soft function. It does so, however, with a modified colour weight, and this is a special case of the previously known fact that in two-parton processes, only those diagrams which are two-particle irreducible (i.e.~webs) exponentiate~\cite{Gatheral:1983cz,Frenkel:1984pz,Sterman:1981jc}, with appropriately modified colour factors. 
In general, we find then the exponentiated soft function can be written as
\begin{align}
\mathcal{S} = {\rm exp}\left[\sum_{D} 
\mathcal{K}(D)\, \tilde{C}(D)\right],
\end{align}
where the argument of the exponent are the so-called web diagrams. 

The above methods generalise to arbitrary multiparton scattering processes in QCD, where the simple criterion of two-particle irreducibility no longer applies. It is nevertheless possible to write a  general structure for the logarithm of the soft function. To do so, note that the ${\cal R}$ operator reorders gluon attachments on all external legs, so that the exponentiated colour factor of a given diagram potentially depends on all other diagrams related by gluon permutations. There is a closed set of such permutations (as exemplified by the diagrams of Fig.~\ref{fig:QCDreplicas}), and Ref.~\cite{Gardi:2010rn} defined a {\it multiparton web} to consist of such a closed set of diagrams. This interpretation survives after the effects of renormalisation are taken into account~\cite{Gardi:2011yz}, and a further refinement of this notion has been introduced in Refs.~\cite{Agarwal:2021him,Agarwal:2022wyk}. In general, the exponentiated colour factor of a diagram $D$ in each web $W$ will be a superposition of all colour factors in the web
\begin{equation}
\tilde{C}(D)=\sum_{D'\in W} R_{DD'}\, C(D')\,.
\label{Rdef}
\end{equation}
Here the quantity $R_{DD'}$ is known as a {\it web mixing matrix}. It has a purely combinatorial definition~\cite{Gardi:2011wa}, of interdisciplinary interest~\cite{Dukes:2013wa,Dukes:2016ger}, and which has been further explored in Refs.~\cite{Dukes:2013gea,Agarwal:2021him,Agarwal:2022wyk}. Combining the exponentiated colour factor of each diagram with its kinematic part, the contribution of a given web to the logarithm of the soft function is 
\begin{equation}
w=\sum_{D,D'\in W}{\cal K}(D)\,R_{DD'}\,C(D')\,.
\label{Rdef2}
\end{equation}

So far we have only considered virtual emissions. Comments regarding the generalisation to real emissions were made in Ref.~\cite{Bahjat-Abbas:2019fqa}, which postulates a soft function at the level of the squared amplitude/cross-section. That is, one may write
\begin{equation}
  {\cal S}(\bar{v}) =\sum_m
    {\rm Tr}\left[\Big\langle 0\Big|\Phi_1\ldots \Phi_n \Big|m\Big\rangle
    \Big\langle m\Big|\Phi_n^\dag\ldots \Phi_1^\dag\Big|0\Big\rangle\right]\delta(v(m)-\bar{v})\,.
  \label{Ssq}
\end{equation}
Here the trace is over the colour indices of the outgoing parton legs,
and the two vacuum expectation values (VEVs) are taken between a vacuum state, and a final state
$|m\rangle$ containing $m$ soft gluons together with a measurement function $\delta(v(m)-\bar{v})$ corresponding to the (dimensionless) observable $\bar{v}$ and the $m$-particle definition of that observable. We use a compact notation for
the sum over final states, which implicitly contains the integral over
the combined phase space of the hard partons and soft gluons for each
$m$.%
\footnote{For particular
scattering processes in the literature, the precise definition of the
soft function may vary, depending on whether it is defined for the
squared amplitude or a normalised differential cross-section.}
It is possible to write a formal generating functional for soft functions such as those in Eq.~(\ref{Ssq}) which allows to apply the replica trick directly, although this was not proven in Ref.~\cite{Bahjat-Abbas:2019fqa}. We thus return to this point in what follows.

In this section, we have reviewed how the replica trick can be used to show how soft gluon emissions exponentiate. It is perhaps more correct to say that the replica trick constitutes a systematic procedure for determining the logarithm of the soft function. Scrutiny of the above arguments, however, reveals that they do not crucially depend on the nature of the Wilson lines in Eq.~(\ref{pathint}). We are free to replace the latter with a generalised soft-emission operator, and indeed Ref.~\cite{Laenen:2008gt} did precisely this in arguing that certain classes of next-to-soft contributions formally exponentiate. Here, we will seek a different generalised soft-emission line, which includes the effects of soft quark emission. This is the subject of the following section.

\section{The replica trick for generalised 
soft-parton emissions}
\label{sec:generalised}

The replica trick may be extended to 
show that not only soft-gluon, but 
also soft-quark emissions can be 
exponentiated. Indeed, in applying 
the replica trick to prove the 
exponentiation of soft gluons, the 
precise definition of the Wilson 
line (and all its gauge-transformation 
properties) never entered. All that 
matters is that we are able to write 
down an operator definition that 
generates the soft emissions in QCD 
up to all orders in the strong coupling. 
When considering also the emission of soft 
quarks, we must apply a similar argument, 
but the difficulty lies in finding a 
suitable replacement for the operator 
that describes the emission of soft 
(anti-)quarks in addition to gluons. 
We will successfully seek such a 
generalised soft-emission operator 
below. Our strategy to prove that 
soft quark emissions exponentiates 
consists of three key ingredients:
\begin{enumerate}
\item We will consider a hard external 
leg of an amplitude, and show that the 
emission of a single soft (anti-)quark 
or gluon can be written in terms of 
universal {\it emission factors} 
(Section~\ref{sec:emission}). The latter 
have been considered in 
Ref.~\cite{vanBeekveld:2019prq}, and 
we build on those results here.
\item These emission factors are used 
to build a generalised soft-emission 
operator for the emission of arbitrary 
numbers of soft partons from a hard 
leg (Section~\ref{sec:multiple}).
\item The generalised soft-emission 
operator is used to define a generalised 
soft function, which captures all soft 
parton emissions. This generalised soft 
function will have a path integral 
representation, allowing the replica 
trick to be used in an analogous 
fashion to the case of soft gluon 
emission (Section~\ref{sec:replica-soft}).
\end{enumerate}
In carrying out this programme, we will also provide a more rigorous treatment of the replica trick for real emissions than has previously appeared in the literature.

\subsection{Emission factors for soft partons}
\label{sec:emission}

\begin{figure}[t]
\begin{center}
    \scalebox{0.8}{\includegraphics{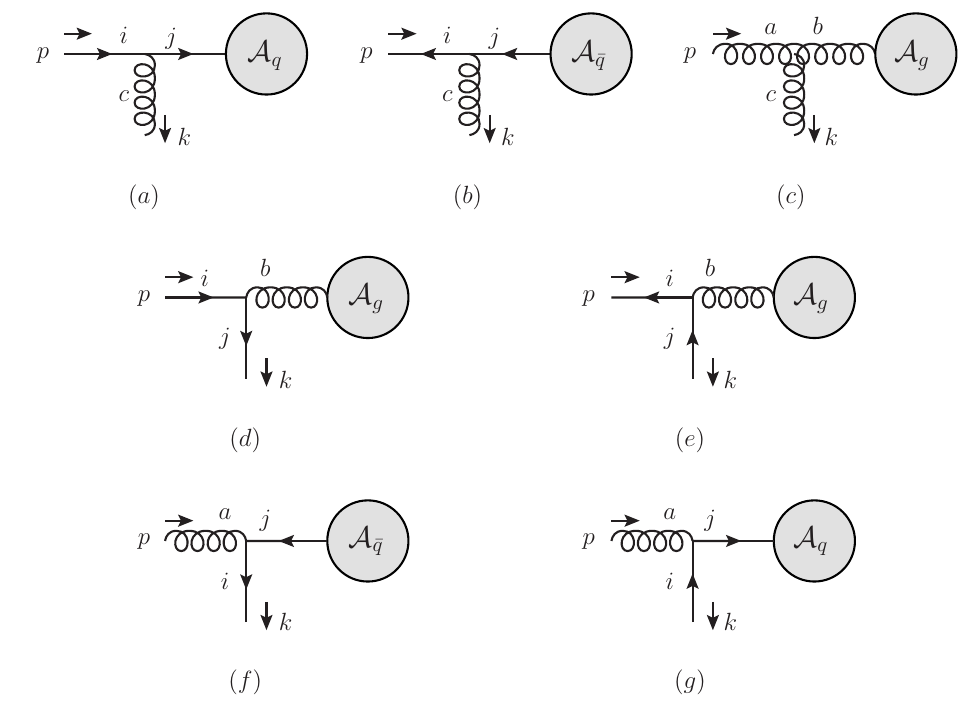}}    
    \caption{Emission of various soft partons 
    from incoming hard legs, where $A_{X}$ 
    denotes the amplitude with an incoming 
    parton $X$, but where the external 
    wavefunction for the particle $X$ 
    is removed. Here $\{i,j,\dots\}$ 
    ($\{a,b,c,\dots\})$ represent
    the fundamental (adjoint) colour 
    indices, while Lorentz and spinor 
    indices are suppressed, though
    explicitly written in the corresponding 
    Eqs. (\ref{qq}) -- (\ref{gq}).} 
    \label{fig:emissions}
\end{center}
\end{figure}
Consider the emission of a soft gluon 
from an incoming quark leg, as shown 
in Fig.~\ref{fig:emissions}(a).
Scaling the gluon momentum with 
$k^\mu \to \lambda k^\mu$ and taking 
$\lambda \to 0$ leads to the well-known 
eikonal Feynman rule, such that the 
matrix element for diagram (a) is 
written as 
\begin{equation}
[\mathcal{M}_{q}^{\rm in}(p)]^c_{\mu} 
= \frac{g_s}{p\cdot k}\,  
[\mathcal{A}_{q}(p)]_{\hat{b}}^{j} \,
t_{ji}^c \left(p_\mu \delta_{\hat{b}\hat{a}}\right) 
 u^i_{\hat{a}}(p)\,.
\label{qq}
\end{equation}
Here $\mathcal{A}_q(p)$ denotes the amplitude 
for an incoming quark leg with the external 
spinor removed. We denote spinor indices by 
hatted lower-case Latin letters 
$\{\hat{a},\hat{b},\dots\}$ (raised or 
lowered spinor indices carry no meaning). 
The fundamental (adjoint) colour indices are 
denoted with $\{i,j,\dots\}$ ($\{a,b,c,\dots\})$. 
The colour generator of the ${\rm SU}(3)$ gauge 
group in the fundamental representation is 
denoted by $t^a_{ij}$. The external wavefunction
of the quark carries a label $i$, which refers to
the colour of the incoming quark, whereas
the amplitude carries the label $j$, referring
to the colour of the internal quark entering
the hard scattering. 
We have not included 
an external particle wavefunction for the 
emitted soft gluon, leaving open the possibility
of this gluon being either real or virtual. 

Similarly, if the emitting particle were an anti-quark 
instead of a quark (diagram (b)) we would have
\begin{equation}
[\mathcal{M}_{\bar{q}}^{\rm in}(p)]^c_{\mu} 
= \frac{g_s}{p\cdot k} \, 
\bar{v}^i_{\hat{a}}(p)\,(-t_{ij}^c)\, 
\left(p_\mu \delta_{\hat{a}\hat{b}} \right) 
\,[\mathcal{A}_{\bar q}(p)]^{j}_{\hat{b}}\,.
\label{qbqb}
\end{equation}
Performing the same exercise
for diagram (c) leads to 
\begin{equation}
[\mathcal{M}_g^{\rm in}(p)]^c_{\mu} 
= \frac{g_s}{p\cdot k} \, i f^{bca} 
\left(p_\mu \eta^{\alpha\beta}\right) 
[\mathcal{A}_{g}(p)]_{\beta}^{b} \, 
\epsilon_{\alpha}^{a}(p)\,,
\label{gg}
\end{equation}
with $f^{abc}$ the colour generator of the 
${\rm SU}(3)$ gauge group in the adjoint 
representation. 

To obtain the Feynman rules for the 
emission of a soft-quark (soft anti-quark) 
from an incoming quark (anti-quark), 
we may, in analogy with the soft-gluon case,  
scale the (anti-)quark momentum with 
$k^\mu \to \lambda k^\mu$ and take $\lambda \to 0$~\cite{vanBeekveld:2019prq}.
The soft (anti-)quark emission contributions from a quark (anti-quark) are shown respectively in diagrams 
(d) and (e) of Fig.~\ref{fig:emissions}.
We obtain
\begin{align}
[\mathcal{M}_{q}^{\rm in}(p)]^j_{\hat{c}} 
& = \frac{g_s}{p\cdot k} \, t^b_{ji}  
\left(-\frac12 \gamma^\beta_{\hat{c}\hat{a}}\right) u^i_{\hat{a}}(p)\, [\mathcal{A}_g(p)]_{\beta}^{b}\,, 
\label{qg} \\
[\mathcal{M}_{\bar{q}}^{\rm in}(p)]^j_{\hat{c}} 
& = \frac{g_s}{p\cdot k} \, 
\bar{v}^i_{\hat{a}}(p) \, t^b_{ij}
\left(-\frac12\gamma^\beta_{\hat{a}\hat{c}}\right)  [\mathcal{A}_g(p)]_{\beta}^{b} \,.
\label{qbg}
\end{align}
Finally, the case of a soft quark (soft 
anti-quark) emission from an incoming 
gluon, shown respectively in diagrams 
(f) and (g) of Fig.~\ref{fig:emissions},
reads
\begin{align}
[\mathcal{M}_g^{\rm in}(p)]^i_{\hat{c}} 
& = \frac{g_s}{p\cdot k}\,(-t^a_{ij})\,\frac12
\big(\gamma^\alpha\slashed{p}\big)_{\hat{c}\hat{b}}
\, [\mathcal{A}_{\bar{q}}(p)]^j_{\hat{b}} \, 
\epsilon^a_{\alpha}(p) \,,
\label{gqb} \\
[\mathcal{M}_g^{\rm in}(p)]^i_{\hat{c}} 
& = \frac{g_s}{p\cdot k}\,  
[\mathcal{A}_{q}(p)]_{\hat{b}}^j 
\, t^a_{ji}  \,\frac12 
\big(\slashed{p}\gamma^\alpha\big)_{\hat{b}\hat{c}}\,
 \epsilon^a_\alpha(p) \,.
\label{gq}
\end{align}
\begin{figure}[t]
\begin{center}
    \scalebox{0.8}{\includegraphics{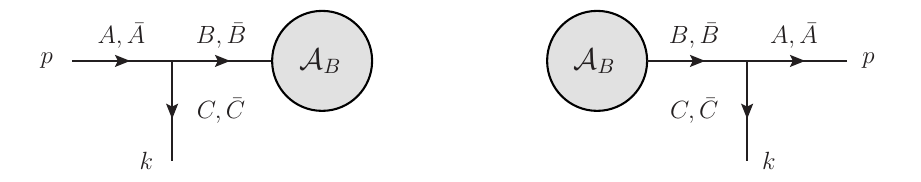}}
    \caption{Emission of a parton with (spinor/vector) 
    index $C$ from two partons with indices $B$ and $A$. 
    The barred letters denote colour indices.}
\label{fig:emissionfac}
\end{center}
\end{figure}
In each case, the additional factors
in the amplitude take the form of an 
eikonal denominator depending on the 
soft momentum $k$, with everything 
else depending only on the hard
momentum $p$. The additional factors 
are sandwiched between the non-radiative amplitude 
on the one-hand, and the external 
wavefunction for the hard leg on the other, 
meaning we can write a general form that
covers all cases. Let us first introduce 
an array of fields
\begin{equation}
\left[\Theta_I\right]^{\bar{A}}_{A} \equiv 
(\psi^i_{\hat{a}},\bar{\psi}^i_{\hat{a}},A^a_\alpha)\,,
  \label{Thetadef}
\end{equation}
where $\psi^{i}$ ($\bar{\psi}^{i}$) denotes 
the (anti-)quark field, and $i$, $a$ are 
colour labels. On the left-hand side, we 
have introduced a flavour index 
$I \in \{q, \bar{q}, g\}$, and an 
index $A$ in Minkowski/spinor space, 
labeling the correct index for the relevant 
fields (i.e.~$A\in\{\hat{a},\hat{a},\alpha\}$). 
Furthermore, the index $\bar{A}$ labels 
the colour index associated with the field 
whose species index is $A$: 
$\bar A\in\{i,i,a\}$. We stress that 
this is purely a book-keeping device that 
is useful for our purposes -- we do not 
wish to imply that these fields form a 
multiplet that is acted upon by any sort 
of symmetry transformation (as would be 
the case in e.g.~a supersymmetric theory). 
Next, using the same basis, let us define an array of external 
initial-state particle wavefunctions that 
go with the fields of Eq.~(\ref{Thetadef})
\begin{equation}
\left[\xi_I\right]^{\bar{A}}_A
=(u^i_{\hat{a}},\bar{v}^i_{\hat{a}},\epsilon^a_\alpha)\,.
\label{wavefns}
\end{equation}
We will also need an array corresponding 
to the various wave-function stripped amplitudes 
occurring in Fig.~\ref{fig:emissions}
\begin{equation}
\left[\mathcal{A}_I\right]^{\bar{A}}_A = \left(
[\mathcal{A}_q(p)]^i_{\hat{a}},
[\mathcal{A}_{\bar{q}}(p)]^i_{\hat{a}},
[\mathcal{A}_g(p)]^a_{\alpha} \right).
\label{AAdef}
\end{equation}
In this notation, we may summarise
Eqs.~(\ref{qq})--(\ref{gq}) in a compact 
notation by introducing a transition matrix 
for soft emissions off incoming hard partons,
$[{\cal T}^{\rm in}_{IJ}(p)]^{\bar C, \bar A\bar B}_{C,AB}$,
such that
\begin{equation}
[\mathcal{M}_{I}^{\rm in}(p)]_{C}^{\bar{C}} 
= \frac{g_s}{p\cdot k} \,
[\xi_{I}(p)]^{\bar{A}}_{A} \,
[{\cal T}^{\rm in}_{IJ}(p)]^{\bar C, \bar A\bar B}_{C,AB}
\, [\mathcal{A}_{J}(p)]^{\bar{B}}_B\,,
\label{emissionfac}
\end{equation}
with\footnote{Note that the indices $I$,$J$ 
in Eq.~\eqref{emissionfacB} are not summed, 
i.e. there is no matrix multiplication in 
the indices $I$,$J$. These simply labels 
the flavour of the transitions, both on the 
l.h.s and the r.h.s of Eq.~\eqref{emissionfacB}.}
\begin{equation}
[{\cal T}^{\rm in}_{IJ}(p)]^{\bar C, \bar A\bar B}_{C,AB}
= [{\bf T}_{IJ}]^{\bar{C}}_{\bar{A}\bar{B}}
\,[Q^{\rm in}_{IJ}(p)]_{C,AB}.
\label{emissionfacB}
\end{equation}
The labeling is defined such that it follows the 
momentum flow of the hard line, which goes from 
$A$ to $B$. One sees that this object consists 
of a kinematic part $[Q^{\rm in}]_{C,AB}$ 
and a colour factor ${\bf T}^{\bar{C}}_{\bar{A}\bar{B}}$. 
In the basis Eq.~\eqref{AAdef} we may write
\begin{equation}\label{TransitionMatrDef}
[{\cal T}^{\rm in}(p)]^{\bar C,\bar A\bar B}_{C,AB}
= \left(\begin{array}{ccc} 
[{\cal T}_{qq}^{\rm in}(p)]^{c,ij}_{\gamma,\hat a\hat b} &
0 & 
[{\cal T}_{qg}^{\rm in}(p)]^{k,ib}_{\hat c,\hat a \beta} \\
0 & 
[{\cal T}_{\bar q \bar q}^{\rm in}(p)]^{c,ij}_{\gamma,\hat a \hat b} & 
[{\cal T}_{\bar q g}^{\rm in}(p)]^{k,ib}_{\hat c,\hat a \beta } \\
({\cal T}_{g q}^{\rm in}(p))^{k,aj}_{\hat c,\alpha \hat b} & 
[{\cal T}_{g\bar q}^{\rm in}(p)]^{k,aj}_{\hat c,\alpha\hat b} & 
[{\cal T}_{gg}^{\rm in}(p)]^{c,ab}_{\gamma,\alpha\beta}
\end{array}\right),
\end{equation}
where we will refer to the entries of this matrix as 
\emph{emission factors}, and where we have explicitly indicated where no valid non-zero entry is possible. The assignment of the colour 
factor  $[{\bf T}_{IJ}]^{\bar{C}}_{\bar{A}\bar{B}}$ 
follows the usual eikonal Feynman rules. In our notation, we have used the colour labels 
$a, b$ ($c$) for colour indices of hard (soft) 
gluons, and $i, j$ ($k$) for that of hard (soft) 
quarks and anti-quarks. The assignment of the 
kinematic factor $[Q^{\rm in}]_{C,AB}$ can be 
inferred by direct comparison with 
Eqs.~(\ref{qq})--(\ref{gq}), and is summarised 
in Appendix~\ref{app:kin-factors}. Combining 
the colour and kinematic factors we find for 
the transition matrix
\begin{equation}
\label{trans-mat-in}
[{\cal T}^{\rm in}(p)]^{\bar C, \bar A\bar B}_{C,AB}
= \left(\begin{array}{ccc} 
t_{ji}^c \left(p_\gamma \delta_{\hat{b}\hat{a}}\right) &
0 & 
t^b_{ki}  
\left(-\frac12 \gamma^\beta_{\hat{c}\hat{a}}\right) \\
0 & 
(-t_{ij}^c)\, 
\left(p_\gamma \delta_{\hat{a}\hat{b}} \right) & 
t^b_{ik}
\left(-\frac12\gamma^\beta_{\hat{a}\hat{c}}\right) \\
t^a_{jk}  \,\frac12 
\big(\slashed{p}\gamma^\alpha\big)_{\hat{b}\hat{c}} & 
(-t^a_{kj})\,\frac12
\big(\gamma^\alpha\slashed{p}\big)_{\hat{c}\hat{b}} & 
i f^{bca} 
\left(p_\gamma \eta^{\beta\alpha}\right) 
\end{array}\right).
\end{equation}
To summarise, in Eq.~\eqref{emissionfac} we 
introduced a compact notation for the emission 
of a single soft parton with Lorentz/spinor 
and colour indices $C, \bar{C}$ off a hard 
incoming line with flavour $I$, space-time 
index $A$, and colour index $\bar{A}$.
This hard parton then continues with internal 
flavour index $J$, space-time index $B$ and 
colour index $\bar{B}$. The notation in 
Eq.~\eqref{emissionfac} is also illustrated 
in Fig.~\ref{fig:emissionfac}. We do not 
include an external wave-function for the 
emitted soft parton, leaving open the 
possibility of this parton being either 
real or virtual. Multiple soft emissions 
off a single hard parton can easily be 
described through repeated insertion of 
the transition matrix, where care of 
course has to be taken that the eikonal 
pre-factor $1/p\cdot k$ is updated 
accordingly. The final product of 
transition matrices may in the end 
then be sandwiched between a hard 
matrix element on one end, and the 
external wave function for the hard 
parton on the other. This will be 
demonstrated in Section~\ref{sec:multiple}.

One may perform the same exercise for a 
final-state emitter with hard momentum 
$p^\mu$. We may summarise the results as
\begin{equation}
[\mathcal{M}_{I}^{\rm out}(p)]_{C}^{\bar{C}} 
= \frac{g_s}{p\cdot k} \,
[\bar{\xi}_{I}(p)]^{\bar{A}}_{A} \,
[{\cal T}^{\rm out}_{IJ}(p)]^{\bar C,\bar A \bar B}_{C,AB}
\, [\mathcal{A}_{J}(p)]^{\bar{B}}_B\,,
\label{emissionfac-out}
\end{equation}
The outgoing 
wave function can be any of
\begin{equation}
 \left[\bar{\xi}_I\right]^{\bar{A}}_A=(\bar{u}^j_{\hat{a}},v^j_{\dot{a}},\epsilon^{\dagger a}_\alpha)\,.
  \label{wavefns-out}
\end{equation}
The transition matrix now describes 
soft emissions off hard outgoing partons, 
and can be written as
\begin{equation}\label{TransitionMatrOutDef}
[{\cal T}^{\rm out}(p)]^{\bar C,\bar A\bar B}_{C,AB}
= \left(\begin{array}{ccc} 
[{\cal T}_{qq}^{\rm out}(p)]^{c,ij}_{\gamma,\hat a\hat b} &
0 & 
[{\cal T}_{qg}^{\rm out}(p)]^{k,ib}_{\hat c,\hat a \beta} \\
0 & 
[{\cal T}_{\bar q \bar q}^{\rm out}(p)]^{c,ij}_{\gamma,\hat a \hat b} & 
[{\cal T}_{\bar q g}^{\rm out}(p)]^{k,ib}_{\hat c,\hat a \beta } \\
({\cal T}_{g q}^{\rm out}(p))^{k,aj}_{\hat c,\alpha \hat b} & 
[{\cal T}_{g\bar q}^{\rm out}(p)]^{k,aj}_{\hat c,\alpha\hat b} & 
[{\cal T}_{gg}^{\rm out}(p)]^{c,ab}_{\gamma,\alpha\beta}
\end{array}\right),
\end{equation}
Again, the colour factor follows the 
rules for eikonal emissions listed above.
Together with the kinematic factor 
$\left[Q^{\rm out}\right]_{C,AB}$ 
(written out explicitly in 
Appendix~\ref{app:kin-factors} 
for all cases) we then obtain 
\begin{equation}
\label{trans-mat-out}
[{\cal T}^{\rm out}(p)]^{\bar C, \bar A\bar B}_{C,AB}
= \left(\begin{array}{ccc} 
(-t_{ij}^c) \left(p_\gamma \delta_{\hat{b}\hat{a}}\right) &
0 & 
t^b_{ik}  
\left(\frac12 \gamma^\beta_{\hat{a}\hat{c}}\right) \\
0 & 
t_{ji}^c\, 
\left(p_\gamma \delta_{\hat{b}\hat{a}} \right) & 
t^b_{ki}
\left(\frac12\gamma^\beta_{\hat{c}\hat{a}}\right) \\
(-t^a_{kj})  \,\frac12 
\big(\gamma^\alpha\slashed{p}\big)_{\hat{c}\hat{b}} & 
(t^a_{jk})\,\frac12
\big(\slashed{p}\gamma^\alpha\big)_{\hat{b}\hat{c}} & 
i f^{acb} 
\left(p_\gamma \eta^{\alpha\beta}\right) 
\end{array}\right).
\end{equation}
It is easy to check that 
$[{\cal T}^{\rm out}(p)]^{\bar C, \bar A\bar B}_{C,AB} = 
[\overline{{\cal T}}^{\rm in}(p)]^{\bar C, \bar B\bar A}_{C,BA}$, as required. We may now work towards a generalisation of the conventional Wilson lines, that includes emission of any soft partons.
Let us start from the case of emission from 
an incoming line. In momentum space, the Feynman 
rule describing this process is given in 
Eq.~(\ref{emissionfac}), i.e.
\begin{equation}
\frac{g_s}{p\cdot k}\,
\left[{\cal T}^{\rm in}\right]^{\bar{C},\bar{A}\bar{B}}_{C,AB},
\label{feynrule}
\end{equation}
which is sandwiched between the external incoming 
wave function $\xi^{\bar{A}}_A$ and the stripped 
hard-scattering amplitude ${\cal A}^{\bar{B}}_B$. 
In position space this Feynman rule reads
\begin{align} 
ig_s
\int_{-\infty}^0\, {\rm d} t \,
\left[{\cal T}^{\rm in}\right]^{\bar{C},\bar{A}\bar{B}}_{C,AB}
\, \Theta^{\bar{C}}_C(tp) \, e^{\varepsilon t}\,.
\label{emissionfac2}
\end{align}
The integral is defined over the 
incoming particle contour (parameterised 
by $x^\mu = t p^\mu$), and the factor 
$e^{\varepsilon t}$ makes sure the integral 
converges. To see the equivalence between 
the two Feynman rules, we may write the 
field in momentum space via the Fourier 
transform
\begin{equation}
\Theta^{\bar{C}}_C(x)
=\int\frac{{\rm d}^d k}{(2\pi)^d}\,
\tilde{\Theta}^{\bar{C}}_C(k) \, e^{ik\cdot x}
=\int\frac{{\rm d}^d k}{(2\pi)^d}\,
\tilde{\Theta}^{\bar{C}}_C(k) \, e^{it (k\cdot p)}.
\label{eq:ThetaFT}
\end{equation}
Inserting this into Eq.~(\ref{emissionfac2}) 
yields
\begin{align}
ig_s \int_{-\infty}^0 {\rm d}t \, 
\left[{\cal T}^{\rm in}\right]^{\bar{C},\bar{A}\bar{B}}_{C,AB}
\int\frac{{\rm d}^d k}{(2\pi)^d}\tilde{\Theta}^{\bar{C}}_C(k)
\left[\frac{e^{it (k\cdot p) 
+ t \varepsilon}}{ik\cdot p + \varepsilon}\right]^0_{-\infty}
=\int\frac{{\rm d}^d k}{(2\pi)^d}\tilde{\Theta}^{\bar{C}}_C(k)
\left[\frac{g_s}{p\cdot k - i\varepsilon} \left[{\cal T}^{\rm in}\right]^{\bar{C},\bar{A}\bar{B}}_{C,AB}
\right],
\label{feynruleint}
\end{align}
where the lower limit of the $t$ integral
vanishes because of the $\varepsilon$ 
regulator. The contents of the square 
bracket is the required momentum-space 
Feynman rule, which indeed agrees with 
Eq.~(\ref{feynrule}). We may define 
Feynman rules for outgoing emissions 
analogously, sending $\varepsilon \to 
-\varepsilon$ to make the integral 
well-defined. To further clarify 
Eq.~(\ref{emissionfac2}), note that 
for the diagonal elements of 
$\left[{\cal T}^{\rm in}\right]^{\bar{C},\bar{A}\bar{B}}_{C,AB}$
(i.e.~soft gluon emission), we may 
simplify it to
\begin{align}
ig_s{\bf T}^{c} \int_{-\infty}^0\, {\rm d} t 
\,p^{\mu}\, A^{c}_{\mu}(tp)\,,
\end{align}
which is the usual exponent of the Wilson-line 
operator (see Eq.~\eqref{Phidef}), omitting the 
explicit identity matrices $\delta^{\hat{a}\hat{b}}$ 
and $\eta^{\alpha\beta}$ in spinor/Lorentz space. 
This is as it should be: the effect of a single 
gluon emission should be given by a Wilson line. 
However, this also suggests an interpretation of Eq.~(\ref{emissionfac2}): it acts as a generalised 
Wilson line exponent that describes the emission 
of soft (anti-)quarks in addition to gluons. That 
we can indeed interpret the object in 
Eq.~(\ref{emissionfac2}) in this way 
is discussed in the following section. 

\subsection{Multiple emissions from a 
generalised soft-emission operator} 
\label{sec:multiple}

\begin{figure}[t]
\begin{center}
    \scalebox{0.7}{\includegraphics{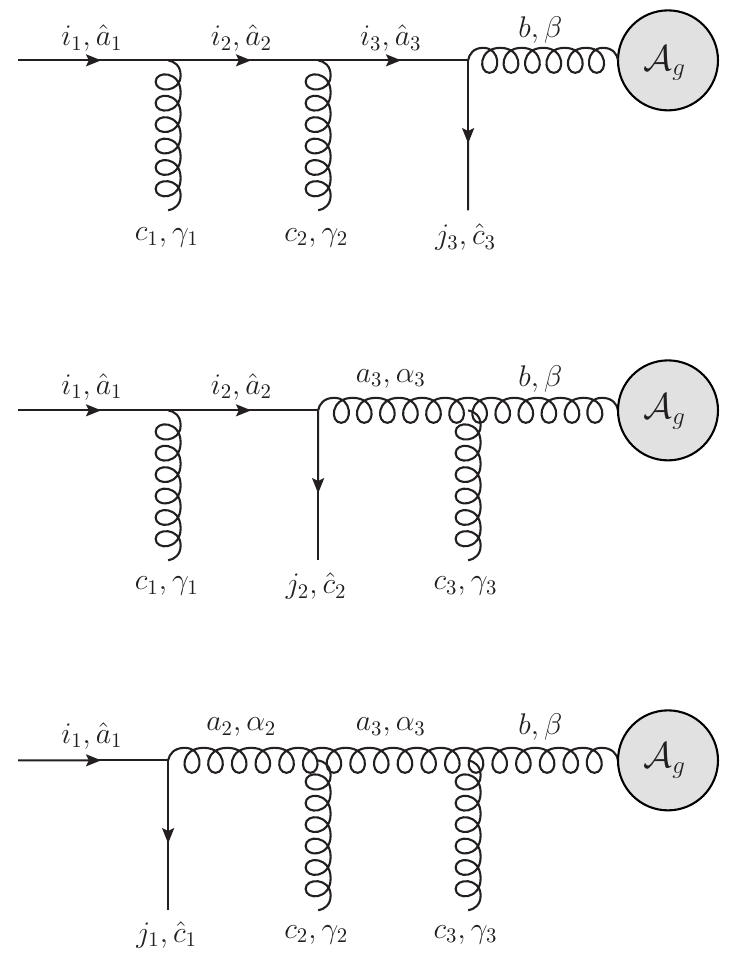}}
    \caption{Diagrams corresponding to the first three
    contributions in Eq. \eqref{qgthirdorderA}, originating 
    from expanding the generalised Wilson line in Eq.~\eqref{Fdef2}
    to third order in $g_s$, and selecting the components
    giving rise to a transition between an initial quark 
    and a gluon incoming into the amplitude ${\cal A}_g$.}
    \label{fig:emissions2a}
  \end{center}
\end{figure}

\begin{figure}[t]
\begin{center}
    \scalebox{0.7}{\includegraphics{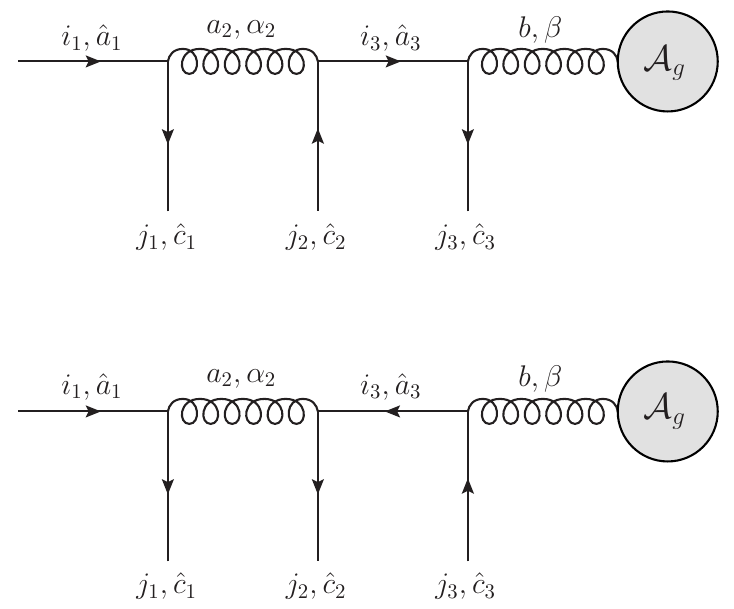}}
    \caption{Diagrams corresponding to the last two
    contributions in Eq. \eqref{qgthirdorderA}, originating 
    from expanding the generalised Wilson line in Eq.~\eqref{Fdef2}
    to third order in $g_s$, and selecting the components
    giving rise to a transition between an initial quark 
    and a gluon incoming into the amplitude ${\cal A}_g$
    These two diagrams are suppressed by a power of 
    $\lambda^2$ w.r.t. the first three diagrams in 
    \ref{fig:emissions2a}.}
    \label{fig:emissions2b}
  \end{center}
\end{figure}

The effect of multiple emissions off a single 
(possibly flavour-changing) hard line with 
momentum $p^\mu$ may be generated by the 
following path-ordered object:
\begin{equation}
{\cal F}^{\bar{A}\bar{B}}_{AB}
={\cal P}\exp\left[ig_s 
\int_{-\infty}^0 {\rm d} t\, 
\left[{\cal T}\right]^{\bar{C}}_{C} 
\,\Theta^{\bar{C}}_{C}(tp)
\right]^{\bar{A}\bar{B}}_{AB}.
\label{Fdef2}
\end{equation}
We will call this a \emph{generalised soft-emission 
operator}. It carries open indices and needs to 
be sandwiched in-between the matrix element 
that the hard line originated from, and its 
external state. The notation of Eq.~\eqref{Fdef2} is appropriate for an incoming 
hard line, and the regularisation is 
implicitly understood. The soft-emission operator
for an outgoing hard line is easily obtained 
by switching the integration boundaries to 
$(0,\infty)$, and adapting the regularisation 
accordingly. To see that the object defined 
in Eq.~\eqref{Fdef2} indeed defines multiple 
insertions of the transition matrix, we may 
expand it and examine the $\mathcal{O}(g_s^l)$ 
contribution, which is given by
\begin{equation}
{\cal F}^{\bar{A}\bar{B}}_{AB}\Big|_{\mathcal{O}(g_s)^l}
= (ig_s)^l \int_{-\infty}^0 {\rm d} t_1 \, 
\left[{\cal T}\right]^{\bar{C}_1,\bar{A}\bar{A}_1}_{C_1, AA_1} 
\Theta^{\bar{C}_1}_{C_1}(t_1 p)\,\,  \dots \,\,  
\int_{t_{l-1}}^0 {\rm d} t_l \, 
\left[{\cal T}\right]^{\bar{C}_l,\bar{A}_{l-1}
\bar{B}}_{C_l, A_{l-1}B} \Theta^{\bar{C}_l}_{C_l}(t_l p)\,.
\label{Fexpansion}
\end{equation}
Writing again the field in momentum space 
through Eq.~\eqref{eq:ThetaFT}, we end up 
with a string of integrals of the form
\begin{equation}\label{multiple}
\int_{-\infty}^0 {\rm d}t_1 \int_{t_1}^0 {\rm d}t_2 
\,\, \dots \,\, \int_{t_{l-1}}^0 {\rm d}t_l \, 
\exp\left[i\sum_{j=1}^l (k_j \cdot p) t_j \right] 
= \frac{(-i)^l}{k_1 \cdot p \, (k_1 + k_2)\cdot p \, 
\dots \, (k_1 + \dots + k_l)\cdot p}\,,
\end{equation}
where $k_j$ is the four-momentum of each 
of the soft fields. With this, the full 
soft-emission operator at $\mathcal{O}(g_s^l)$ 
becomes
\begin{equation}
{\cal F}^{\bar{A}\bar{B}}_{AB}\Big|_{\mathcal{O}(g_s)^l}
= \prod_{j=1}^l \left[\int \frac{{\rm d}^dk_j}{(2\pi)^d}
\widetilde{\Theta}^{\bar{C}_l}_{C}(k_j) \right] 
\frac{g_s^l}{k_1 \cdot p \, \dots \, (k_1 + \dots k_l)\cdot p} 
\left[{\cal T}\right]^{\bar{C}_1,\bar{A}\bar{A}_1}_{C_1, AA_1} 
\, \dots \, \left[{\cal T}\right]^{\bar{C}_l,
\bar{A}_{l-1}\bar{B}}_{C_l, A_{l-1}B} \,.
\label{FexpansionMomSpace}
\end{equation}
We may now see how one may use this 
object to generate soft emissions.
For instance, we can consider a 
configuration involving an incoming 
quark with colour label $i_1$ and 
spinor label $\hat{a}_1$, transitioning 
into a gluon of colour index $b$, 
and space-time index $\beta$, entering 
into the hard scattering ${\cal A}_g$, 
at third order in the strong coupling 
$g_s$. The expansion of the generalised
Wilson line up to third order
automatically generates all 
contributions given in 
Figs.~\ref{fig:emissions2a} 
and~\ref{fig:emissions2b}, 
given that the transitions 
operators ${\cal T}_{IJ}$
are matrices in the flavour 
space $IJ$, and matrix multiplication
generates all possible transitions. 
We obtain 
\begin{align}\label{qgthirdorderA}
\frac{g_s^3}{p\cdot k_1 \, p \cdot (k_1 + k_2) 
\, p \cdot (k_1 + k_2 + k_3)}  u^{i_1}_{\hat{a}_1}(p) 
\Big\{ 
\left[{\cal T}^{{\rm in}}_{qq}(p)
\right]^{c_1, i_1i_2}_{\gamma_1, \hat{a}_1 \hat{a}_2} \, 
\left[{\cal T}^{\rm in}_{qq}(p)
\right]^{c_2, i_2i_3}_{\hat{c}_2, \hat{a}_2 \hat{a}_3} \, 
\left[{\cal T}^{\rm in}_{qg}(p)
\right]^{j_3, i_3 b}_{\hat{c}_3,  \hat{a}_3 \beta} & \\ 
\nonumber
&\hspace{-8.5cm}
+\,\left[{\cal T}^{{\rm in}}_{qq}(p)
\right]^{c_1, i_1 i_2}_{\gamma_1, \hat{a}_1 \hat{a}_2} \, 
\left[{\cal T}^{\rm in}_{qg}(p)
\right]^{j_2, i_2 a_3}_{\hat{c}_2, \hat{a}_2 \alpha_3} \, 
\left[{\cal T}^{\rm in}_{gg}(p)
\right]^{c_3, a_3 b}_{\gamma_3, \alpha_3 \beta} \\
\nonumber
&\hspace{-8.5cm}
+\,\left[{\cal T}^{{\rm in}}_{qg}(p)
\right]^{j_1, i_1 a_2}_{\hat{c}_1, \hat{a}_1 \alpha_2} \, 
\left[{\cal T}^{\rm in}_{gg}(p)
\right]^{c_2, a_2 a_3}_{\gamma_2, \alpha_2\alpha_3} \, 
\left[{\cal T}^{\rm in}_{gg}(p)
\right]^{c_3, a_3 b}_{\gamma_3, \alpha_3 \beta} \\ 
\nonumber
&\hspace{-8.5cm}
+\,\left[{\cal T}^{{\rm in}}_{qg}(p)
\right]^{j_1, i_1a_2}_{\hat{c}_1, \hat{a}_1 \alpha_2} \, 
\left[{\cal T}^{\rm in}_{gq}(p)
\right]^{j_2, a_2i_3}_{\hat{c}_2, \alpha_2\hat{a}_3} \, 
\left[{\cal T}^{\rm in}_{qg}(p)
\right]^{j_3, i_3 b}_{\hat{c}_3, \hat{a}_3 \beta} \\ 
&\hspace{-8.5cm}
+\,\left[{\cal T}^{{\rm in}}_{qg}(p)
\right]^{j_1, i_1a_2}_{\hat{c}_1, \hat{a}_1 \alpha_2} \, 
\left[{\cal T}^{\rm in}_{g\bar{q}}(p)
\right]^{j_2, a_2i_3}_{\hat{c}_2, \alpha_2\hat{a}_3} \, 
\left[{\cal T}^{\rm in}_{\bar{q}g}(p)
\right]^{j_3, i_3 b}_{\hat{c}_3, \hat{a}_3 \beta}
\Big\}\left[{\cal A}_g(p)\right]^{b}_{\beta} \, .  \nonumber
\end{align}
Filling in the transition-matrix elements we get
\begin{align}\label{qgthirdorderB} \nonumber
\frac{g_s^3}{p\cdot k_1 \, p \cdot (k_1 + k_2) 
\, p \cdot (k_1 + k_2 + k_3)} 
\Big\{ 
-\,\frac{1}{2} [t^bt^{c_2}t^{c_1}]_{j_3 i_1} 
p_{\gamma_1} p_{\gamma_2} 
[\gamma^{\beta} u(p)]^{i_1}_{\hat c_3} & \\ 
\nonumber
&\hspace{-6.0cm}
-\,\frac{i}{2} f^{bc_3a_3} [t^{a_3}t^{c_1}]_{j_2 i_1} 
p_{\gamma_1} p_{\gamma_3} 
[\gamma^{\beta} u(p)]^{i_1}_{\hat c_2} & \\ 
\nonumber
&\hspace{-6.0cm}
+\,\frac{1}{2} f^{bc_3a_3} f^{a_3c_2a_2} 
[t^{a_2}]_{j_1 i_1} p_{\gamma_2} p_{\gamma_3} 
[\gamma^{\beta} u(p)]^{i_1}_{\hat c_1} & \\ 
\nonumber
&\hspace{-6.0cm}
+\, \frac{1}{4} [t^{a_2}]_{j_1 i_1} 
[t^{a_2}t^{b}]_{j_3 j_2} 
[\gamma^{\beta} \slashed p \gamma^{\alpha_2}]_{\hat c_3\hat c_2}
[\gamma_{\alpha_2} u(p)]^{i_1}_{\hat c_1} \\ 
&\hspace{-6.0cm}
-\,\frac{1}{4} [t^{a_2}]_{j_1 i_1} 
[t^{a_2}t^{b}]_{j_2 j_3} 
[\gamma^{\alpha_2} \slashed p \gamma^{\beta}]_{\hat c_2\hat c_3}
[\gamma_{\alpha_2} u(p)]^{i_1}_{\hat c_1}
\Big\} \left[{\cal A}_g(p)\right]^{b}_{\beta} \, . 
\end{align}
Each line correctly reproduces the corresponding 
diagrams in Figs.~\ref{fig:emissions2a} 
and~\ref{fig:emissions2b}, at the lowest order 
at which each diagram contributes within the soft 
expansion. In this respect, it is important to 
highlight that the diagrams in 
Figs.~\ref{fig:emissions2a} and~\ref{fig:emissions2b} 
contribute at different orders in the soft expansion, 
although this is not apparent from the expressions 
in Eq. \eqref{qgthirdorderB}. The correct power 
counting emerges when inserting the expressions 
above in a full squared matrix element. In the virtual case,
the soft gluons/quarks are contracted to produce a soft propagator,
whereas in the real-emission case, the sum over physical polarisation/spinor
degrees-of-freedom produces additional terms.
In both cases, the contribution of a 
soft quark produces an additional factor of 
the soft momentum $k_i$ w.r.t.~the emission 
of a soft gluon, i.e., it is power suppressed. 
Thus, in the example at hand the three 
diagrams in Fig.~\ref{fig:emissions2a}
involve the emission of a soft quark, 
so contribute at NLP compared to 
diagrams where only soft gluons 
appear. The remaining two 
diagrams in Fig.~\ref{fig:emissions2b}
involve the emission of three soft 
quarks, so contribute at N$^3$LP
in a squared matrix element. This 
observation can be encoded directly 
in the matrices of Eqs.~\eqref{trans-mat-in}
and \eqref{trans-mat-out}, by noticing that 
the off-diagonal elements give rise to 
power-suppressed contributions.
In this way one has a definite 
power counting, that can be used 
to determine at which order in the 
power expansion a given term in 
Eq.~\eqref{FexpansionMomSpace} will 
contribute.

As a last comment, note that, by 
taking only the diagonal elements of 
${\cal T}$, we directly see that the 
generalised soft-emission operator 
defined in Eq.~\eqref{Fdef2} is equal 
to the standard Wilson line. However, 
one must bear in mind that although the 
normal Wilson line is an object that 
transforms covariantly under the gauge 
group, our soft-emission operator does 
not straightforwardly share this property. Whilst it may be possible to make rigorous the gauge-transformation properties of the generalised soft emission operator (e.g.~by exploiting the property that it lives in a reducible representation of the gauge group), this is unnecessary for our purposes. Instead, we can simply consider our operator as a convenient book-keeping device, 
casting soft-quark emissions in the same 
framework as soft-gluon emissions, thereby 
allowing us to use the replica trick.  

\subsection{The replica trick for emissions of soft partons}
\label{sec:replica-soft}

\begin{figure}[t]
\begin{center}
    \includegraphics[width=0.35\textwidth]{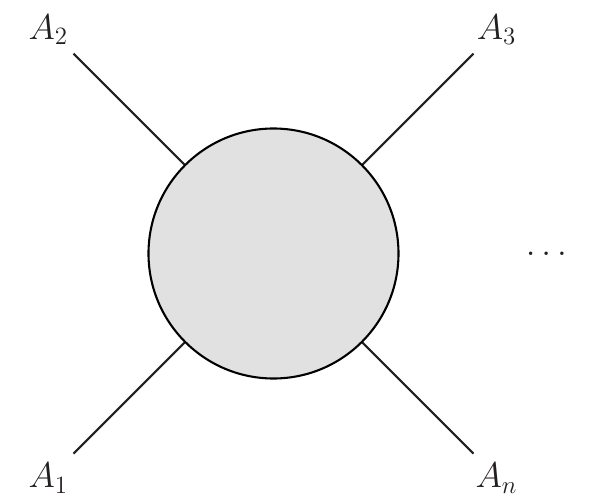}
    \caption{Amplitude with $n$ external parton legs, 
    each of which carries an abstract index $A_i$ 
    in species space.}
    \label{fig:ampN}
\end{center}
\end{figure}
Let us consider a scattering process with 
$n$ external parton legs, carrying hard 
momenta $\{p_i\}$. Each external line $i$ 
will carry an index $A_i$ in the flavour/spin space that has been introduced above. 
Following Fig.~\ref{fig:ampN} 
(c.f. Fig~\ref{fig:emissionfac}), we label 
the amplitude with external wavefunctions 
removed as ${\cal A}^{\bar{A}_1\ldots\bar{A}_n}_{A_1\ldots A_n}$, where barred indices are colour indices as 
before. The full hard-scattering amplitude 
then takes the form
\begin{equation}
{\cal A}(\{p_i\})=
{\cal A}^{\bar{A}_1\ldots\bar{A}_n}_{A_1\ldots A_n}(\{p_i\}) \prod_{i=1}^n \xi^{\bar{A}_i}_{A_i}(p_i)\,,
\end{equation}
where we again use $\xi^{\bar{A}}_{A}$ to 
denote the external wave function for a 
particle with species index $A$ and colour 
index $\bar{A}$. We may then describe the 
emission of any number of virtual soft 
partons from the external lines by dressing 
each line of the stripped amplitude with 
the generalised soft-emission operator of
Eq.~(\ref{Fdef2}), such that this is
sandwiched between the amplitude and the 
external wavefunctions. That is, we may 
define a generalised (and flavour-changing) 
soft function
\begin{equation}
{\cal A}(\{p_i\}) =
\left(\prod_{i=1}^n \xi^{\bar{A}_i}_{A_i}(p_i)\right) 
{\cal S}^{\bar{A}_1\ldots \bar{A}_n,\bar{B}_1\ldots 
\bar{B}_n}_{A_1\ldots A_n,B_1\ldots B_n}\,
{\cal A}^{\bar{B}_1\ldots \bar{B}_n}_{B_1 \ldots B_n}\,,
\label{Sfdef}
\end{equation}
where now ${\cal A}$ is the full amplitude 
dressed by soft parton emission, and the 
generalised soft function is given by
\begin{equation}
{\cal S}^{\bar{A}_1\ldots\bar{A}_n,\bar{B}_1
\ldots\bar{B}_n}_{A_1\ldots A_n,B_1\ldots B_n}
=\Big\langle 0\Big|\prod_{i=1}^n
{\cal F}^{\bar{A}_i\bar{B}_i}_{A_iB_i}(p_i)
\Big|0\Big\rangle\,.
\label{Sfdef2}
\end{equation}
That is, the soft function is defined 
similarly to the pure gluon emission 
case, by a VEV of generalised soft-emission 
operators acting along each incoming particle 
contour. We may note that it is matrix-valued 
in the tensor product space of the flavour/spin 
and colour indices associated with each hard line. 
Note that this is directly analogous to how the 
usual soft function is matrix-valued in the 
tensor product space of colour indices, so 
the definition in Eq.~\eqref{Sfdef2} presents 
no significant additional complication from 
the conceptual point of view. As in 
Eq.~(\ref{pathint}), we may write a path-integral 
formula for the definition of Eq.~(\ref{Sfdef2}). 
Let us first introduce the compact notation
\begin{equation}
{\cal D}\Theta^{\bar{C}}_C \equiv 
{\cal D}\psi^i\, {\cal D}\bar{\psi}^j\, 
{\cal D}A_\mu^a\,,
\label{DTheta}
\end{equation}
i.e.~the path-integral over the individual 
parton fields appearing in Eq.~(\ref{Thetadef}) 
can be simply written as a single measure involving 
the array of fields $\Theta_C^{\bar{C}}$. Then
Eq.~(\ref{Sfdef2}) may be written as
\begin{equation}
{\cal S}_{A_1 \ldots A_n,B_1\ldots 
B_n}^{\bar{A}_1\ldots \bar{A}_n,\bar{B}_1\ldots\bar{B}_n}
=\int{\cal D}\Theta^{\bar{C}}_C\, \left(\prod_{i=1}^n 
{\cal F}^{\bar{A}_i\bar{B}_i}_{A_i B_i}(p_i)\right)
e^{iS[\Theta^{\bar{C}}_C]}\,,
\label{pathintf}
\end{equation}
which is similar in form to the pure gluon case of Eq.~(\ref{pathint}), and  
\begin{align}
S[\Theta^{\bar{C}}_C] = 
\int {\rm d}^d x\, \mathcal{L}_{\rm QCD}\,
\end{align}
is the QCD action.
Consequently, the replica trick argument 
for exponentiation of arbitrary soft parton 
corrections proceeds by direct analogy to 
the soft gluon case. Carrying out the path 
integral over the compound soft parton field 
$\Theta^{\bar{C}}_C$ generates all possible 
Feynman diagrams in which the external hard 
partons are connected by soft partons. We 
may then introduce $N$ identical replicas 
of the original soft parton field, such 
that Eq.~(\ref{SNdef2}) is replaced by  
\begin{align} \label{SNdef3}
\left[{\cal S}_N\right]_{A_1 \ldots A_n,B_1\ldots 
B_n}^{\bar{A}_1\ldots \bar{A}_n,\bar{B}_1\ldots\bar{B}_n}
&= \int{\cal D}[\Theta^{(1)}]^{\bar{C}}_C 
\ldots {\cal D}[\Theta^{(n)}]^{\bar{C}}_C \,
\exp\left\{i\sum_{j=1}^N S\Big[
[\Theta^{(j)}]^{\bar{C}}_C\Big]\right\} \\
&\quad \times\, \prod_{k=1}^n {\cal R}
\left\{{\cal P}\exp
\left(ig_s{\bf T}^{\bar{C}}\,
\sum_{j=1}^N \int {\rm d}t_k \,
{\bf Q}^C[\Theta^{(j)}]^{\bar{C}}_C(t_kp_k)\right)
\right\}^{\bar{A}_i\bar{B}_i}_{A_i B_i}. \notag
\end{align}
There is now a product of generalised 
soft-emission operators on each external 
leg, associated with the different soft 
parton replicas that may be emitted. As 
before, there is a replica-ordering 
operator, which in any given diagram 
acts to reorder parton emissions as 
necessary. We must bear in mind, 
however, that the action of the 
replica-ordering operator ${\cal R}$ 
in Eq.~(\ref{SNdef3}) is more complicated 
than for pure soft gluon emission. In the 
present case, the replica-ordering operator 
acts on the full transition matrix 
$[{\cal T}_{IJ}]_{C,AB}^{\bar{C},\bar{A}\bar{B}}$, 
instead of only on the colour matrices as 
was the case for soft-gluon emissions. 
This leads us to introduce 
$[{\cal T}_{N,IJ}(D)]_{C,AB}^{\bar{C},\bar{A}\bar{B}}$, 
which is the transition matrix of a replicated 
theory. This matrix will determine both the 
colour factor and the kinematic numerator 
of a given diagram, whereas the denominator 
is untouched. To see this, let us decompose 
the kinematic part of a given soft diagram 
$D$ in momentum space as
\begin{equation}
{\cal K}(D)=\frac{{\cal N}(D)}{{\cal D}(D)}\,,
\label{FDdecomp}
\end{equation}
where ${\cal N}(D)$ is a kinematic numerator 
factor, which results from combining the 
emission factors $\{Q^{C,AB}\}$ of the transition 
functions for each emitted soft parton, as well 
as coupling factors, external wavefunctions etc. 
The denominator ${\cal D}$ arises from performing 
the integrals over the distance parameters 
$\{t_k\}$, which results in a product of nested 
momentum factors as exemplified in 
Eq.~(\ref{multiple}). While the numerator 
depends on the spinor structure of the diagrams, 
the denominator only depends on the kinematic 
assignment. If only soft gluons are emitted, 
the kinematic numerator for a diagram $D$ with 
$n_j$ soft emissions on a given line $j$ takes 
the simple form
\begin{displaymath}
{\cal N}(D)\propto \prod_{j} 
p^{\mu_1} p^{\mu_2}\ldots p^{\mu_j}\,.
\end{displaymath}
That is, there is a simple factor of the 
hard momentum for each emission, such that 
the emission factors for individual emissions 
commute with each other, and the numerator 
for {\it any} diagram with the same number 
of emissions on each line is the same. This 
indeed follows straightforwardly from the 
conventional Wilson line definition of 
Eq.~(\ref{Phidef}), where the only 
non-commuting quantity in the exponent 
is the colour generator ${\bf T}_i$, 
and it is for this reason that the 
replica ordering operator ${\cal R}$ 
acts to modify the colour factors of 
graphs $C(D)\rightarrow \tilde{C}(D)$, 
but not their kinematics. This ceases 
to be the case for the generalised soft emission 
operator of Eq.~\eqref{Fdef2}. In this 
case the emission factors are not mutually 
commuting, and so the action of ${\cal R}$ 
is to reorder them, in addition to the 
colour generators. In the replicated 
theory, we denote by ${\cal N}_N(D)$ 
the kinematic numerator of diagram $D$, 
such that its full contribution is
\begin{equation}
C_N(D)\, {\cal K}_N(D)
=\frac{{\cal N}_N(D)\,C_N(D)}{{\cal D}(D)}\,,
\label{FDN}
\end{equation}
where the replicated colour factor 
$C_N(D)$ is defined as in the case 
of pure gluon emission. Despite this 
additional complication, the replica 
trick argument proceeds as before, 
where the logarithm of the soft 
amplitude can be directly obtained 
by taking the ${\cal O}(N)$ part of 
each soft diagram. Defining 
exponentiated numerators and 
colour factors via
\begin{equation}
\tilde{{\cal N}}(D)=
{\cal N}_N(D)\Big|_{{\cal O}(N)}\,,
\quad
\tilde{C}(D)=C_N(D)\Big|_{{\cal O}(N)}\,,
\label{ECN}
\end{equation}
we may write the contribution of a 
given soft diagram to the logarithm 
of the amplitude as
\begin{equation}
\tilde{E}(D)=\frac{\tilde{\cal N}(D)\,
\tilde{C}(D)}{{\cal D}(D)}\,.
\label{ED}
\end{equation}
An exponentiated form of the soft 
function is then simply found as
\begin{align}\label{SExpDef}
\mathcal{S} = 
{\rm exp}\left[\sum_{D} \tilde{E}(D) \right],
\end{align}
where $\tilde{E}(D)$ are the generalised webs. 

\begin{figure}[t]
\begin{center}
    \scalebox{0.6}{\includegraphics{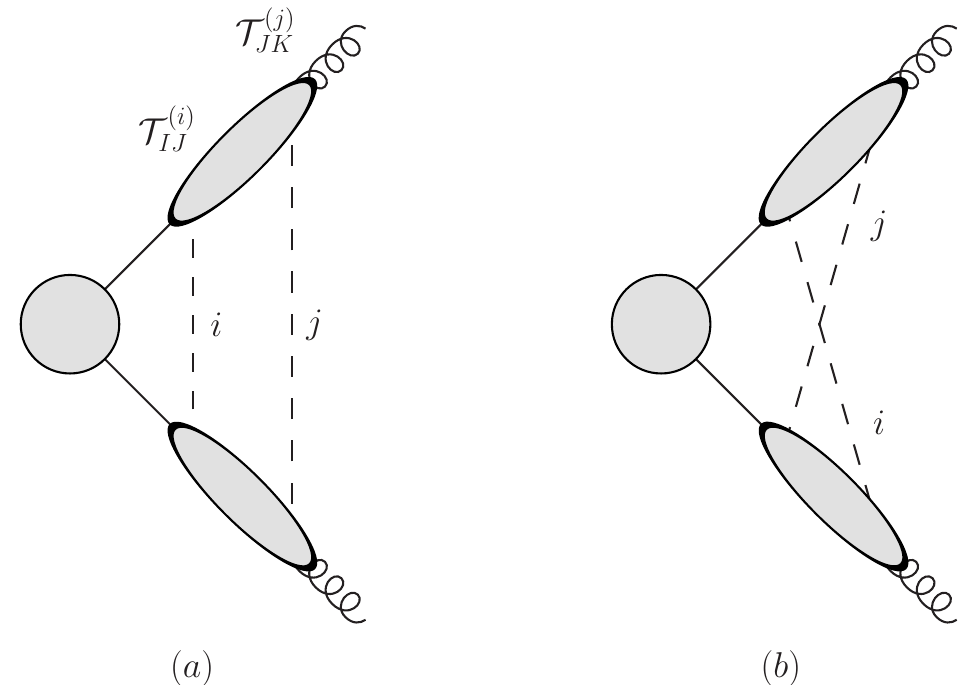}}
    \caption{Diagrams (a) and (b) constitute a 
    generalization at NLP of the web in 
    Fig.~\ref{fig:QCDreplicas}.}
    \label{fig:replicaNLP}
\end{center}
\end{figure}
As in the pure gluon emission case, the replica 
trick provides a completely systematic way of 
calculating the logarithm of the soft amplitude 
directly, rather than the amplitude itself. 
However, the abstract nature of the generalised 
soft function of Eq.~(\ref{SNdef3}), and the 
exponentiated numerators appearing in 
Eq.~(\ref{ED}), will doubtless look rather 
baffling to the reader. To explain our notation, 
let us consider an example. We take an amplitude 
with external partons $q(p_1)$ and $\bar{q}(p_2)$, 
which define two hard outgoing legs. At  
one-loop level, these fermions can either remain 
the same flavour (LP), or transition into gluons 
(NLP). When allowing for two loops, we may either 
consider two gluon exchanges (LP), one gluon and 
one fermion exchange (NLP), or two fermionic 
exchanges (NNLP). For this example we focus 
on the NLP contribution to the two-loop diagram, 
meaning that the external particles will have 
to be gluons. We thus consider two 
insertions of the replicated transition matrix 
on each leg, as shown in Fig.~\ref{fig:replicaNLP}. 
For diagram (a) we denote the ordering of 
transition matrices as
\begin{align}
\left[{\cal T}^{(i)}(p_1)\right]^{\bar{C}_1, \bar{A}_1\bar{a}_1}_{C_1, A_1a_1}
\left[{\cal T}^{(j)}(p_1)\right]^{\bar{C}_2, \bar{a}_1\bar{B}_1}_{C_2, a_1B_1} 
\left[{\cal T}^{(i)}(p_2)\right]^{\bar{C}_1, \bar{A}_2\bar{a}_2}_{C_1, A_2a_2}
\left[{\cal T}^{(j)}(p_2)\right]^{\bar{C}_2, \bar{a}_2\bar{B}_2}_{C_2, a_2B_2}\,.
\label{eq:trans-ord-diagrama}
\end{align}
This object is sandwiched 
in between two external wave functions 
$\epsilon^{\dagger}_{B_1}(p_1)$ and 
$\epsilon^{\dagger}_{B_2}(p_2)$, and 
a hard-scattering matrix $\mathcal{A}
(p_1,p_2)_{A_1 A_2}^{\bar{A}_1\bar{A}_2}$. 
In Eq.~\eqref{eq:trans-ord-diagrama}, 
lower-case letters correspond to intermediate particles on the hard lines. The replica-reordering operator 
leaves the order the same for all choices 
of $i,j$. This is not true for diagram (b), 
for which the ordering of transition 
matrices is 
\begin{align}
{\cal T}^{(i)}(p_1)
{\cal T}^{(j)}(p_1)
{\cal T}^{(j)}(p_2)
{\cal T}^{(i)}(p_2)\,.
\label{eq:trans-ord-diagramb}
\end{align}
Here we leave the indices implicit for 
ease of notation. The reordering acts 
as follows:
\begin{align}
{\cal R}\left[
{\cal T}^{(i)}(p_1)
{\cal T}^{(j)}(p_1)
{\cal T}^{(j)}(p_2)
{\cal T}^{(i)}(p_2)\right]= \begin{cases}
{\cal T}^{(i)}(p_1)
{\cal T}^{(j)}(p_1)
{\cal T}^{(i)}(p_2)
{\cal T}^{(j)}(p_2) & \, i < j \quad \left(\frac{N(N-1)}{2}\right), \\
{\cal T}^{(j)}(p_1)
{\cal T}^{(i)}(p_1)
{\cal T}^{(j)}(p_2)
{\cal T}^{(i)}(p_2) & \, j < i \quad \left(\frac{N(N-1)}{2}\right),  \\
{\cal T}^{(j)}(p_1)
{\cal T}^{(j)}(p_1)
{\cal T}^{(j)}(p_2)
{\cal T}^{(i)}(p_2) & \, i = j \quad (N)\,,  \\ \end{cases}
\end{align}
where we have also denoted the multiplicity of each replica assignment in the round brackets. 
In direct analogy with the soft-gluon case, 
we see that the first two expressions 
correspond to the ordering for diagram (a), 
whereas the last entry is the ordering for 
diagram (b). When combined with the external 
wave functions, we can write the contribution 
of diagram (b) to the logarithm of the 
generalised soft function as 
\begin{align}\label{LogEb}
\widetilde{E}(b) = 
\frac{{\cal N}(b)C(b)-{\cal N}(a)C(a)}{{\cal D}(b)}\,.
\end{align}
The projection onto external wave functions 
and the hard-scattering amplitude will turn 
the matrix notation into a sum expression, 
and will contract the open indices of the 
transition matrices. 

So far we have discussed the use of the replica trick for virtual corrections only. Let us now generalise this to include real emissions. 

\subsection{The replica trick for real emissions}
\label{sec:replicareal}

The starting point for using the replica trick in the case of virtual exchanges is to consider a generating functional for virtual soft parton diagrams (Eq.~(\ref{pathintf}) in the general case). In the replicated theory, the corresponding functional is simply given by the original one raised to a power, as in Eq.~(\ref{SNdef}). It is this fact that allows one to conclude that the ${\cal O}(N)$
contribution to diagrams in the replicated theory formally
exponentiates. To include real emissions, we must therefore identify a
suitable generating functional, such that the appropriate analogue of Eq.~(\ref{SNdef}) can be applied. To this end, we simplify our arguments for the moment by considering a toy model of a single scalar field $\phi$, and a generating functional of the form
\begin{equation}
  Z=\int{\cal D}\phi\,\Psi[\phi] e^{iS[\phi]}\,,
  \label{Zdef}
\end{equation}
where $\Psi[\phi]$ is some function of the field that is the analogue
of the product of generalised soft emission operators in Eq.~(\ref{pathintf}). As is well-known,
functionals such as those of Eqs.~(\ref{pathintf}, \ref{Zdef}) generate
all possible Green's functions, or vacuum expectation values of
(time-ordered) products of fields. Usually in scalar field theory, one
introduces a current $J$ conjugate to $\phi$, such that
differentiation with respect to this current can be used to pick out
Green's functions of arbitrary multiplicity. Here, we will instead use
the alternative approach of Ref.~\cite{Siegel:1999ew}. First, we may separate the
action into its free-field part and an interaction term
\begin{equation}
  S=S_0+S_I=\int {\rm d}^d x\, \left(\frac{1}{2}\phi K\phi+S_I\right)\,.
  \label{Ssum}
\end{equation}
Here $K$ is the operator appearing in the quadratic term in the
Lagrangian. To carry out the path integral, one may use the
integral identity
\begin{equation}
\int {\rm d}u \, e^{-uMu/2}f(u+v) \, \sim \, e^{\partial_v M^{-1}\partial_v/2}f(v)\,,
\label{intid}
\end{equation}
where we have neglected an overall factor. By applying the appropriate
functional generalisation of this around $v=0$, the path integral in
Eq.~(\ref{Zdef}) can be carried out (after substituting
Eq.~(\ref{Ssum})) to give
\begin{equation}
  Z=\left.\exp\left[-\int {\rm d}^d x\,{\rm d}^d x'\,\frac{1}{2}\frac{\delta}
    {\delta \phi(x)}\Delta(x-x')\frac{\delta}{\delta \phi(x')}\right]
  e^{iS_I[\phi]}\Psi[\phi]\right|_{\phi=0}\,.
    \label{Zform}
\end{equation}
Here $\Delta(x-x')$ is the Feynman propagator, corresponding (up to a
numerical factor) with the inverse of the operator $K$. This is a less
standard form of the generating functional, but nevertheless generates
the usual Feynman diagram expansion. To see this, note that Taylor
expanding the factors on the right-hand side yields a sum of products
of the field $\phi$, where these come manifestly from interaction
vertices of the theory. The latter may be contained in the interaction
Lagrangian appearing in $S_I$, or the function $\Psi[\phi]$. In the
case of soft parton emission, this corresponds to vertices off or on
the hard particle lines respectively. The prefactor in Eq.~(\ref{Zform})
takes pairs of fields appearing in each term, and strips them off in
favour of a propagator. After setting $\phi\rightarrow 0$, one is left
with a sum of contributions in which interaction vertices are joined
by propagators. These are the usual Feynman diagrams, and the
exponential dependence in the prefactor turns out to be precisely such
as to yield the required combinatorial factors associated with the
usual Feynman rules. Examples can be found in
Ref.~\cite{Siegel:1999ew} for conventional scalar field theory,
although we have here modified the argument slightly so that it
provides a scalar analogue of Eq.~(\ref{pathintf}). All Feynman diagrams
will then correspond to hard particle lines which are connected by
conventional propagators and interaction vertices. That is,
Eq.~(\ref{Zform}) generates virtual diagrams only, as expected from
the fact that the original generating function of Eq.~(\ref{Zform}) is
a vacuum-to-vacuum transition amplitude.

So much for virtual corrections. Next, we can include the effect of
real corrections, and before doing so we note that the exponential
prefactor appearing in Eq.~(\ref{Zform}) can be written in momentum
space as~\cite{Siegel:1999ew}
\begin{equation}
  \exp\left[-\int {\rm d}^d x\,{\rm d}^d x' \, \frac{1}{2}\frac{\delta}
    {\delta \phi(x)}\Delta(x-x')\frac{\delta}{\delta \phi(x')}\right]
  =\exp\left[
    \int \frac{{\rm d}^d p}{(2\pi)^d}
    \frac{1}{2}\frac{\delta}{\delta \phi(p)}\Delta(p)
    \frac{\delta}{\delta\phi(-p)}
    \right],
  \label{expmomspace}
\end{equation}
where we have used the same symbol for the field and propagator in
position or momentum space, but where the argument resolves any
ambiguity. To include real corrections, we may consider the cut
propagator
\begin{equation}
  \Delta_+(p)=2\pi\Theta(p^0)\delta\left[(p^2+m^2)\right].
  \label{Delta_+}
\end{equation}
Then, a generating functional for the squared amplitude is
\begin{equation}
  Z_{\rm sq}=Z\exp\left[-\int\frac{{\rm d}^d p}{(2\pi)^d}
    \frac12 \frac{\overleftarrow{\delta}}{\delta\phi(p)}
    \Delta_+(p)\frac{\overrightarrow{\delta}}{\phi(-p)}
    \right] Z^\dag.
  \label{Zsq}
\end{equation}
Here the arrows on the derivatives indicate that they act to the left
or right, and we have introduced the generating functionals for
virtual diagrams $Z$ and $Z^\dag$ on either side of the final-state
cut. By direct analogy with Eqs.~(\ref{Zform}, \ref{expmomspace}), the
factor in the middle acts to strip off pairs of fields, and replace
them with a propagator. Now, however, this is a cut propagator, and
connects fields on different sides of the cut, as is required for a
real particle. We have here considered the simplified case of a scalar field to
simplify our arguments. However, it is straightforward to generalise
this to the soft parton case. Let ${\cal S}$ be the (virtual) soft function of Eq.~(\ref{pathintf}) with colour and species indices suppressed. A suitable generating functional for the squared amplitude is
\begin{equation}
  Z_{{\rm sq}}={\cal S}
  \exp\left[
    -\int\frac{{\rm d}^d p}{(2\pi)^d}\frac12
    \frac{\overleftarrow{\delta}}{\delta \Theta^{\bar{C}}_C(p)}
    \Delta_{+\phantom{,}CD}^{\phantom{+,}\bar{C}\bar{D}}(p)\frac{\overrightarrow{\delta}}
    {\delta \bar{\Theta}^{\bar{D}}_D(-p)}
    \right]  {\cal S}^\dag,
  \label{Zsqparton}
\end{equation}
where we have been careful to note that the functional differentiation acting to the right involves the conjugate field to that acting to the left.%
\footnote{
The Keldysh formalism 
\cite{Schwinger:1960qe,Keldysh:1964ud} provides 
provide an alternative approach to arrive at an equivalent form of Eq.~\eqref{Zsqparton} (also see Appendix~C of Ref.~\cite{Becher:2007ty} for an application 
to the Drell-Yan Soft function).}

We can now apply the replica trick. Upon replicating the soft parton theory, Eq.~(\ref{Zsqparton}) is replaced by 
\begin{equation}
  Z_{{\rm sq},N}={\cal S}_N\left\{\prod_{i=1}^N
  \exp\left[
    -\int\frac{d^d p}{(2\pi)^d}\frac12
    \frac{\overleftarrow{\delta}}{\delta \Theta^{(i)\bar{C}}_C(p)}
    \Delta_{+\phantom{,}CD}^{\phantom{+,}\bar{C}\bar{D}}(p)\frac{\overrightarrow{\delta}}
    {\delta \bar{\Theta}^{(i)\bar{D}}_D(-p)}
    \right] \right\} {\cal S}^\dag_N,
  \label{ZsqNparton}
\end{equation}
where $\Theta_C^{(i)\bar{C}}$ is the compound soft parton field associated with replica $i$. Furthermore, we have introduced the replicated generating functional for virtual soft-parton exchange diagrams of Eq.~(\ref{SNdef3}), again with colour and flavour indices suppressed. Given that this functional satisfies ${\cal S}_N={\cal S}^N$, we obtain
\begin{equation}
  Z_{{\rm sq},N}=(Z_{\rm sq})^N=1+N\log(Z_{\rm sq})+{\cal O}(N^2)\,.
  \label{ZsqN2}
\end{equation}
Thus, the usual replica trick arguments apply,%
\footnote{In extending this statement to a given observable, one must make sure that the corresponding measurement function does not differentiate between the exchanged soft partons. This will indeed be the case for infrared-safe observables.}
and the ${\cal O}(N)$ piece of an arbitrary diagram contributes to the logarithm of the squared soft parton amplitude. This succeeds in proving the exponentiation of either real or virtual soft parton corrections, as well as providing a systematic method for deriving how the colour and/or kinematic dependence of Feynman diagrams becomes modified upon their entering the logarithm of the soft function. In the following sections,  we illustrate the use of the abstract arguments of this and the preceding sections, by applying them to processes of phenomenological interest.

\section{Real emission corrections in DIS}
\label{sec:DISreal}

\begin{figure}[t]
\centering
\includegraphics[width=0.32\textwidth]{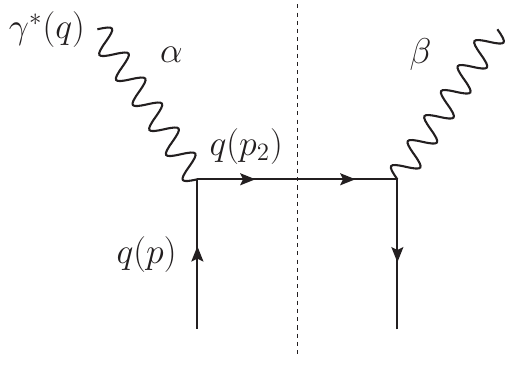}
\caption{Squared Feynman diagram for the DIS process at LO.}
\label{fig:DISLO}
\end{figure}
As a testing ground for applying the 
replica trick to the generalised soft 
function, we will consider photon-initiated 
deep-inelastic scattering (DIS) process. This serves 
two purposes: we will reproduce a known 
exponentiation property of soft quark and 
gluon emissions, which was recently used in Ref.~\cite{vanBeekveld:2021mxn} (see also Ref.~\cite{Beneke:2020ibj}) to resum NLP 
logarithms in the off-diagonal DGLAP 
splitting functions. Secondly, the discussion 
in this section paves the way for the proof (in the following section)
of a conjecture made in Refs.~\cite{Beneke:2020ibj,Moult:2019uhz},
namely that the leading virtual corrections 
in the quark channel should formally 
exponentiate. 

First we introduce some notation. 
At LO, the incoming virtual 
photon couples to a valence quark, 
leading to the process
\begin{equation}
q(p)+\gamma^*(q)=q(p_2)\,,
\label{DISLO}
\end{equation}
The LO squared Feynman diagram is 
shown in Fig.~\ref{fig:DISLO}, and 
it is conventional to define
\begin{equation}
Q^2=-q^2>0\,,\qquad \qquad 
x=\frac{Q^2}{2\,p\cdot q}\,,
\label{Q2x}
\end{equation}
where the latter is the so-called 
{\it Bj\"{o}rken} $x$ variable. In 
terms of these parameters, one may 
define the structure function 
\begin{equation}
F_2(x,Q^2) = \int {\rm d}\Phi_1 \, 
T_2^{\alpha\beta}\, 
\overline{|{\cal A}|^2_{\alpha\beta}}\,,
\label{F2def}
\end{equation}
with ${\rm d}\Phi_1$ the one-particle phase-space. The integrand contains the 
squared amplitude summed (averaged) 
over final (initial) colours and 
spins respectively, with photon 
Lorentz indices as in Fig.~\ref{fig:DISLO}. 
We have also introduced the projector
\begin{equation}
T_2^{\alpha\beta} = 
-\frac{1}{4\pi}\frac{1}{1-\epsilon}
\left(\eta^{\alpha\beta}
+(3-2\epsilon)\frac{q^2}{(p\cdot q)^2}
p^\alpha p^\beta\right),
\label{T2def}
\end{equation}
where we work in $d=4-2\epsilon$ dimensions.
The LO amplitude and 
complex conjugate amplitude read 
\begin{subequations}
\label{eq:zero-emsn-amplitudes}
\begin{align}
\big[\mathcal{A}^{(0)}\big]^{\alpha}
&= \bar u_{\hat b}^j(p_2) 
\, i e \gamma_{\hat b\hat a}^{\alpha} \delta^{ji}
\, u_{\hat a}^i(p_2), \\ 
\big[\mathcal{A}^{\dag(0)}\big]^{\beta}
&= \bar u_{\hat a'}^{i'}(p) 
\,(-i) e \gamma_{\hat a'\hat b'}^{\beta} \delta^{i'j'}
\, u_{\hat b'}^{j'}(p_2),
\end{align}
\end{subequations}
leading to  
\begin{equation}\label{MsqTree}
\overline{|{\cal A}^{(0)}|^2_{\alpha\beta}}
= \frac{e^2}{2} {\rm Tr}\big[\slashed{p}
\gamma_{\beta}\slashed{p}_2\gamma_{\alpha}\big]. 
\end{equation}
The matrix element squared is then 
inserted in Eq.~\eqref{F2def}, where at 
LO one has 
\begin{equation}\label{1PPS}
\int {\rm d}\Phi_1 =
\int \frac{{\rm d}^dp_2}{(2\pi)^{d-1}}
\, (2\pi)^d \delta^{(d)}(p+q-p_2) 
\, \delta(p_2^2)\theta(p_2^0) 
= \frac{2 \pi}{Q^2} \delta(1-x) \theta(Q).
\end{equation}
With the normalization given by the projector 
in Eq. \eqref{T2def}, one easily finds
\begin{equation}
F^{(0)}_2(x,Q^2) = \delta(1-x),  
\end{equation}
where the superscript (0) indicates 
that this is the leading order (Born) 
result.

At higher orders the emission of
soft particles can be expressed 
in terms of the generalised soft emission 
operators introduced in Section~\ref{sec:multiple}. To this end, 
let us define the vector of 
amplitudes with stripped-off
spinors/polarisation vectors:
\begin{equation}
\big[\mathcal{A}^{(0)}_{\alpha}\big]^{\bar{B}\bar A}_{BA} = 
\left([\mathcal{A}^{(0)}_{q,\alpha}]^{ji}_{\hat{b}\hat{a}}, 
0, 0 \right), \qquad \qquad \big[
\mathcal{A}^{\dag(0)}_{\beta}\big]^{\bar{A}'\bar{B}'}_{A'B'} = 
\left([\mathcal{A}^{\dag(0)}_{q,\beta}]^{i'j'}_{\hat{a}'\hat{b}'},
0, 0 \right).
\label{AA0def}
\end{equation}
The stripped-off amplitudes can 
be obtained by direct comparison 
with \eqref{eq:zero-emsn-amplitudes}, and we obtain
\begin{equation}
\label{eq:strip-zero-emsn-amplitudes}
\big[\mathcal{A}_q^{(0)\alpha}\big]^{ji}_{\hat b\hat a}
= i e \gamma_{\hat b\hat a}^{\alpha} \delta^{ji}, 
\qquad \qquad  
\big[\mathcal{A}_q^{\dag(0)\beta}\big]^{i'j'}_{\hat a' \hat b'}
= -i e \gamma_{\hat a'\hat b'}^{\beta} \delta^{i'j'}.
\end{equation}
In this notation, the tree level matrix
element squared reads 
\begin{eqnarray}\label{MsqTreeB} \nonumber
\overline{|{\cal A}^{(0)}|^2_{\alpha\beta}} 
&=& \frac{1}{2N_c} \bar \xi^{\bar{A}'}_{A'}(p)
\big[\mathcal{A}^{\dag(0)\beta}\big]^{\bar{A}'\bar{B}'}_{A'B'}
\xi^{\bar{B}'}_{B'}(p_2) \,
\bar \xi^{\bar{B}}_{B}(p_2)
\big[\mathcal{A}^{(0)\alpha}\big]^{\bar{B}\bar A}_{BA}
\xi^{\bar{A}}_{A}(p)\,,
\end{eqnarray}
where the vector of spinors/polarisation vectors
in the first line has been defined in Eqs.
\eqref{wavefns}, \eqref{wavefns-out}, and we use the summation convention for repeated species indices. Inserting the expressions from 
\eqref{eq:strip-zero-emsn-amplitudes}
in the second line above one readily 
reproduces Eq.~\eqref{MsqTree}.

\begin{figure}[t]
\begin{center}
\includegraphics[width=0.70\textwidth]{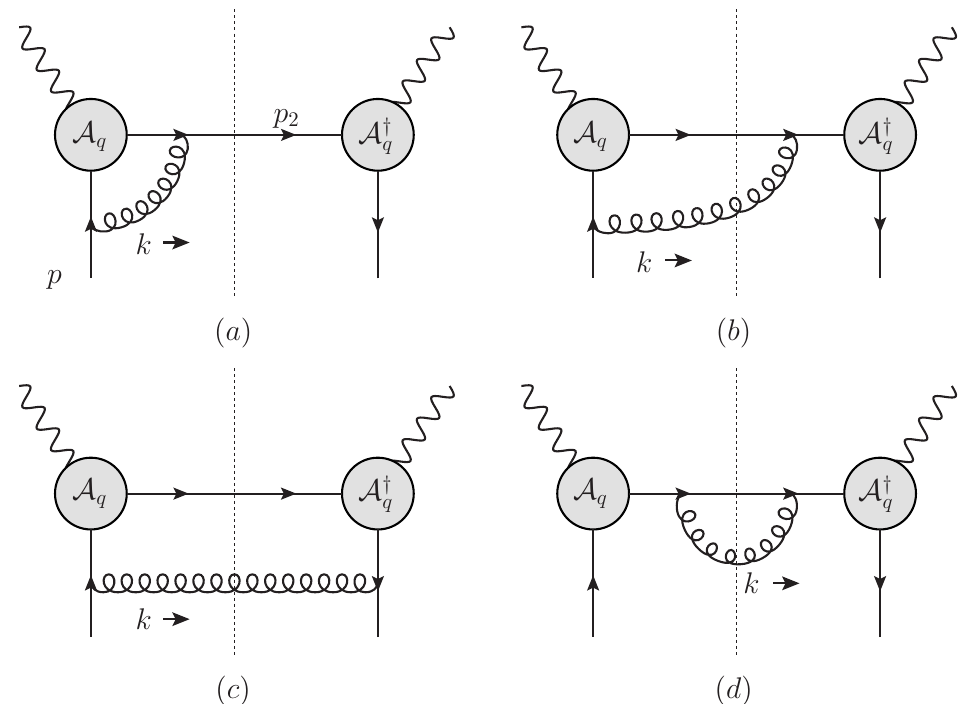}
  \caption{Diagrams corresponding to the LP NLO soft 
  function involving the exchange of a virtual (diagram a)
  or real (diagrams b, c, d) soft gluon. The corresponding
  soft function is given in Eq.~\eqref{SNLOqqqqLPv},  Eq.~\eqref{SNLOqqqqLPr} and Eq.~\eqref{SNLOqqqqLPv}. 
  The complex conjugate of diagrams 
  (a) and (b) are not shown in this figure, but are 
  taken into account in Eqs.~\eqref{SNLOqqqqLPv} and 
  \eqref{SNLOqqqqLPr}.}
  \label{fig:softcalcLP}
\end{center}
\end{figure}

\begin{figure}[t]
\begin{center}
\includegraphics[width=0.70\textwidth]{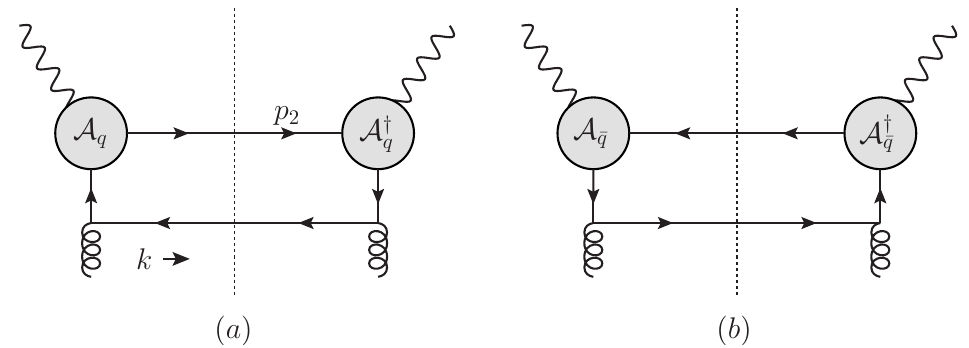}
  \caption{Diagrams corresponding to the NLP NLO soft 
  function arising from the exchange of a soft (anti-)quark.
  The corresponding soft functions ${\cal S}_{\bar q}^{(1)}$ 
  and ${\cal S}_{q}^{(1)}$ are given in Eq.~\eqref{SNLOqbargNLP}
  and \eqref{SNLOqgNLP} respectively. 
  Notice that the label $\bar q$, $q$ refers to the particle that 
  crosses the cut, not to the particle in the hard scattering 
  amplitude ${\cal A}_i$, where the label is actually opposite 
  compared to the label in the soft function. There are additional 
  diagrams where the soft (anti-)quark is emitted from the line 
  with momentum $p_2$. These however contribute beyond LL, as explained in the main text, and therefore
  are not shown here. }
  \label{fig:softcalcNLP}
\end{center}
\end{figure}

The hard-scattering amplitude may now be dressed with 
emissions of soft particles through adding generalised 
soft emission operators:
\begin{eqnarray}\label{MsqSoftGen} \nonumber
\overline{|{\cal A}|^2_{\alpha\beta}} &=&
\sum_m \frac{1}{{\cal N}_s {\cal N}_c} 
\left\langle 0\left|
\Big( \bar \xi^{\bar{A}_1'}_{A_1'}(p)\,
\big[{\cal F}_{1,\rm out}^{\dag}
\big]^{\bar{A}'_1\bar{A}'_2}_{A'_1 A'_2}
\, \big[\mathcal{A}^{\dag(0)\beta}
\big]^{\bar{A}_2'\bar{B}_2'}_{A_2'B_2'} \, 
\big[{\cal F}_{2,\rm in}
\big]^{\bar{B}'_2\bar{B}'_1}_{B'_2 B'_1}
\,\xi^{\bar{B}'_1}_{B'_1}(p_2) \Big) 
\right|m\right\rangle   \\ 
&&\hspace{1.5cm} \times \, 
\left \langle m\left| \Big(
\bar \xi^{\bar{B}_1}_{B_1}(p_2)\,
\big[{\cal F}_{2,\rm out}^{\dag}
\big]^{\bar{B}_1\bar{B}_2}_{B_1 B_2} \, 
\big[\mathcal{A}^{(0)\alpha}\big]^{\bar{B}_2\bar A_2}_{B_2A_2}
\, \big[{\cal F}_{1,\rm in}
\big]^{\bar{A}_2\bar{A}_1}_{A_2 A_1}
\, \xi^{\bar{A}_1}_{A_1}(p)\Big)
\right|0 \right\rangle,
\end{eqnarray}
where the factors ${\cal N}_s$, ${\cal N}_c$
represent the spin and color average over 
the initial-state particle. At tree level the 
initial state particle is a quark, thus 
${\cal N}_{s_q} = 2$, ${\cal N}_{c_q} = N_c$, 
as in Eq. \eqref{MsqTreeB}. At higher orders it 
is possible to have gluons in the initial state, 
leading to ${\cal N}_{s_g} = (d-2)$, 
${\cal N}_{c_g} = N_c^2-1$. Introducing 
a soft function with $m$ real gluon emissions: 
\begin{equation}\label{Sxsec}
\big({\cal S}_m\big)^{\bar{A}'_1\bar{A}'_2\, 
\bar{B}'_2\bar{B}'_1\, 
\bar{B}_1\bar{B}_2\, 
\bar{A}_2\bar{A}_1}_{A'_1 A'_2
\, B'_2 B'_1\,B_1 B_2\, A_2 A_1}
= \left\langle 0\left|
\big[{\cal F}_{1,\rm out}^{\dag}
\big]^{\bar{A}'_1\bar{A}'_2}_{A'_1 A'_2} \,
\big[{\cal F}_{2,\rm in}
\big]^{\bar{B}'_2\bar{B}'_1}_{B'_2 B'_1}
\right|m\right\rangle \left \langle m\left|
\big[{\cal F}_{2,\rm out}^{\dag}
\big]^{\bar{B}_1\bar{B}_2}_{B_1 B_2} \, 
\big[{\cal F}_{1,\rm in}
\big]^{\bar{A}_2\bar{A}_1}_{A_2 A_1}
\right|0 \right\rangle,
\end{equation}
we can write Eq.~\eqref{MsqSoftGen} as 
\begin{eqnarray} \nonumber
\overline{|{\cal A}|^2_{\alpha\beta}} &=&
\frac{1}{{\cal N}_s {\cal N}_c} 
\sum_m \,\Big( 
\bar \xi^{\bar{A}_1'}_{A_1'}(p) \, 
\big[\mathcal{A}^{\dag(0)\beta}
\big]^{\bar{A}_2'\bar{B}_2'}_{A_2'B_2'} 
\, \xi^{\bar{B}'_1}_{B'_1}(p_2) \Big) \\ 
&&\hspace{2.0cm}\times \,
\big({\cal S}_m\big)^{\bar{A}'_1\bar{A}'_2\, 
\bar{B}'_2\bar{B}'_1\, 
\bar{B}_1\bar{B}_2\, 
\bar{A}_2\bar{A}_1}_{A'_1 A'_2
\, B'_2 B'_1\,B_1 B_2\, A_2 A_1} \, 
\Big(\bar \xi^{\bar{B}_2}_{B_2}(p_2) \,
\big[\mathcal{A}^{(0)\alpha}
\big]^{\bar{B_2}\bar A_2}_{B_2A_2}
\,\xi^{\bar{A}_1}_{A_1}(p)\Big).
\label{MsqSoftGen2}
\end{eqnarray}
Upon inserting this into Eq.~\eqref{F2def} and performing the 
relevant phase-space integrals, 
one obtains the LL result for the structure function at any power. 
Let us stress that 
the above amplitude only should reproduces
the maximally soft momentum configuration, i.e.\
the LL contribution. 
Corrections beyond LL could in principle be taken into 
account by modifying the diagonal terms 
${\cal T}_{ii}^{\rm in}(p)$ of the transition 
matrix in Eq.~\eqref{TransitionMatrDef}, with 
$i= q$, $\bar q$, $g$, in such a way to 
include NLP corrections to these transitions,
but are not considered here.%
\footnote{They have been obtained in 
\cite{Laenen:2008gt,Laenen:2010uz}, see 
in particular Eq. 5.7 of \cite{Laenen:2010uz},
which provide the NLP contribution to the 
diagonal terms of the generalised soft emission operator ${\cal F}$.}
In Appendix~\ref{App:NLOsoft} we show explicitly 
how to calculate $F_2(x,Q^2)$ at NLO, for which one
needs to consider the diagrams in Figs.~\ref{fig:softcalcLP}
and \ref{fig:softcalcNLP}.

We now turn our attention to the exponentiation of the real 
emission diagrams by means of the replica 
trick. First of all, let us notice that the 
structure function $F_2$
is gauge invariant. We can exploit 
this by choosing a gauge where the 
calculation of higher-order contributions 
simplifies. To this end it proves 
useful~\cite{Gribov:1972ri,Gribov:1972rt,Dokshitzer:1977sg}
to consider physical gluon polarisation 
vectors defined via the 
requirements
\begin{equation}
k\cdot \epsilon(k)=c\cdot\epsilon(k)=0\,,
\label{physpols}
\end{equation}
where $c$ is a null vector ($c^2=0$).
This means that the sum over physical gluon 
polarisation states take the form
\begin{equation}
{\cal P}_{\mu\nu}(k) = -\eta_{\mu\nu}
+\frac{k_\mu c_\nu + k_\nu c_\mu}{c\cdot k}\,.
\label{polsum}
\end{equation}
In what follows we consider 
\begin{equation}
c=p_2,
\label{c=p2}
\end{equation}
a choice which is particularly convenient,
as it eliminates the need to consider ${\cal F}_2$ in the
case of soft gluon emissions. For soft quark emissions,
at NLP, ${\cal F}_2$  will contribute to at most NLL, which is also beyond what we are considering here. 
That is, we only need to take into account emissions from 
the generalised Wilson line ${\cal F}_{1,\rm in}$, 
${\cal F}_{1,\rm out}^{\dag}$. Indeed, it is easy to check that by setting 
$c=p_2$, only diagram (c) in Fig.~\ref{fig:softcalcLP} 
contributes to the real emission at LP, while diagrams 
(b) and (d) are zero. 
\begin{figure}[t]
\begin{center}
\includegraphics[width=0.80\textwidth]{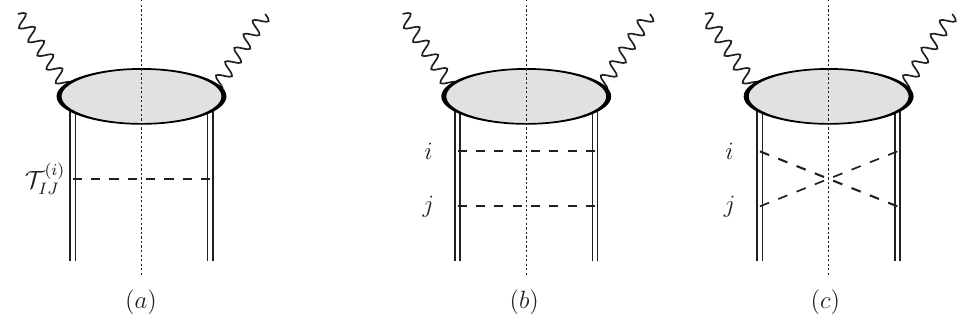}
  \caption{(a) Single soft topology corresponding to each 
  of the partonic assignments in Figs.~\ref{fig:softcalcLP}
  (c) and~\ref{fig:softcalcNLP} (a); 
  (b) and (c) soft topologies generated at NNLO in the 
  squared amplitude, with replica indices indicated.}
  \label{fig:softgraph}
\end{center}
\end{figure}
With this, we note that the diagrams of Fig.~\ref{fig:softcalcLP}(c)
and Fig.~\ref{fig:softcalcNLP}(a), (b) are all 
given by the single topology in 
Fig.~\ref{fig:softgraph}(a). At higher orders, using
the gauge of Eq.~\eqref{physpols},
the LL of the real emission is given in terms of 
(crossed-)ladder diagrams, 
where emission of any 
partonic species is expressed by the transition 
matrix ${\cal T}_{IJ}$. 
In the case of DIS, at NNLO in the gauge defined by 
Eqs.~(\ref{polsum})~(\ref{c=p2}) one needs to take 
into account the two real topologies (b) and (c) 
in Fig.~\ref{fig:softgraph}, namely an uncrossed 
and a crossed ladder, with replica index $i$ and $j$. 
One immediately sees that diagrams (b) and (c) of 
Fig.~\ref{fig:softgraph} are the analog of the diagrams 
in Fig.~\ref{fig:replicaNLP}. Therefore,
following the discussion in Section~\ref{sec:replica-soft}, the contributions of diagrams (b) and (c) to the 
logarithm of the generalised soft function in 
Eq.~\eqref{SExpDef} read (Cf. Eq.~\eqref{LogEb})
\begin{align}\label{LogEc}
\widetilde{E}(b) = 0, \qquad \qquad
\widetilde{E}(c) = 
\frac{{\cal N}(c)C(c)-{\cal N}(b)C(b)}{{\cal D}(c)}\,.
\end{align}

This result is consistent with exponentiation 
of the generalised soft-emission operator in the 
generalised soft function. To see this, let us explicitly construct the $\mathcal{O}(\alpha_s^2)$ expansion
of the soft function using the reduced diagrams of Eq.~\eqref{LogEc}.
With only one emission present, there is only one 
choice of replica index, and thus all graphs with 
one soft emission are ${\cal O}(N)$. They 
thus contribute to the logarithm of the soft 
function unmodified. When two partons are emitted, 
the only contribution that needs to be added 
according to Eq.~\eqref{LogEc} is $\widetilde{E}(c)$, 
the rest is already generated by the one-emission 
diagram. Therefore the logarithm of the soft function 
becomes 
\begin{equation}
\ln{\cal S} = E(a)+\tilde{E}(c)\,,
\label{logS}
\end{equation}
where the first term corresponds to diagram (a)
in Fig.~\ref{fig:softgraph}. 
Focusing on the NNLO contribution, obtained after exponentiating the above result and then expanding $\cal S$ up to $\mathcal{O}(\alpha_s^2)$, we obtain
\begin{equation}
\exp\left[\ln {\cal S}\right]\Big|_{\mathcal{O}(\alpha_s^2)} = \frac12 \Big[E(a)\Big]^2
+\frac{{\cal N}(c)C(c)-{\cal N}(b)C(b)}{{\cal D}(c)}\,.
\label{logS2}
\end{equation}
In the first term, squaring the one-emission 
contribution will generate kinematic numerators, which are products of two distinct emissions 
on the left-and right-hand side of the cut. 
Selecting the correct partonic assignment 
will yield the kinematic numerator and 
colour factor of diagram (b). However, 
the denominator arising from squaring 
the one-emission graphs will have 
uncoupled momenta
\begin{align}
\Big[D(a)\Big]^2 = (p\cdot k_1)^2(p\cdot k_2)^2\,,
\end{align}
rather than the actual denominator of 
diagram (b), which is
\begin{align}
{\cal D}(b)=(p\cdot k_1)^2[p\cdot (k_1+k_2)]^2\,.
\end{align}
The denominator for diagram (c) is
\begin{align}
{\cal D}(c)=(p\cdot k_1)(p\cdot k_2)[p\cdot (k_1+k_2)]^2\,.
\end{align}
Using the eikonal identity
\begin{equation}
\frac{1}{(p\cdot k_1)[p\cdot (k_1+k_2)]}
+\frac{1}{(p\cdot k_2)[p\cdot (k_1+k_2)]}=\frac{1}{(p\cdot k_1)}
\frac{1}{(p\cdot k_2)}\,,
\label{eikid}
\end{equation}
we can write
\begin{equation}
\frac{1}{(p\cdot k_1)^2(p\cdot k_2)^2}=2\left[
\frac{1}{{\cal D}(b)}+\frac{1}{{\cal D}(c)}\right],
\label{DABrel}
\end{equation}
where we have used the freedom to relabel 
$k_1\leftrightarrow k_2$ to identify 
$2{\cal D}(b)$, which is allowed given that 
all emission momenta are integrated over. 
We thus find
\begin{equation}
\frac12\Big[E(a)\Big]^2 =
{\cal N}(b)C(b)\left[
\frac{1}{{\cal D}(b)}
+\frac{1}{{\cal D}(c)}\right].
\label{D1sq}
\end{equation}
Combining this with the second term in 
Eq.~(\ref{logS2}) gives
\begin{equation}
\frac12 \Big[E(a)\Big]^2+\tilde{E}(c)
=\frac{{\cal N}(b)C(b)}{{\cal D}(b)}+\frac{{\cal N}(c)C(c)}{{\cal D}(c)}\,.
\label{NNLOsum}
\end{equation}
That is, exponentiating the logarithm of the 
generalised soft function and extracting a 
given partonic assignment does indeed 
reproduce the conventional diagrams in 
the amplitude, with their correct kinematic 
numerators and colour factors. Similar 
arguments can be made at higher orders.

We now use the replica trick to obtain the LL contribution of the emission of a soft parton, in which case we need only 
consider the first term of $\ln \cal S$, namely the one consisting of 
a sum over all possible one real-emission 
contributions. In the gauge defined by 
Eq.~(\ref{c=p2}), we may therefore simply exponentiate the NLO soft function $S^{(1)}$.
The corresponding
matrix element squared becomes
\begin{eqnarray}\label{MsqSoftGen2exp} \nonumber
\overline{|{\cal A}|^2_{\alpha\beta}}\big|_{\rm real} 
&=& \frac{1}{{\cal N}_s {\cal N}_c} \,\Big( 
\bar \xi^{\bar{A}_1'}_{A_1'}(p) \, 
\big[\mathcal{A}^{\dag(0)\beta}
\big]^{\bar{A}_2'\bar{B}_2'}_{A_2'B_2'} 
\, \xi^{\bar{B}'_1}_{B'_1}(p_2) \Big) \\ 
&&\hspace{0.5cm}\times \,
\exp \Big[{\cal S}_{\rm}^{(1)}\Big]^{\bar{A}'_1\bar{A}'_2\, 
\bar{B}'_2\bar{B}'_1\, 
\bar{B}_1\bar{B}_2\, 
\bar{A}_2\bar{A}_1}_{A'_1 A'_2
\, B'_2 B'_1\,B_1 B_2\, A_2 A_1} \, 
\Big(\bar \xi^{\bar{B}_2}_{B_2}(p_2) \,
\big[\mathcal{A}^{(0)\alpha}
\big]^{\bar{B_2}\bar A_2}_{B_2A_2}
\,\xi^{\bar{A}_1}_{A_1}(p)\Big),
\end{eqnarray}
Using this expression, we make contact with 
the known results at NLP for the quark-gluon
channel in photon-induced DIS~\cite{vanBeekveld:2021mxn}. 
To this end, let us recall that expansion of 
the exponential in Eq.~\eqref{MsqSoftGen2exp} 
generates contributions at all subleading power, 
but we can select the NLP term by associating 
a book-keeping parameter $\lambda \ll1$ to the 
quark-gluon transitions in the matrix Eq.~\eqref{TransitionMatrDef}. The parameter 
$\lambda$ is chosen such that the soft momentum 
scales as $k \sim \lambda^2$, with $\lambda^2 \sim (1-x)$
for $x\to 1$. Then it is easy to see that the 
LP and NLP soft functions scale respectively 
as $\lambda^{-4n}$ and $\lambda^{-4n+2}$
at order $\alpha_s^{n}$. To select 
the gluon-initiated DIS channel we need 
at least the exchange of a single soft fermion, 
which is suppressed by a power of $\lambda^2$
compared to the exchanged a soft gluon.
Therefore, 
at ${\cal O}(\alpha_s^n)$ the NLP contributions 
is obtained by allowing 
only one insertion of the soft function 
${\cal S}_{\bar gq}$, $m$ insertions of ${\cal S}_{gg}$, 
and $n-1-m$ insertions of ${\cal S}_{qq}$.
Given its length, the explicit expression is given in Eq.~\eqref{MsqSoftGen2iterated}.
Substituting the results for the partonic emission 
factors, contracting these with the LO amplitudes,
and applying the eikonal identity as discussed e.g.\
in Sec~2.2.1 of \cite{vanBeekveld:2021mxn} one 
finds for the squared NLP amplitude
\begin{align} \label{Sexp} \nonumber
\overline{|{\cal A}|^2_{\alpha\beta}}\big|_{\rm qg,\,real,\,LL}^{(n)} 
&= \frac{g_s^{2n}\, e^2 \, n_f\, T_R}{n! (d-2)} \left[\prod_{i=1}^{n-1} 
\frac{p^{\mu_i}p^{\nu_i}}{(p\cdot k_i)^2} {\cal P}_{\mu_i\nu_i}(k_i)\right]
\sum_{m=0}^{n-1} C_A^{m}\, C_F^{n-1-m} \\
&\quad \times\, \bigg\{ {\rm Tr} \big[ 
\slashed{k}_q \gamma^{\alpha_1'} \slashed{p} \gamma_{\beta} 
\slashed{p}_2 \gamma_{\alpha} \slashed{p} \gamma^{\alpha_1}\big] 
+ {\rm Tr} \big[ 
\slashed{k}_q \gamma^{\alpha_1} \slashed{p} \gamma_{\alpha} 
\slashed{p}_2 \gamma_{\beta} \slashed{p} \gamma^{\alpha'_1}\big]
\bigg\} \frac{{\cal P}_{\alpha_1\alpha_1'}(p)}{(2 p\cdot k_q)^2},
\end{align}
This indeed agrees with the known form of the 
squared amplitude with one soft quark emission 
and $n-1$ soft gluon emissions, as conjectured 
in Ref.~\cite{Vogt:2010cv} and proven in 
Ref.~\cite{vanBeekveld:2021mxn}. 

As discussed 
above, in Eqs.~\eqref{MsqSoftGen2iterated},
\eqref{Sexp} we have only considered those 
graphs which are non zero in the gauge defined
by Eqs.~\eqref{polsum} and \eqref{c=p2}. One 
could of course choose a different gauge, in 
which case the remaining diagrams would have 
to be considered. These will exponentiate by 
the same arguments as above, leading the to 
same final (gauge-invariant) result.

Whilst the arguments of this section may take 
some getting used to, they greatly streamline 
the derivation of the all-order structure of 
the leading real emission corrections at 
arbitrary order in perturbation theory. 
That is, exponentiating the one-emission 
contribution to the generalised soft function 
leads immediately to the fact that at NLP, 
the soft quark emission process is dressed 
by $n-1$ soft gluons at ${\cal O}(\alpha_s^{n})$, 
where the emission momenta are uncoupled, 
and where there is a combinatorial factor 
of $1/n!$. The correct colour factor 
is also obtained, namely that associated 
with a sum over all possible uncrossed 
ladder graphs, in which gluons are emitted 
either before or after the soft quark emission.

To summarise, in this section we have used 
the structure of real emission corrections 
in DIS to illustrate the use of the replica 
trick in determining the leading singular 
behaviour of scattering amplitudes in the 
soft limit. As well as recovering previously 
proven results, however, we may also use the 
replica trick to prove a recent conjecture 
regarding virtual corrections in the same 
process. This is the subject of the 
following section.

\section{Endpoint contributions in deep inelastic scattering}
\label{sec:DIS}

Ref.~\cite{Beneke:2020ibj} considered the so-called off-diagonal splitting functions in deep-inelastic scattering, arising from kinematically subleading partonic channels that open up for the first time at NLO. The authors then derived the all-order structure of leading logarithmic terms, agreeing with the previous conjectures of Ref.~\cite{Vogt:2010cv}. However, in doing so, it was assumed that the leading IR divergent virtual corrections to each subleading partonic channel exponentiate. A related conjecture was made in Ref.~\cite{Moult:2019uhz}, namely that leading virtual corrections to a certain form factor involving soft quark emission also exponentiate, and indeed this is the form factor that would replace the usual quark form form factor as applied to DIS, in generating subleading partonic emissions at NLO. 

\begin{figure}[t]
    \centering
    \includegraphics[width=0.25\textwidth]{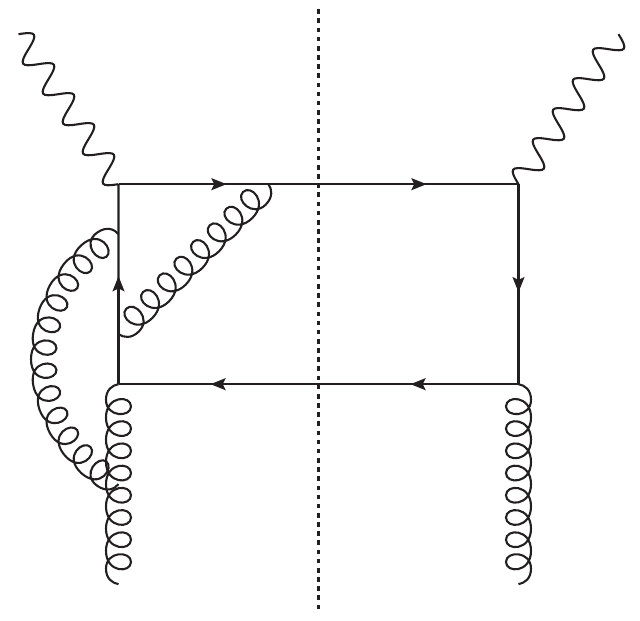}
    \caption{Example higher-order diagram with a single soft quark crossing the final state cut.}
    \label{fig:softquarkdiags}
\end{figure}

The aim of this section is to show how the above results naturally arise from using the replica trick arguments of this paper. To this end, we must again consider the generating functional for soft parton diagrams at the level of the squared amplitude, as constructed in Eq.~(\ref{Zsqparton}). As explained after Eq.~(\ref{ZsqNparton}), this generating functional implies that the ${\cal O}(N)$ part of an arbitrary diagram 
enters the logarithm of the squared soft parton amplitude. However, to confirm the above conjecture, we must consider only those diagrams that have a single soft quark crossing the final state cut. This is straightforward using the generating functional of Eq.~(\ref{ZsqNparton}), but unfortunately does not lead directly to a proof (or otherwise) of the desired conjecture. To see why, it is sufficient to note that diagrams containing a single soft real quark will arise at {\it any} order in perturbation theory, such that some of them will indeed exponentiate. An example is shown in Fig.~\ref{fig:softquarkdiags}, and we can clearly add more soft gluons to this as desired. Thus, a na\"{i}ve application of the replica trick results in a complicated form of the soft function for the squared amplitude, in which single soft quark corrections enter the logarithm at arbitrary orders. What the above conjecture then states is that it must somehow be possible to reorder the diagrammatic expansion, so that the soft quark emission factorises from the remaining soft gluon emissions, leading to a simple exponential form for the latter. Put more simply, the conjecture states that the {\it additional} soft gluon emissions {\it dressing} a single soft quark emission exponentiate. What is desirable is to isolate the single quark emission from the outset, such that the replica trick can be applied to the remaining soft gluon emissions alone.

It is indeed possible to achieve this by appealing again to Eq.~(\ref{Zsqparton}). The exponential operator in the middle of the diagram contracts pairs of (conjugate) fields on either side of the final state cut. If we are to have a single soft quark as the only real emission, then the exponential may be expanded to first-order only, retaining only the term involving the (anti-)quark fields. We may also simplify the generating functional ${\cal S}$ that appears on the left-hand side. This contains a generalised soft emission operator associated with each hard particle. However, only one soft quark emission needs to be included, from the incoming parton leg. On the latter, the operator of Eq.~(\ref{Fdef2}) takes the form
\begin{equation}
    {\cal P}\exp\left[
ig_s \int {\rm d}t \left(
{\bf Q}_\Psi {\bf T}_\Psi \Psi(tp_1)+{\bf Q}_{\bar{\Psi}} 
{\bf T}_{\bar{\Psi}} \bar{\Psi}(tp_1)+{\bf T}^a p^\mu\sum_{i=1}^N A_{\mu}^{(i),a}(tp_1)\right)\right],
\label{Fform}
\end{equation}
where ${\bf Q}_\Psi$ (${\bf Q}_{\bar{\Psi}}$) and ${\bf T}_\Psi$ (${\bf T}_{\bar{\Psi}}$) are the kinematic and colour factors associated with soft (anti-)quark emission respectively. To NLP order, inclusion of a single soft quark emission amounts to expanding Eq.~(\ref{Fform}) to first order in ${\bf Q}_\Psi$, such that Eq.~(\ref{Fform}) becomes
\begin{equation}
\int_{-\infty}^0 {\rm d}t_q\,\Phi_1(-\infty, t_q)\left[{\bf Q}_\Psi{\bf T}_{\Psi}\Psi(t_q p)\right]\Phi_1(t_q,0),
\label{Fform2}
\end{equation}
where $\Phi_1(a,b)$ is the conventional Wilson line operator on parton leg 1, between distance parameters $a$ and $b$. We have also introduced $t_q$ as the distance parameter at which the soft quark is emitted. All remaining soft emission operators in the squared amplitude can then be replaced with conventional Wilson lines, such that the generating functional for diagrams containing a single real soft quark emission becomes
\begin{align}
Z_{sq}\Big|_{q}&={\cal S}_q\left[-\int\frac{{\rm d}^d k_q}{(2\pi)^d}\frac12
\frac{\overleftarrow{\delta}}{\delta\Psi(k_q)}\Delta^{(q)}_+(k_q)
\frac{\overrightarrow{\delta}}{\delta\bar{\Psi}(-k_q)}
\right]{\cal S}^\dag_q,
\label{Zsq|q}
\end{align}
where $\Delta^{(q)}_+$ is the appropriate cut propagator for a quark of momentum $k_q$, and we have introduced the generating functional for diagrams containing a single soft real quark:
\begin{align}
    {\cal S}_q&= \int_{-\infty}^0 {\rm d}t_q\int{\cal D}\Psi\int{\cal D}\bar{\Psi}
    \,{\bf Q}_\Psi {\bf T}_{\Psi}\Psi(t_q p_1) {\cal S}_A;\notag\\
    {\cal S}_A&=\left[
    \int{\cal D}A_\mu \Phi_1(-\infty,t_q)\Phi_1(t_q,0)\Phi_2(0,\infty)
    e^{iS[A_\mu,\Psi,\bar{\Psi}]}\right].
    \label{Sqdef}
\end{align}
Here we have isolated the path integral over the soft gluon field, together with the Wilson lines that source the soft gluons.%
\footnote{Although it appears that we may combine the two Wilson lines associated with the incoming parton leg with momentum $p_1$, it must be remembered that they contain colour matrices in different representations, due to acting on either side of the soft quark emission.} 
This path integral acts to dress the amplitude for the emission of a single soft quark with additional gluon emissions, thus generating all possible virtual corrections to the single soft quark emission process. 
The soft gluons may interact either with the hard parton legs (through the Wilson lines), or with the emitted soft quark (via the interaction Lagrangian in the soft parton action). That these additional contributions exponentiate can now be surmised by replicating the soft gluon field $A_\mu$, but {\it not} the soft (anti-)quark fields. Then, in the replicated theory with $N$ replicas $A_\mu^{(i)}$, Eq.~(\ref{Sqdef}) becomes
\begin{align}
    {\cal S}_{q,N}&= \int_{-\infty}^0 {\rm d}t_q\int{\cal D}\Psi\int{\cal D}\bar{\Psi}
    \,{\bf Q}_\Psi {\bf T}_{\Psi}\Psi(t_q p_1) {\cal S}_A^N.
\end{align}
Following similar arguments to our previous cases, the ${\cal O}(N)$ piece of the soft gluon dressings exponentiates, precisely confirming the desired conjecture that virtual corrections to the soft quark emission process exponentiate.

For completeness, we mention a subtlety in the above argument. At higher orders, there will be diagrams in which fermion bubbles occur off the hard parton lines. These are generated by pulling down multiple interaction vertices from the action $S[A_\mu,\Psi,\bar{\Psi}]$, and a potential problem occurs given that we have only replicated the soft gluons: gluons of different replica number may interact with each other through a fermion bubble, thus contradicting the assumptions that go into the replica trick. That is, one would no longer be able to conclude that the factor ${\cal S}_A$ gets replaced, in the replicated theory, by ${\cal S}_A^N$. One way around this would be to introduce an additional set of soft (anti-)quark fields, that enter the action, but cannot be sourced by the hard parton legs. In any case, this issue is irrelevant for proving the conjecture of Refs.~\cite{Beneke:2020ibj,Moult:2019uhz}, namely that {\it leading} virtual corrections to the quark channel in DIS exponentiate: the relevant diagrams contain only soft gluons in addition to the single radiated soft (anti-)quark.

In this section, we have shown that the additional soft gluon corrections to the squared amplitude with a single soft quark emission exponentiate. Given the rather combinatorial nature of the above analysis, however, it is useful to make things more concrete. In particular, we may explicitly confirm that our generalised soft-emission lines do indeed match the results of a complete calculation of virtual corrections, where the soft expansion is performed {\it after} the integral over the virtual gluon momentum has been carried out. 

\subsection{Calculation of endpoint contributions}
\label{sec:endpointcalc}
\begin{figure}[t]
\begin{center}
    \includegraphics[width=0.80\textwidth]{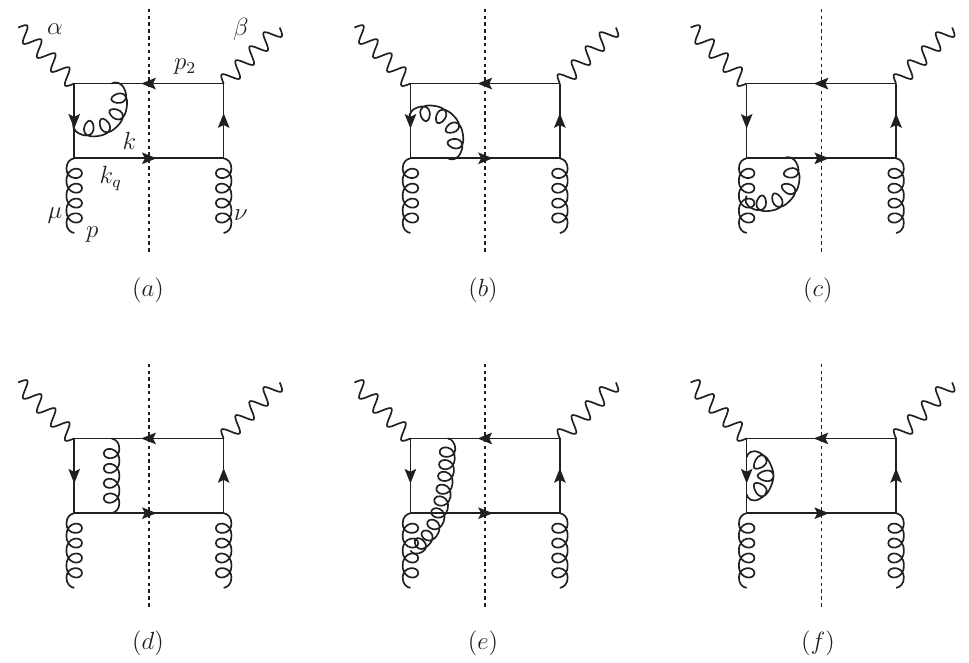}
    \caption{Diagrams contributing to the one-real, one-virtual contributions to DIS, in which a soft quark is emitted in the final state.}
    \label{fig:DIS1r1v}
\end{center}
\end{figure}
The various diagrams contributing to the leading virtual corrections to the quark channel of DIS are shown in Fig.~\ref{fig:DIS1r1v}. Dressing the appropriate NLO squared amplitude with additional factors from the generalised soft-emission line, the kinematic part of diagram (a) in Fig.~\ref{fig:DIS1r1v} is given (in the Feynman gauge) by 
\begin{align}
{\cal K}(a)&=
T_{2\alpha\beta}\,P_{\mu\nu}(p)
\int\frac{{\rm d}^d k}{(2\pi)^d}
\frac{1}{k^2}\left(\frac{p^\sigma}{p\cdot (k-k_q)}\right)
\left(\frac{p_{2\sigma}}{p_2\cdot k}\right){\rm Tr}
\left[\slsh{k}_q\left(\frac{\gamma^\mu\slsh{p}}{2p\cdot k_q}\right)
\gamma^\alpha\slsh{p}_2\gamma^\beta \left(\frac{\slsh{p}\gamma^\nu}
{2p\cdot k_q}\right)\right]\notag\\
&=4p\cdot p_2 \,T_1(p,p_3;k_q) T_{2\alpha\beta}\,P_{\mu\nu}(p){\rm Tr}
\left[\slsh{k}_q\left(\frac{\gamma^\mu\slsh{p}}{2p\cdot k_q}\right)
\gamma^\alpha\slsh{p}_2\gamma^\beta\left(\frac{\slsh{p}\gamma^\nu}
{2p\cdot k_q}\right)
\right],
\label{Facalc1}
\end{align}
where we have also included the projector for the structure function $F_2(x,Q^2)$, as given in Eq.~(\ref{T2def}). Furthermore, in the second line of Eq.~(\ref{Facalc1}) we have introduced the soft triangle integral
\begin{equation}
T_1(p_1,p_2;k_2)=\int\frac{{\rm d}^d k_1}{(2\pi)^d}\frac{1}
{k_1^2\,2p\cdot(k_1-k_2)\,2p_2\cdot k_1}.
\label{T1def}
\end{equation}   
We calculate this integral in Appendix~\ref{app:integrals}, and the result is given in Eq.~(\ref{T1res}). To express the so-called endpoint contribution at which all emitted partons are soft, we can define the Mandelstam invariants
\begin{equation}
s=(k_q+p_2)^2,\quad t=(p-k_q)^2,\quad u=(p-p_2)^2
\label{mandies}
\end{equation}
satisfying
\begin{equation}
s+t+u=q^2\equiv -Q^2.
\label{momsum}
\end{equation}
We may then parameterise these as
\begin{equation}
s=\frac{Q^2(1-x)}{x},\quad
t=-\frac{z}{x}{Q^2},\quad u=-\frac{(1-z)Q^2}{x}.
\label{mandiesparam}
\end{equation}
The first of these follows from the definition of the Bj\"{o}rken variable in Eq.~(\ref{Q2x}). The second corresponds to introducing an additional parameter $z$, such that Eq.~(\ref{momsum}) is respected, and such that the threshold limit in which the final-state quark (rather than anti-quark) is soft corresponds to
\begin{equation}
x\rightarrow 1,\quad z\rightarrow 0.
\label{xzlim}
\end{equation}  
Substituting the results of Eqs.~(\ref{mandies}, \ref{mandiesparam}, \ref{T1res}) into Eq.~(\ref{Facalc1}) and expanding about $d=4-2\epsilon$ dimensions in the limit of Eq.~(\ref{xzlim}), one finds
\begin{equation}
{\cal K}(a)=-\frac{i}{4\pi^3}\frac{1}{z}\frac{1}{\epsilon^2}
\left[\left(\frac{\mu^2}{Q^2}
\right)^\epsilon-\left(\frac{\mu^2}{zQ^2}\right)^\epsilon\right]
+{\cal O}(\epsilon^{-1})+{\cal O}[(1-x)]+{\cal O}(z^0).
\label{Fares}
\end{equation}  
We see that this diagram has a leading $\epsilon$ pole, so will indeed contribute to the leading logarithm in $(1-x)$ after integrating over the final state phase space. Furthermore, we may note the non-analytic dependence on $z$ in the second term, which is important to keep track of, given that this will affect the leading logarithmic threshold behaviour after integration over the final state phase space. 

We may apply similar methods to the other diagrams in Fig.~\ref{fig:DIS1r1v}. First, we may make things easier by noting that diagram (f) will have at most a bubble integral, which cannot produce a leading $\epsilon$ pole: it is ${\cal O}(\epsilon^{-1})$ rather than ${\cal O}(\epsilon^{-2})$. Furthermore, diagram (b) also fails to contribute a leading pole. To see this, we may first evaluate its kinematic part, which turns out to yield
\begin{align}
{\cal K}(b)&=\int\frac{{\rm d}^d k}{(2\pi)^d}\frac{p_\sigma}{[-p\cdot(k-k_q)]}
\frac{P_{\mu\nu}(p)\,T_{2\alpha\beta}}{k^2}\notag\\
&\quad\times{\rm Tr}\left[
\slsh{k}_q\left(\frac{\gamma^\sigma(\slsh{k}_q-\slsh{k})}{(k_q-k)^2}\right)\left(\frac{\gamma^\mu\slsh{p}}{(2p\cdot k_q)}\right)
\gamma^\alpha\slsh{p}_2\gamma^\beta \left(\frac{\gamma^\beta\slsh{p}}{p\cdot k_q}\right)
\right],
\end{align}
which involves the virtual integral
\begin{equation}
\int\frac{{\rm d}^d k_1}{(2\pi)^d}\frac{1}{k^2\,[2p\cdot(k-k_q)](k-k_q)^2}.
\label{diagbint}
\end{equation}
However, the Dirac trace (after contraction with polarisation and projection tensors) evaluates to
\begin{equation}
-\frac{64 (p\cdot k_q)(p\cdot p_2)[p\cdot(k-k_q)]}
{\pi}+{\cal O}(\epsilon),
\end{equation}
such that the $k$-dependent factor in the numerator cancels the similar factor in the denominator of Eq.~(\ref{diagbint}), leaving a bubble integral which is necessarily ${\cal O}(\epsilon^{-1})$. 

Carrying out a similar analysis for the remaining diagrams yields
\begin{subequations}
\begin{align}
{\cal K}(c)&=\int\frac{{\rm d}^d k}{(2\pi)^d}\frac{2p^\sigma}{(2p\cdot k_q)^2}\left\{
T_2(p,p_2;k_q){\rm Tr}\left[
\slsh{k}_q\gamma^\sigma\slsh{k_q}\gamma^\mu\slsh{p}\gamma^\alpha\slsh{p}_2
\gamma^\beta\slsh{p}\gamma^\nu\right]\right.\notag\\
&\left.\hspace{2.0cm}-\,T_2^\delta(p,p_2;k_q){\rm Tr}\left[
\slsh{k}_q\gamma^\sigma\gamma_\delta\gamma^\mu\slsh{p}\gamma^\alpha
\slsh{p}_2\gamma^\beta \slsh{p}\gamma^\nu\right]
\right\} P_{\mu\nu}(p) ;\\
{\cal K}(d)&=\int\frac{{\rm d}^d k}{(2\pi)^d}\frac{p_{2\sigma}}{p\cdot k_q}{\rm Tr}\left[
\slsh{k}_q\gamma^\sigma\gamma_\delta\gamma^\mu\slsh{p}\gamma^\alpha
\slsh{p}_2\gamma^\beta\slsh{p}\gamma^\nu\right]
\left[
B_2(p,p_2;k_q)k_q^\delta-B_2^\delta(p,p_2;k_q)\right]
P_{\mu\nu}(p);\\
{\cal K}(e)&=\int\frac{{\rm d}^d k}{(2\pi)^d}\frac{p\cdot p_2}{p\cdot k_q}B_1(p,p_2;-k_q)
{\rm Tr}\left[\slsh{k}_q\gamma^\mu\slsh{p}\gamma^\alpha\slsh{p}_2
\gamma^\beta\slsh{p}\gamma^\nu\right] P_{\mu\nu}(p),
\end{align}
\label{Fdiags}
\end{subequations}
where we have introduced various integrals, which are defined and calculated in Appendix~\ref{app:integrals}. Substituting those results into Eq.~(\ref{Fdiags}) before expanding in the limit of Eq.~(\ref{xzlim}) gives
\begin{subequations}
\begin{align}
{\cal K}(c)&=\frac{i}{4\pi^3}\frac{1}{z}\frac{1}{\epsilon^2}
\left(\frac{\mu^2}{zQ^2}\right)^\epsilon; \\
{\cal K}(d)&=-\frac{i}{4\pi^3}\frac{1}{z}\frac{1}{\epsilon^2}
\left(-\frac{\mu^2}{Q^2 z(1-x)}\right)^\epsilon; \\
{\cal K}(e)&=-\frac{i}{4\pi^3}\frac{1}{z}\frac{1}{\epsilon^2}
\left(\frac{\mu^2}{zQ^2}\right)^\epsilon.
\end{align} 
\label{diagresults}
\end{subequations}

To verify our results, we may compare Eqs.~(\ref{Fares}, \ref{diagresults}) with the result of carrying out an exact calculation of each diagram in Fig.~\ref{fig:DIS1r1v}, before expanding in the threshold limit after the integration over the virtual gluon momentum has already been carried out. First, we may note results for the kinematic part of each diagram, using the same conventions as above. That is, we ignore overall coupling and spin averaging factors, but nevertheless contract with polarisation and projection tensors. We then find kinematic parts:
\begin{subequations}
\begin{align}
(a):&\quad  \int\frac{{\rm d}^d k}{(2\pi)^d}\frac{{\rm Tr}\left[
\slsh{k}_q\gamma^\mu(\slsh{k}_q-\slsh{p})\gamma^\sigma
(\slsh{k}_q-\slsh{k}-\slsh{p})\gamma^\alpha(-\slsh{p}_2-\slsh{k})
\gamma_\sigma\slsh{p}_2\gamma^\beta(\slsh{k}_q-\slsh{p})\gamma^\nu
\right]}{k^2(k_q-p)^2(k_q-k-p)^2(k+p_2)^2(k_q-p)^2};\\
(b):&\quad \int\frac{{\rm d}^d k}{(2\pi)^d}\frac{{\rm Tr}\left[
\slsh{k}_q\gamma^\sigma(\slsh{k}_q-\slsh{k})\gamma^\mu
(\slsh{k}_q-\slsh{k}-\slsh{p})\gamma_\sigma(\slsh{k}_q-\slsh{p})
\gamma^\alpha\slsh{p}_2\gamma^\beta(\slsh{k}_q-\slsh{p})\gamma^\nu
\right]}{k^2(k_q-k-p)^2(k_q-k)^2(k_q-p)^4};\\
(c):& \quad \int\frac{{\rm d}^d k}{(2\pi)^d}\frac{{\rm Tr}\left[
\slsh{k}_q\gamma^\sigma(\slsh{k}_q-\slsh{k})\gamma^\tau(\slsh{p}-\slsh{k}_q)\gamma^\alpha\slsh{p}_2\gamma^\beta(\slsh{p}-\slsh{k}_q)
\gamma^\nu\right]}{k^2(k_q-k)^2(p-k)^2(p-k_q)^4}
V^{\tau\sigma\mu}(p-k,k,-p]);\\
(d):& \quad \int\frac{{\rm d}^d k}{(2\pi)^d}\frac{{\rm Tr}\left[
\slsh{k}_q\gamma^\sigma(\slsh{k}_q-\slsh{k})\gamma^\mu(\slsh{k}_q-\slsh{k}-\slsh{p})\gamma^\alpha\slsh{p}_2\gamma_\sigma\slsh{p}_2\gamma^\beta(\slsh{p}-\slsh{k}_q)\gamma^\nu\right]}
{k^2(k_q-k)^2(p+k-k_q)^2(p_2+k)^2(p-k_q)^2};\\
(e):&\quad \int\frac{{\rm d}^d k}{(2\pi)^d}\frac{{\rm Tr}\left[
\slsh{k}_q\gamma^\tau(\slsh{k}_q+\slsh{k}-\slsh{p})\gamma^\alpha
(-\slsh{p}_2+\slsh{k})\gamma_\sigma\slsh{p}_2\gamma^\beta(\slsh{p}-\slsh{k}_q)\gamma^\nu\right]}{k^2(p-k)^2(k_q+k-p)^2(k-p_2)^2(p-k_2)^2}
V_{\tau\sigma\mu}(p-k,k,-p),
\end{align}
\label{diagsfull}
\end{subequations}
where
\begin{equation}
V_{\alpha\beta\gamma}(p_1,p_2,p_3)=(p_{1\gamma}-p_{2\gamma})\eta_{\alpha\beta}+(p_{2\alpha}-p_{3\alpha})\eta_{\beta\gamma}
+(p_{3\beta}-p_{1\beta})\eta_{\alpha\gamma}\,.
\label{Vdef}
\end{equation}
Each individual expression can be reduced to scalar integrals using the well-known Passarino-Veltman reduction algorithm~\cite{Passarino:1978jh}. To carry out this and also Dirac algebra, we use the \texttt{FeynCalc}~\cite{Mertig:1990an,Shtabovenko:2016sxi,Shtabovenko:2020gxv} plugin for \texttt{Mathematica}~\cite{Mathematica}. Explicit results for scalar integrals (including both triangles and boxes, both of which contribute leading poles) are taken from Ref.~\cite{Ellis:2007qk}. We may then substitute the Mandelstam invariants of Eqs.~(\ref{mandies}), parameterised according to Eq.~(\ref{mandiesparam}). Finally, one may expand to in the limit of Eq.~(\ref{xzlim}), keeping leading terms only. We find that the results of all non-zero diagrams explicitly agree with Eqs.~(\ref{Fares}, \ref{diagresults}). 

To summarise, we have obtained two properties: (i) the generalised soft-emission lines introduced in Section~\ref{sec:generalised} correctly capture the leading virtual corrections to the off-diagonal partonic channel in DIS at NLO; (ii) the additional gluon corrections to this partonic channel exponentiate, and hence the leading virtual corrections at this order do as well. This confirms the conjectures presented in Refs.~\cite{Moult:2019uhz,Beneke:2020ibj}, which we thus obtain as a result of the replica trick.

\section{Discussion}
\label{sec:discuss}

In this paper, we have addressed whether certain perturbative corrections in QCD can be shown to exponentiate, and thus be resummed to all orders in perturbation theory. It is very well-known that this is true for soft gluon emissions, and a recent series of papers has developed an elegant method for proving exponentiation properties in (multi-)parton scattering~\cite{Gardi:2010rn,Gardi:2013saa,Gardi:2011yz,Gardi:2011wa,Dukes:2013gea,Falcioni:2014pka,Gardi:2021gzz,Almelid:2017qju,Almelid:2015jia}, based on a variant of the replica trick that originated in statistical physics~\cite{Replica}. Here, we have applied this technique to the exponentiation of soft (anti-)quark emissions, in addition to soft gluons. Previous replica trick arguments for gluon emissions have relied upon the fact that the soft function describing soft gluon emissions is expressed in terms of vacuum expectation values of Wilson lines, where one may define the soft function at either amplitude or cross-section level. One may then write a generating functional for these VEVs, whose path integral description provides the basis for applying the replica trick. The latter involves considering a modified theory containing $N$ identical non-interacting copies, or {\it replicas}, of the soft gluon field. One may then show that the ${\cal O}(N)$ part of each diagram in the replicated theory exponentiates, such that upon setting $N\rightarrow 1$ one establishes exponentiation in the original (non-replicated) theory. Crucially, this argument can be applied at either amplitude or cross-section level, and we have here provided more rigorous arguments than have appeared before, regarding how the replica trick works for real emissions.

As demonstrated in this paper, applying the replica trick to soft (anti-)quark emission is also possible, provided that one first identifies a generalised soft-emission line operator that can be used to describe the emission of any soft parton. The latter may be described using so-called {\it emission factors}~\cite{vanBeekveld:2019prq}, valid only in the leading soft limit. Assembling these into something that resembles a Wilson line is ultimately nothing more than a book-keeping exercise. It differs from the formal definition of a Wilson line in that it does not share the usual gauge transformation properties, and that the resulting object is matrix-valued in not only colour space, but also parton species/spin. Nevertheless, it allows us to successfully apply the replica trick, which has some interesting additional features when compared with its previous incarnation for soft gluons. As in the previous case, one must introduce a replica-ordering operator as a means to elucidate which diagrams exponentiate. In the gluon case, this affects the colour factors of each graph, such that the colour factor in the logarithm of the soft amplitude is different to the colour factor in the amplitude itself. This remains true in the case of general soft parton emission, but with the additional complication that the replica-ordering operator also modifies the kinematic numerator of each graph. We verified in a simple example that this is indeed what is required to reproduce the structure of the soft amplitude, upon expanding the exponential form of the soft function.

To illustrate our arguments, we rederived the exponentiation properties of certain real emission corrections in DIS, agreeing with the results of Refs.~\cite{Vogt:2010cv,vanBeekveld:2021mxn}. Having verified our approach in a known example, we then turned to the unproven conjectures of Refs.~\cite{Moult:2019uhz,Beneke:2020ibj}, namely that the leading virtual corrections to the NLO off-diagonal partonic channel in DIS exponentiate. This necessitated a slightly modified form of the replica trick, in which some but not all soft parton fields are replicated. This in turn coincides with the fact that only a single soft quark emission is needed to generate leading logarithmic threshold effects at NLP in the threshold expansion. We confirm, however, that the exponentiation of the relevant virtual contributions indeed follows straightforwardly from this modified replica argument.

 We expect that our results are much more general than the specific examples presented here. That is, the replica trick in principle implies that soft parton emissions completely exponentiate, such that certain perturbative contributions of leading logarithmic type would also exponentiate at NNLP order and beyond in the threshold expansion. It would then be important to ascertain whether these types overlap with other sources of leading logarithms. Furthermore, although we have considered DIS in this paper, our arguments can be applied to any scattering process, including those of direct phenomenological interest for the ongoing LHC experiments. We thus foresee a programme of work in applying our methods to resum NLP contributions in a variety of contexts. We hope our results also inspire further investigations of what the replica trick may be useful for.

\section*{Acknowledgments}
This work has been supported by the UK Science and Technology Facilities Council (STFC) Consolidated Grant ST/P000754/1 "String Theory, gauge theory and duality" (CDW), and ST/T000864/1 (MvB). 
The work of LV has been partly supported by Fellini - Fellowship for Innovation at INFN, funded by the European Union's Horizon 2020 research programme under the Marie Sk\l{}odowska-Curie Cofund Action, grant agreement no. 754496.
LV thanks the Erwin-Schrödinger International Institute for Mathematics and Physics at the University of Vienna for partial support during the Programme ``Quantum Field Theory at the Frontiers of the Strong Interactions", July 31 - September 1, 2023, and the Galileo Galilei Institute for Theoretical Physics of INFN, Florenze, for partial support during the Programme ``Theory Challenges in the Precision Era of the Large Hadron Collider", August 28 - October 13, 2023.

\appendix

\section{Kinematic factors for soft emissions}
\label{app:kin-factors}

Here we summarise our results for the kinematic 
factors that feature in the transition matrix elements
defined in Eq.~\eqref{emissionfacB}. 
For an incoming hard parton we have
\begin{align}
\label{Qvals}
Q_{\rm in}^{C, BA}(p) = \begin{cases}
p^{\mu}\delta^{\hat{b}\hat{a}} &
\text{for $q^{\hat{a}}(p)
\to q^{\hat{b}}(p-k)\,g^{\mu}(k)$ 
\,\,\,(diagram (a)),} \\
p^{\mu}\delta^{\hat{a}\hat{b}} &
\text{for $\bar{q}^{\hat{a}}(p)
\to \bar{q}^{\hat{b}}(p-k)\,g^{\mu}(k)$ 
\,\,\,(diagram (b)),} \\
p^{\mu}\eta^{\alpha\beta} &
\text{for $g^{\alpha}(p)
\to g^{\beta}(p-k)\,g^{\mu}(k)$ 
\,\,\,(diagram (c)),} \\
-\frac{1}{2}(\gamma^\beta)_{\hat{c}\hat{a}} &
\text{for $q^{\hat{a}}(p)
\to g^{\beta}(p-k)\,q^{\hat{c}}(k)$ 
\,\,\,(diagram (d)),} \\
-\frac{1}{2}(\gamma^\beta)_{\hat{a}\hat{c}} &
\text{for $\bar{q}^{\hat{a}}(p)
\to g^{\beta}(p-k)\,\bar{q}^{\hat{c}}(k)$ 
\,\,\,(diagram (e)),} \\
\frac{1}{2}(\gamma^\alpha\slsh{p})_{\hat{c}\hat{b}} &
\text{for $g^{\alpha}(p)
\to \bar{q}^{\hat{b}}(p-k)\,q^{\hat{c}}(k)$
\,\,\,(diagram (f)),} \\
\frac{1}{2}(\slsh{p}\gamma^\alpha)_{\hat{b}\hat{c}} &
\text{for $g^{\alpha}(p)
\to q^{\hat{b}}(p-k)\,\bar q^{\hat{c}}(k)$ 
\,\,\,(diagram (g)),} \\
0 & \text{otherwise.}
\end{cases}
\end{align}
For an outgoing hard parton the kinematic factor reads
\begin{align}
\label{Qvals-out}
Q_{\rm out}^{C, AB}(p) = \begin{cases}
p^{\mu}\delta^{\hat{b}\hat{a}} 
&\text{for $q^{\hat{b}}(p+k)
\to q^{\hat{a}}(p)\,g^{\mu}(k)$,} \\
p^{\mu}\delta^{\hat{a}\hat{b}} 
&\text{for $\bar{q}^{\hat{b}}(p+k)
\to \bar{q}^{\hat{a}}(p)\,g^{\mu}(k)$,} \\
p^{\mu}\eta^{\alpha\beta} 
&\text{for $g^{\beta}(p+k)
\to g^{\alpha}(p)\,g^{\mu}(k)$,} \\
\frac{1}{2}(\gamma^\alpha\slsh{p})_{\hat{c}\hat{b}} 
&\text{for $q^{\hat{b}}(p+k)
\to g^{\alpha}(p)\,q^{\hat{c}}(k)$,} \\
\frac{1}{2}(\slsh{p}\gamma^\alpha)_{\hat{b}\hat{c}} 
&\text{for $\bar{q}^{\hat{b}}(p+k)
\to g^{\alpha}(p)\,\bar{q}^{\hat{c}}(k)$,} \\
\frac{1}{2}(\gamma^\beta)_{\hat{a}\hat{c}} 
&\text{for $g^{\beta}(p+k)
\to q^{\hat{a}}(p)\,\bar q^{\hat{c}}(k)$,}\\
\frac{1}{2}(\gamma^\beta)_{\hat{c}\hat{a}} 
&\text{for $g^{\beta}(p+k)
\to \bar{q}^{\hat{a}}(p)\, q^{\hat{c}}(k)$,} \\
0 & \text{otherwise.}
\end{cases}
\end{align}

\section{One-loop soft function}
\label{App:NLOsoft}
In this appendix we show how to obtain the structure function
for DIS at LL.
We start with our expression for the structure function
\begin{eqnarray} \label{F2defHiger} \nonumber
F_2(x,Q^2) &=& \frac{1}{{\cal N}_s {\cal N}_c} 
\sum_m \int {\rm d}\Phi_{m+1} \, 
T_2^{\alpha\beta}\,\Big( 
\bar \xi^{\bar{A}_1'}_{A_1'}(p) \, 
\big[\mathcal{A}^{\dag(0)\beta}
\big]^{\bar{A}_2'\bar{B}_2'}_{A_2'B_2'} 
\, \xi^{\bar{B}'_1}_{B'_1}(p_2) \Big) \\ 
&&\hspace{2.0cm}\times \,
\big({\cal S}_m\big)^{\bar{A}'_1\bar{A}'_2\, 
\bar{B}'_2\bar{B}'_1\, 
\bar{B}_1\bar{B}_2\, 
\bar{A}_2\bar{A}_1}_{A'_1 A'_2
\, B'_2 B'_1\,B_1 B_2\, A_2 A_1} \, 
\Big(\bar \xi^{\bar{B}_1}_{B_1}(p_2) \,
\big[\mathcal{A}^{(0)\alpha}
\big]^{\bar{B_2}\bar A_2}_{B_2A_2}
\,\xi^{\bar{A}_1}_{A_1}(p)\Big),
\end{eqnarray}
obtained after inserting Eq.~\eqref{MsqSoftGen2} into Eq.~\eqref{F2def},
and accounting for the $m+1$-body phase space. 

\subsection{NLO DIS soft function}
To see how Eq.~\eqref{F2defHiger} can be used in practice
let us first consider the NLO contribution, 
which is obtained by expanding the 
generalised soft emission operators 
at NLO. Matrix multiplication with 
the vector of tree level amplitudes 
in Eq.~\eqref{AA0def} then selects 
which term in the transition matrices 
of Eqs.~\eqref{trans-mat-in} and
\eqref{trans-mat-out} contributes 
at NLO. Furthermore, as discussed
at the end of Section~\ref{sec:multiple}, 
we can keep track of the counting of 
subleading-power contributions
by recalling that emissions on the 
diagonal of the transition operator 
${\cal T}$ count as LP (we
do not consider next-to-soft gluon emissions), while off-diagonal
emissions count as $\sqrt{\rm NLP}$.
Fermion number conservation requires 
off-diagonal emissions to appear always 
in an even number, such that two subsequent 
soft quark emissions give an overall 
contribution at NLP. 
Given these considerations, it is easy 
to check that at NLO we get contributions 
up to NLP, given by the diagrams in 
Figs.~\ref{fig:softcalcLP} and~\ref{fig:softcalcNLP} in Section~\ref{sec:DISreal}.

To be explicit, we introduce a 
soft emission matrix, corresponding to the 
transition matrix in Eqs.~\eqref{TransitionMatrDef} 
and \eqref{TransitionMatrOutDef}: 
\begin{equation} \label{STransitionMatrDef}
\big[{\cal S}^{\rm in/out}(p,k)\big]^{\bar C,\bar A\bar B}_{C,AB} = 
\frac{g_s}{p\cdot k} \big[{\cal T}^{\rm in/out}(p)\big]^{\bar C,\bar A\bar B}_{C,AB}.
\end{equation}
Upon inserting the transition matrices, there 
are a number of contributions that appear. 
Considering for now just the LP contribution, 
we need to compute the virtual exchange, 
corresponding to diagram (a) in Fig.~\ref{fig:softcalcLP}:
\begin{eqnarray} 
\label{SNLOqqqqLPv}\nonumber
\big[{\cal S}_{0}^{(1)}
\big]^{i'_1 i'_2\, j'_2 j'_1\, 
j_1 j_2\, i_2 i_1}_{\hat a'_1 \hat a'_2
\, \hat b'_2 \hat b'_1\, 
\hat b_1 \hat b_2\, \hat a_2 \hat a_1} && \\ \nonumber
&&\hspace{-3.5cm}=\,  \Bigg[ 
\Big(\delta_{\hat a'_1 \hat a'_2} \delta^{i'_1 i'_2}\Big) 
\Big(\delta_{\hat b'_2 \hat b'_1} \delta^{j'_2 j'_1}\Big) 
\big[{\cal S}_{qq}^{\rm out}(p_2,-k)\big]^{c',j_1 j_2}_{\mu',\hat b_1 \hat b_2} 
\big[{\cal S}_{qq}^{\rm in}(p,k)\big]^{c,i_1 i_2}_{\mu,\hat a_1 \hat a_2} \\  \nonumber
&&\hspace{-2.8cm}\,+
\big[{\cal S}_{qq}^{\rm in \dag}(p,-k)\big]^{c',i'_1 i'_2}_{\mu',\hat a'_1 \hat a'_2} 
\big[{\cal S}_{qq}^{\rm out \dag}(p_2,k)\big]^{c,j'_1 j'_2}_{\mu,\hat b'_2 \hat b'_1}  
\Big(\delta_{\hat b_1 \hat b_2} \delta^{j_1 j_2}\Big) 
\Big(\delta_{\hat a_2 \hat a_1} \delta^{i_2 i_1}\Big) \Bigg] 
\frac{i\, \varepsilon^{\dag c}_{\mu}(k)
\varepsilon^{c'}_{\mu'}(k)}{k^2} \\ \nonumber
&&\hspace{-3.5cm}=\, 
g_s^2 \Bigg[ 
\Big(\delta_{\hat a'_1 \hat a'_2} \delta^{i'_1 i'_2}\Big) 
\Big(\delta_{\hat b'_2 \hat b'_1} \delta^{j'_2 j'_1}\Big) 
\Bigg(\frac{p_2^{\mu'}\,\delta_{\hat b_1 \hat b_2} 
\, T_{j_1 j_2}^{c'} }{p_2\cdot k}\Bigg)
\Bigg(\frac{p^{\mu}\,\delta_{\hat a_2 \hat a_1} 
\, T_{i_2 i_1}^{c} }{p\cdot k} \Bigg) \\ 
&&\hspace{-2.8cm}\,+
\Bigg(\frac{p^{\mu'}\,\delta_{\hat a'_1 \hat a'_2} 
\, T_{i'_1 i'_2}^{c'} }{p\cdot k}\Bigg)
\Bigg(\frac{p_2^{\mu}\,\delta_{\hat b'_2 \hat b'_1} 
\, T_{j'_2 j'_1}^{c} }{p_2\cdot k} \Bigg) 
\Big(\delta_{\hat b_1 \hat b_2} \delta^{j_1 j_2}\Big) 
\Big(\delta_{\hat a_2 \hat a_1} \delta^{i_2 i_1}\Big) \Bigg] 
\frac{i\, \varepsilon^{\dag c}_{\mu}(k)
\varepsilon^{c'}_{\mu'}(k)}{k^2},
\end{eqnarray}
but also the one-real emission contribution (diagrams (b), (c), (d) in Fig.~\ref{fig:softcalcLP}),
\begin{eqnarray} 
\label{SNLOqqqqLPr}\nonumber
\big[{\cal S}_{g}^{(1)}
\big]^{i'_1 i'_2\, j'_2 j'_1\, 
j_1 j_2\, i_2 i_1}_{\hat a'_1 \hat a'_2
\, \hat b'_2 \hat b'_1\, 
\hat b_1 \hat b_2\, \hat a_2 \hat a_1} && \\ \nonumber
&&\hspace{-3.5cm}=\, \Bigg[ 
\Big(\delta_{\hat a'_1 \hat a'_2} \delta^{i'_1 i'_2}\Big) 
\big[{\cal S}_{qq}^{\rm out \dag}(p_2,-k)\big]^{c',j'_1 j'_2}_{\mu',\hat b'_1 \hat b'_2} 
\Big(\delta_{\hat b_1 \hat b_2} \delta^{j_1 j_2}\Big) 
\big[{\cal S}_{qq}^{\rm in}(p,k)\big]^{c,i_1 i_2}_{\mu,\hat a_1 \hat a_2} \\ \nonumber
&&\hspace{-2.8cm}\,+
\big[{\cal S}_{qq}^{\rm in \dag}(p,-k)\big]^{c',i'_1 i'_2}_{\mu',\hat a'_1 \hat a'_2}
\Big(\delta_{\hat b'_2 \hat b'_1} \delta^{j'_2 j'_1}\Big) 
\Big(\delta_{\hat b_1 \hat b_2} \delta^{j_1 j_2}\Big) 
\big[{\cal S}_{qq}^{\rm in}(p,k)\big]^{c,i_1 i_2}_{\mu,\hat a_1 \hat a_2} \\ \nonumber
&&\hspace{-2.8cm}\,+
\Big(\delta_{\hat a'_1 \hat a'_2} \delta^{i'_1 i'_2}\Big) 
\big[{\cal S}_{qq}^{\rm out \dag}(p_2,-k)\big]^{c',j'_1 j'_2}_{\mu',\hat b'_1 \hat b'_2} 
\big[{\cal S}_{qq}^{\rm out}(p_2,k)\big]^{c,j_1 j_2}_{\mu',\hat b_1 \hat b_2} 
\Big(\delta_{\hat a_2 \hat a_1} \delta^{i_2 i_1}\Big) \\ \nonumber
&&\hspace{-2.8cm}\,+
\big[{\cal S}_{qq}^{\rm in \dag}(p,-k)\big]^{c',i'_1 i'_2}_{\mu',\hat a'_1 \hat a'_2}
\Big(\delta_{\hat b'_2 \hat b'_1} \delta^{j'_2 j'_1}\Big) 
\big[{\cal S}_{qq}^{\rm out}(p_2,k)\big]^{c,j_1 j_2}_{\mu',\hat b_1 \hat b_2} 
\Big(\delta_{\hat a_2 \hat a_1} \delta^{i_2 i_1}\Big)\Bigg] 
\varepsilon^{\dag c}_{\mu}(k)
\varepsilon^{c'}_{\mu'}(k) \\  \nonumber
&&\hspace{-3.5cm}=\, 
g_s^2 \Bigg[ 
\Big(\delta_{\hat a'_1 \hat a'_2} \delta^{i'_1 i'_2}\Big) 
\Bigg(\frac{-p_2^{\mu'}\,\delta_{\hat b'_2 \hat b'_1} 
\, T_{j'_2 j'_1}^{c'} }{p_2\cdot k} \Bigg) 
\Big(\delta_{\hat b_1 \hat b_2} \delta^{j_1 j_2}\Big) 
\Bigg(\frac{p^{\mu}\,\delta_{\hat a_2 \hat a_1} 
\, T_{i_2 i_1}^{c} }{p\cdot k}\Bigg) \\ \nonumber
&&\hspace{-2.8cm}\,+
\Bigg(\frac{p^{\mu'}\,\delta_{\hat a'_1 \hat a'_2} 
\, T_{i'_1 i'_2}^{c'} }{p\cdot k}\Bigg)
\Big(\delta_{\hat b'_2 \hat b'_1} \delta^{j'_2 j'_1}\Big) 
\Big(\delta_{\hat b_1 \hat b_2} \delta^{j_1 j_2}\Big) 
\Bigg(\frac{p^{\mu}\,\delta_{\hat a_2 \hat a_1} 
\, T_{i_2 i_1}^{c} }{p\cdot k} \Bigg) \\ \nonumber
&&\hspace{-2.8cm}\,+
\Big(\delta_{\hat a'_1 \hat a'_2} \delta^{i'_1 i'_2}\Big) 
\Bigg(\frac{-p_2^{\mu'}\,\delta_{\hat b'_2 \hat b'_1} 
\, T_{j'_2 j'_1}^{c'} }{p_2\cdot k} \Bigg) 
\Bigg(\frac{-p_2^{\mu}\,\delta_{\hat b_1 \hat b_2} 
\, T_{j_1 j_2}^{c} }{p_2\cdot k}\Bigg)
\Big(\delta_{\hat a_2 \hat a_1} \delta^{i_2 i_1}\Big) \\
&&\hspace{-2.8cm}\,+
\Bigg(\frac{p^{\mu'}\,\delta_{\hat a'_1 \hat a'_2} 
\, T_{i'_1 i'_2}^{c'} }{p\cdot k}\Bigg)
\Big(\delta_{\hat b'_2 \hat b'_1} \delta^{j'_2 j'_1}\Big) 
\Bigg(\frac{-p_2^{\mu}\,\delta_{\hat b_1 \hat b_2} 
\, T_{j_1 j_2}^{c} }{p_2\cdot k}\Bigg)
\Big(\delta_{\hat a_2 \hat a_1} \delta^{i_2 i_1}\Big)\Bigg] 
\varepsilon^{\dag c}_{\mu}(k)
\varepsilon^{c'}_{\mu'}(k).
\end{eqnarray}

We used the subscript ($g$) to refer to the exchanged soft 
particle, whereas the superscript ($(1)$) indicates the perturbative order of the soft function. The indices represent colour/flavour indices of either the hard-scattering amplitude or the external lines. 
Inserting Eqs. (\ref{SNLOqqqqLPv}) and (\ref{SNLOqqqqLPr}) 
into Eq. (\ref{F2defHiger}) we obtain the LP contribution 
to the structure function:
\begin{eqnarray} \label{F2NLOLP} 
F_2^{(1)}(x,Q^2)|_{\rm LP,LL} &=& g_s^2 \, C_F\,  
T_2^{\alpha\beta} \bigg\{   
\int {\rm d}\Phi_{1} \, [d^d k] \,
\overline{|{\cal A}^{(0)}|^2_{\alpha\beta}} \,
\frac{2i}{k^2} \, \frac{p^{\mu}  p_2^{\mu'}  
{\cal P}_{\mu\mu'}(k)}{p\cdot k \, p_2\cdot k} \\ \nonumber
&&+\,\int {\rm d}\Phi_{2} \,
\overline{|{\cal A}^{(0)}|^2_{\alpha\beta}}\,
\Big[\frac{p^{\mu} p^{\mu'} 
{\cal P}_{\mu\mu'}(k)}{(p\cdot k)^2}
- \frac{2 p^{\mu} p_2^{\mu'} 
{\cal P}_{\mu\mu'}(k)}{p\cdot k \, p_2\cdot k}
+ \frac{p_2^{\mu} p_2^{\mu'} 
{\cal P}_{\mu\mu'}(k)}{(p_2\cdot k)^2} \Big]\bigg\},
\end{eqnarray}
where the one-particle phase 
space ${\rm d}\Phi_{1}$ has been defined 
in Eq. \eqref{1PPS}, the loop integral 
$[{\rm d}^d k]$ is defined as 
\begin{equation}
[{\rm d}^d k] \equiv 
\bigg(\frac{\mu^2 e^{\gamma_E}}{4\pi}\bigg)^{\epsilon} 
\int \frac{{\rm d}^d k}{(2\pi)^d},     
\end{equation}
and the two-particle phase space reads 
\begin{equation}\label{2PPS}
\int {\rm d}\Phi_2 = 
\int \frac{{\rm d}^dp_2}{(2\pi)^{d-1}}\, [{\rm d}^d k]\, \delta^{(d)}(p+q-p_2-k) 
\, \delta(p_2^2)\theta(p_2^0)
\, \delta(k^2)\theta(k^0).
\end{equation}
Furthermore, we have introduced 
the gluon polarisation sum 
\begin{equation}\label{polsum2}
{\cal P}_{\mu\nu}(k) = \sum_{\rm pol.}
\epsilon^\dag_\mu(k) \, \epsilon_\nu(k).
\end{equation}
The integrals in Eq.~\eqref{F2NLOLP} can be 
calculated with standard methods. 
The integration over the phase space 
${\rm d}\Phi_{2}$ can be performed 
by for example introducing 
the Sudakov decomposition, like was
done in Ref.~\cite{vanBeekveld:2021mxn}, i.e.
\begin{equation}\label{Sudakovk}
k_{\mu} = \alpha p_{\mu} + \beta q'_{\mu} 
+ k_{\perp\mu}, \qquad \qquad 
k_{\perp}\cdot p = k_{\perp}\cdot q' = 0,
\end{equation}    
where the vector $q'$ is defined as follows: 
\begin{equation}\label{qprimeDef}
q' = q + x p, \qquad \qquad
p\cdot q' = p\cdot q \neq 0. 
\end{equation}
With this decomposition the phase 
space in Eq.~\eqref{2PPS} reads 
\begin{equation}\label{2PPSbis}
\int {\rm d}\Phi_2 =
\frac{1}{8\pi} \frac{e^{\epsilon \gamma_E} 
x^{\epsilon}}{\Gamma(1-\epsilon)}
\bigg(\frac{\mu^2}{Q^2} \bigg)^{\epsilon}  
\int_0^1 {\rm d}\alpha\, {\rm d}\beta\, (\alpha \beta)^{-\epsilon} 
\delta\big( (1-x)(1-\beta)-\alpha\big).
\end{equation}
Using this we obtain for the real soft gluon emission
\begin{eqnarray} \label{F2NLOLPRealRes} 
F_2^{(1)}(x,Q^2)|^{\rm real}_{\rm LP,LL} 
&=&  \frac{\as\, C_F}{4\pi} \bigg\{ 
-\frac{4}{\epsilon} + \ldots \bigg\} (1-x)^{-\epsilon},
\end{eqnarray}
where the dots indicate contributions beyond LL. 

For the NLP contribution one needs
to consider the diagrams in Fig.~\ref{fig:softcalcNLP}.
We have 
\begin{eqnarray} 
\label{SNLOqbargNLP} \nonumber
\big[{\cal S}_{\bar q}^{(1)}
\big]^{a'_1 i'_2\, 
j'_2 j'_1\, 
j_1 j_2\, 
i_2 a_1}_{\alpha'_1 \hat a'_2
\, \hat b'_2 \hat b'_1\,
\hat b_1 \hat b_2\, \hat a_2 \alpha_1} &=& 
\bigg( \bar v_{\hat a_3'}^{i_3'}(k) \,
\big[{\cal S}_{gq}^{\rm in \dag}(p,-k)\big]^{i_3',a_1'  i_2' }_{\hat a_3',\alpha_1' \hat a_2'} \bigg) 
\bigg(\delta_{\hat b'_2 \hat b'_1} \delta^{j'_2 j'_1}\bigg) \\ \nonumber
&&\hspace{1.0cm}\times\, 
\bigg(\delta_{\hat b_1 \hat b_2} \delta^{j_1 j_2}\bigg) 
\bigg( \big[{\cal S}_{gq}^{\rm in}(p,k)\big]^{i_3,a_1 i_2}_{\hat a_3, \alpha_1 \hat a_2}
\,v_{\hat a_3}^{i_3}(k) \bigg) \\ \nonumber
&=& g_s^2 
\Bigg( \bar v_{\hat a_3'}^{i_3'}(k) \, T^{a_1'}_{i_3'i_2'}
\frac{(\gamma^{\alpha_1'} \slashed{p})_{\hat a_3'\hat a_2'}}{2p\cdot k} \Bigg) 
\bigg(\delta_{\hat b'_2 \hat b'_1} \delta^{j'_2 j'_1}\bigg) \\
&&\hspace{1.0cm}\times\, 
\bigg(\delta_{\hat b_1 \hat b_2} \delta^{j_1 j_2}\bigg) 
\Bigg( \frac{(\slashed{p}\gamma^{\alpha_1})_{\hat a_2\hat a_3}}{2p\cdot k} 
T^{a_1}_{i_2i_3} \,v_{\hat a_3}^{i_3}(k) \Bigg),
\end{eqnarray}
and 
\begin{eqnarray} 
\label{SNLOqgNLP}\nonumber
\big[{\cal S}_{q}^{(1)}
\big]^{a'_1 i'_2\, 
j'_2 j'_1\, 
j_1 j_2\, 
i_2 a_1}_{\alpha'_1 \hat a'_2
\, \hat b'_2 \hat b'_1\,
\hat b_1 \hat b_2\, \hat a_2 \alpha_1} &=&  
\bigg(\delta_{\hat b'_1 \hat b'_2} \delta^{j'_1 j'_2}\bigg)
\bigg( \big[{\cal S}_{gq}^{\rm in \dag}(p,-k)\big]^{i_2',a_1'  i_3' }_{\hat a_2',\alpha_1' \hat a_3'}  
\, u_{\hat a_3'}^{i_3'}(k) \bigg) \\ \nonumber
&&\hspace{1.0cm}\times\, 
\bigg(\bar u_{\hat a_3}^{i_3}(k) \,
\big[{\cal S}_{gq}^{\rm in}(p,k)\big]^{i_2,a_1 i_3}_{\hat a_2, \alpha_1 \hat a_3} \bigg)
\bigg(\delta_{\hat b_2 \hat b_1} \delta^{j_2 j_1}\bigg) \\ \nonumber
&=& g_s^2 
\bigg(\delta_{\hat b'_1 \hat b'_2} \delta^{j'_1 j'_2}\bigg)
\Bigg(\frac{(\slashed{p}\gamma^{\alpha_1'})_{\hat a_2'\hat a_3'}}{2p\cdot k}
T^{a_1'}_{i_2'i_3'} \, u_{\hat a_3'}^{i_3'}(k) \Bigg) \\ 
&&\hspace{1.0cm}\times\, 
\Bigg(\bar u_{\hat a_3}^{i_3}(k) \,
\frac{(\gamma^{\alpha_1}\slashed{p})_{\hat a_3\hat a_2}}{2p\cdot k} 
T^{a_1}_{i_3i_2} \Bigg)
\bigg(\delta_{\hat b_2 \hat b_1} \delta^{j_2 j_1}\bigg).
\end{eqnarray}
There is an additional off-diagonal 
contribution, where the soft quark is 
emitted from the line with momentum $p_2$. 
This does not contribute at LL accuracy, 
and so we do not list it here. 
After inserting Eqs.~(\ref{SNLOqgNLP}) and 
(\ref{SNLOqbargNLP}) into Eq.~(\ref{F2defHiger})
we obtain the NLP contribution for the
emission of a soft (anti-) quark:
\begin{eqnarray} \label{F2NLOqNLP} \nonumber
F_{2;q+\bar q}^{(1)}(x,Q^2)|_{\rm NLP,LL} &=& 
\frac{g_s^2\, e^2 \, T_R\, n_f}{d-2} 
\int {\rm d}\Phi_{2} \, T_2^{\alpha\beta} \,  
\bigg\{
{\rm Tr} \big[ 
\slashed{k} \gamma^{\alpha_1'} \slashed{p} \gamma_{\beta} 
\slashed{p}_2 \gamma_{\alpha} \slashed{p} \gamma^{\alpha_1}\big] \\ 
&&\hspace{3.0cm}+\, {\rm Tr} \big[ 
\slashed{k} \gamma^{\alpha_1} \slashed{p} \gamma_{\alpha} 
\slashed{p}_2 \gamma_{\beta} \slashed{p} \gamma^{\alpha'_1}\big]
\bigg\} \frac{{\cal P}_{\alpha_1\alpha_1'}(p)}{(2 p\cdot k)^2},
\end{eqnarray}
where the contribution in the first line 
corresponds to the term proportional to
\eqref{SNLOqbargNLP}, the contribution
in the second line originates from 
\eqref{SNLOqgNLP}, and we have included a 
factor $n_f$ to take into account the sum
over quark flavours. 
After integration over phase space we obtain
\begin{eqnarray} \label{F2NLOgqqqNLP} 
F_{2;q+\bar q}^{(1)}(x,Q^2)|_{\rm NLP,LL} 
&=&  \frac{2 \as\, T_R\, n_f}{4\pi} \bigg\{ 
-\frac{2}{\epsilon} + \ldots \bigg\} (1-x)^{-\epsilon}.
\end{eqnarray}
\subsection{NLP real emission contributions with $n-1$ soft-gluon emissions}
\label{app:sub:all-order-dis}
Here we report the result obtained after using Eq.~\eqref{MsqSoftGen2exp} 
and Eq.~\eqref{TransitionMatrDef} in Eq.~\eqref{MsqSoftGen2} with the gauge choice of Eq.~\eqref{physpols}. We focus on the NLP contribution (i.e.\ the emission of one soft quark).
\begin{figure}[t]
\begin{center}
\includegraphics[width=0.90\textwidth]{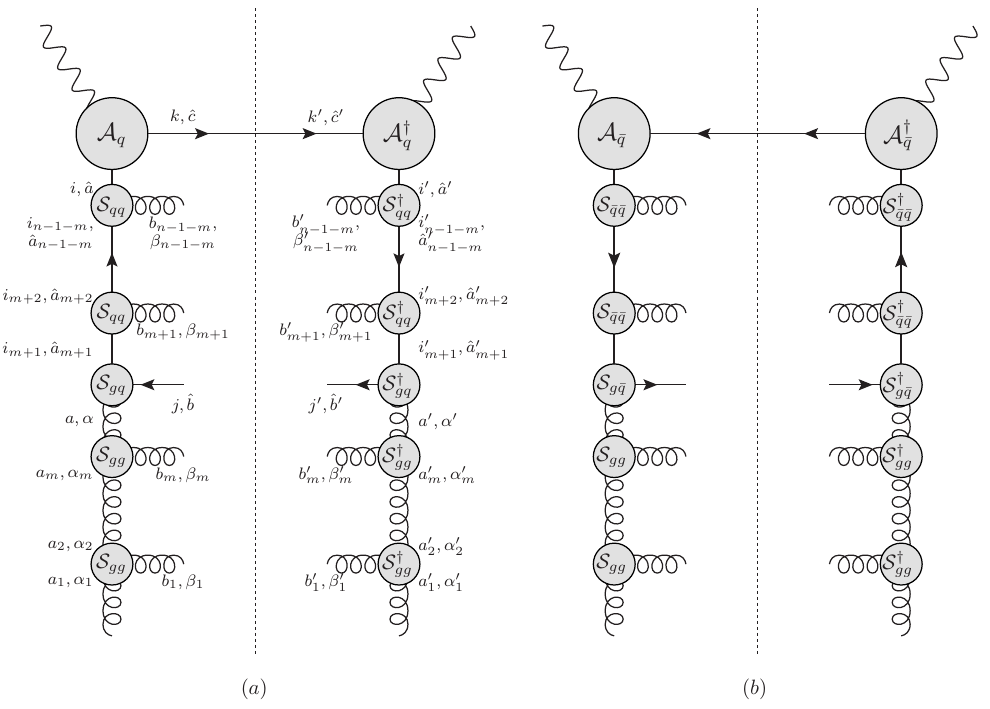}
  \caption{Ladder diagrams corresponding to Eq.~\eqref{MsqSoftGen2iterated}.}
  \label{fig:ladder}
\end{center}
\end{figure}
We obtain
\begin{eqnarray}\label{MsqSoftGen2iterated} \nonumber
\overline{|{\cal A}|^2_{\alpha\beta}}\big|_{\rm qg,\,real,\,LL}^{(n)} 
&=& \frac{g_s^{2n}\, e^2 \, n_f}{n!(d-2)(N_c^2-1)}  \\ \nonumber
&&\hspace{-2.5cm} \times \, \sum_{m=0}^{n-1}
\varepsilon_{a'_1}^{\alpha_1'}(p) 
\Big(\varepsilon_{b'_1}^{\beta'_1}(k_1)
\big[{\cal S}_{gg}^{\rm in \dag}(p,-k_1)\big]^{b'_1,a'_1 a'_2}_{\beta_1',\alpha'_1 \alpha'_2} \Big) \ldots
\Big(\varepsilon_{b'_m}^{\beta'_{m}}(k_{m})
\big[{\cal S}_{gg}^{\rm in \dag}(p,-k_{m})\big]^{b'_{m},a'_{m} a'}_{\beta_{m}',\alpha'_{m} \alpha'}\Big)
\\ \nonumber
&&\hspace{-2.5cm} \times \, \bigg[ \Big( \bar v_{\hat b'}^{j'}(k_q) \,
\big[{\cal S}_{gq}^{\rm in \dag}(p,-k_q)\big]^{j',a'  i_{m+1}'}_{\hat b',\alpha' \hat a_{m+1}'} \Big)  
\Big(\varepsilon_{b'_{m+1}}^{\beta'_{m+1}}(k_{m+1})
\big[{\cal S}_{qq}^{\rm in \dag}(p,-k_{m+1})\big]^{b'_{m+1},i'_{m+1} i'_{m+2}}_{\beta_{m+1}',\hat a'_{m+1} \hat a'_{m+2}}\Big)
\ldots \\ \nonumber
&&\hspace{-1.5cm} \times \,  
\Big(\varepsilon_{b'_{n-1-m}}^{\beta'_{n-1-m}}(k_{n-1-m})
\big[{\cal S}_{qq}^{\rm in \dag}(p,-k_{n-1-m})\big]^{b'_{n-1-m},i'_{n-1-m} i'}_{\beta_{n-1-m}',\hat a'_{n-1-m} \hat a'}\Big)
\big[\mathcal{A}_q^{\dag(0)\beta} \big]^{i'k'}_{\hat a' \hat c'} 
u^{k'}_{\hat c'}(p_2) \\[0.2cm] \nonumber
&&\hspace{-1.5cm} \times \, \bar u^{k}_{\hat c}(p_2)
\big[\mathcal{A}_q^{(0)\alpha} \big]^{ki}_{\hat c\hat a}
\Big(\varepsilon_{b_{n-1-m}}^{\beta_{n-1-m}}(k_{n-1-m})
\big[{\cal S}_{qq}^{\rm in}(p,k_{n-1-m})\big]^{b_{n-1-m},i_{n-1-m} i}_{\beta_{n-1-m},\hat a_{n-1-m} \hat a}\Big)
\ldots \\ \nonumber
&&\hspace{-1.5cm} \times \, 
\Big(\varepsilon_{b_{m+1}}^{\beta_{m+1}}(k_{m+1})
\big[{\cal S}_{qq}^{\rm in}(p,k_{m+1})\big]^{b_{m+1},i_{m+1} i_{m+2}}_{\beta_{m+1},\hat a_{m+1} \hat a_{m+2}}\Big)
\Big( \big[{\cal S}_{gq}^{\rm in}(p,k_q)\big]^{j,a  i_{m+1}}_{\hat b,\alpha \hat a_{m+1}} 
\, \bar v_{\hat b}^{j}(k_q) \Big) \\[0.2cm] \nonumber
&&\hspace{-2.0cm} +\,
\Big( \bar u_{\hat b}^{j}(k_q) \,
\big[{\cal S}_{g\bar q}^{\rm in}(p,k_q)\big]^{j,a  i_{m+1}}_{\hat b,\alpha \hat a_{m+1}} \Big)
\Big(\varepsilon_{b_{m+1}}^{\beta_{m+1}}(k_{m+1})
\big[{\cal S}_{\bar q\bar q}^{\rm in}(p,k_{m+1})\big]^{b_{m+1},i_{m+1} i_{m+2}}_{\beta_{m+1},\hat a_{m+1} \hat a_{m+2}}\Big)
\ldots \\ \nonumber
&&\hspace{-1.5cm} \times \, 
\Big(\varepsilon_{b_{n-1-m}}^{\beta_{n-1-m}}(k_{n-1-m})
\big[{\cal S}_{\bar q\bar q}^{\rm in}(p,k_{n-1-m})\big]^{b_{n-1-m},i_{n-1-m} i}_{\beta_{n-1-m},\hat a_{n-1-m} \hat a}\Big)
\big[\mathcal{A}_{\bar q}^{(0)\alpha} \big]^{ki}_{\hat c\hat a} v^{k}_{\hat c}(p_2) \\[0.2cm] \nonumber
&&\hspace{-1.5cm} \times \, \bar v^{k'}_{\hat c'}(p_2)
\big[\mathcal{A}_{\bar q}^{\dag(0)\beta}\big]^{i'k'}_{\hat a' \hat c'} 
\Big(\varepsilon_{b'_{n-1-m}}^{\beta'_{n-1-m}}(k_{n-1-m})
\big[{\cal S}_{\bar q\bar q}^{\rm in \dag}(p,-k_{n-1-m})\big]^{b'_{n-1-m},i'_{n-1-m} i'}_{\beta_{n-1-m}',\hat a'_{n-1-m} \hat a'}\Big)
\ldots \\ \nonumber  
&&\hspace{-1.5cm} \times \,  
\Big(\varepsilon_{b'_{m+1}}^{\beta'_{m+1}}(k_{m+1})
\big[{\cal S}_{\bar q\bar q}^{\rm in \dag}(p,-k_{m+1})\big]^{b'_{m+1},i'_{m+1} i'_{m+2}}_{\beta_{m+1}',\hat a'_{m+1} \hat a'_{m+2}}\Big)
\Big(\big[{\cal S}_{g\bar q}^{\rm in \dag}(p,-k_q)\big]^{j',a'  i_{m+1}'}_{\hat b',\alpha' \hat a_{m+1}'}  
u_{\hat b'}^{j'}(k_q) \Big) \bigg] \\ 
&&\hspace{-2.0cm} \times \, 
\Big(\varepsilon_{b_m}^{\beta_{m}}(k_{m})
\big[{\cal S}_{gg}^{\rm in}(p,k_{m})\big]^{b_{m},a_{m} a}_{\beta_{m},\alpha_{m} \alpha}\Big) 
\ldots \Big(\varepsilon_{b_1}^{\beta_1}(k_1)
\big[{\cal S}_{gg}^{\rm in}(p,k_1)\big]^{b_1,a_1 a_2}_{\beta_1,\alpha_1 \alpha_2} \Big)
\varepsilon_{a_1}^{\alpha_1}(p),
\end{eqnarray}
where the two terms correspond respectively to 
the ladder diagrams (a) and (b) in Fig.~\ref{fig:ladder}.

\section{Results for soft integrals}
\label{app:integrals}

In this appendix, we collect results for the various soft integrals used in section~\ref{sec:DIS} to derive the form of the endpoint contributions in deep-inelastic scattering. To show how our results have been obtained, we first consider the triangle integral
\begin{align}
    \label{T1calc1}
    T_1(p,p_2;k_q)&=\int \frac{{\rm d}^d k}{(2\pi)^d}\frac{1}
    {k^2\,\,2p\cdot(k-k_q)\,\,2 p_2\cdot k}\\
    &=2\int\frac{{\rm d}^d l}{(2\pi)^d}\left[\prod_{i=1}^3\int_0^1  {\rm d}\alpha_i
    \right]\frac{\delta(\sum_i \alpha_i-1)}
    {\alpha_1^3 (l^2 - M^2)^3}, \notag
\end{align}
where we have applied the Feynman parameter trick
\begin{align}
\frac{1}{A_1^{a_1} A_2^{a_2}\ldots A_n^{a_n}}&=
\frac{\Gamma(a_1+a_2+\ldots+a_n)}
{\Gamma(a_1)\Gamma(a_2)\ldots \Gamma(a_n)}
\left[\prod_{i=1}^n \int_0^1  {\rm d}\alpha_i\right]\delta
\left(1-\sum_{i=1}^n \alpha_i\right)\notag\\
&\quad\times \frac{\alpha_1^{a_1-1}
\alpha_2^{a_2-1}\ldots \alpha_n^{a_n-1}}{(A_1\alpha_1+A_2\alpha_2
+\ldots+A_n\alpha_n)^{a_1+a_2+\ldots+a_n}},
    \label{Feynman}
\end{align}
and also defined (see Eq.~\eqref{mandies} for the definition of $t$ and $u$)
\begin{align}
    l^\mu=k_q^\mu+\frac{\alpha_2 p^\mu+\alpha_3 p_2^\mu}{\alpha_1}\,,\quad
    M^2=\frac{-\alpha_1\alpha_2\, t-\alpha_2\alpha_3
    \,u}{\alpha_1^2}.
    \label{T1calc2}
\end{align}
The momentum integral may be carried out using the identity
\begin{equation}
    \int\frac{ {\rm d}^dl}{(2\pi)^d}\frac{1}{(l^2-M^2)^n}
    =\frac{(-1)^n\, i}{(4\pi)^{d/2}}\frac{\Gamma(n-d/2)}{\Gamma(n)}
    (M^2)^{\frac{d}{2}-n},
    \label{momint}
\end{equation}
such that we find
\begin{align}
    T_1(p_1,p_2;k_2)&=-\frac{i\,\Gamma(3-d/2)}{(4\pi)^{d/2}}
\left[\prod_{i=1}^3\int_0^1{\rm d}\alpha_i\right]
    \frac{\delta\left(\sum_i \alpha_i-1\right)
    \alpha_1^{3-d}}{\left(-\alpha_1\alpha_2\,t-\alpha_2\alpha_3
    \,u\right)^{-d/2+3}}\,.
    \label{T1calc3}
\end{align}
We may now transform variables according to
\begin{equation}
(\alpha_1,\alpha_2,\alpha_3)=\left(1-x_1,x_1(1-x_2),x_1x_2\right),
    \label{T1vartrans}
\end{equation}
such that Eq.~(\ref{T1calc3}) becomes
\begin{align}
    \label{T1calc4}
    T_1(p,p_2;k_q)&=-\frac{i\,\Gamma(3-d/2) (-t)^{d/2-3}}{(4\pi)^{d/2}}
\int_0^1{\rm d}x_1 
   (1-x_1)^{-d/2} x_1^{d/2-2} \\
   & \qquad \times \int_0^1{\rm d}x_2(1-x_2)^{d/2-3}\left(1+\frac{x_1x_2
    u}{(1-x_1)t}\right)^{d/2-3}\,. \notag
\end{align}
Using the integral representation of the hypergeometric function
\begin{align}
\int_0^1{\rm d}x\, x^{b-1}(1-x)^{c-b-1}(1-x\,z)^{-a} = \frac{\Gamma(c-b)\Gamma(b)}{\Gamma(c)}{_2}F_1(a,b,c;z)\,,
\end{align}
valid for ${\rm Re}(c) > {\rm Re}(b) > 0$ and $|{\rm arg}(1-z)| < \pi$, we get
\begin{align}
    \label{T1calc5}
    T_1(p,p_2;k_q)&=-\frac{i\,\Gamma(3-d/2) \Gamma(d/2-2)(-t)^{d/2-3}}{(4\pi)^{d/2}\Gamma(d/2-1)} \\&\qquad 
\int_0^1{\rm d}x_1 
   (1-x_1)^{-d/2} x_1^{d/2-2} {_2}F_1\left(3-\frac{d}{2},1,\frac{d}{2}-1;\frac{-x_1
    u}{(1-x_1)t}\right)\,. \notag
\end{align}
We aim to expand the result around $k_q \to 0$ (i.e.~$t\to 0$), as we are only keeping the terms where the quark is soft. However, the hypergeometric function is not in a form that allows us to easily expand in this limit. We may proceed via use of the contiguous relation
\begin{align}
    {_2}F_1(a,b,c;z)&=(1-z)^{-a}\frac{\Gamma(c)\Gamma(b-a)}
    {\Gamma(b)\Gamma(c-a)}{_2}F_1\left(a,c-b,a-b+1;\frac{1}{1-z}
    \right)\notag\\
    &\quad+(1-z)^{-b}\frac{\Gamma(c)\Gamma(a-b)}{\Gamma(a)\Gamma(c-b)}
    {_2}F_1\left(b,c-a,b-a+1;\frac{1}{1-z}\right),
    \label{contiguous}
\end{align}
after which the $k_q \to 0$ limit may safely be taken setting the hypergeometric functions to 1.
After substituting this result into Eq.~\eqref{T1calc5} we obtain
\begin{align}
    \label{T1calc6}
    T_1(p,p_2;k_q)&=\frac{i\,\Gamma(3-d/2)}{(4\pi)^{d/2}} \frac{1}{u}\Bigg\{\frac{\Gamma^2(d/2-2)}{\Gamma(d-4)}(-u)^{d/2-2}\int_0^1{\rm d}x_1\, (1-x_1)^{3-d}x_1^{d-5} \\
    &\qquad + \frac{\Gamma(2-d/2)}{\Gamma(3-d/2)}(-t)^{d/2-2}\int_0^1{\rm d}x_1\,(1-x_1)^{1-d/2} x_1^{d/2-3}\Bigg\} + \mathcal{O}(k_q).\notag
\end{align}
The two $x_1$ integrals appearing here are formally zero. Physically, this arises due to the fact that linearising propagators in the soft limit leads to the presence of a spurious UV divergence in the momentum integral of Eq.~(\ref{T1calc1}). This precisely cancels the IR divergence we are hoping to extract. We can isolate both contributions by inserting 
\begin{align}
1=x_1+(1-x_1)
\end{align}
into the $x_1$ integral, such that one obtains
\begin{align}
    \int_0^1 {\rm d}x_1 x_1^{d-4}(1-x_1)^{3-d}+\int_0^1 {\rm d}x_1 x_1^{d-5}(1-x_1)^{4-d}&= \Gamma(d-3)\Gamma(4-d)+\Gamma(d-4)\Gamma(5-d)\,.
    \label{x1int2}
\end{align}
Each term on the right-hand side is now well-defined, where the first and second terms contain the divergences as $x_1\rightarrow 1$ and $x_1\rightarrow 0$ respectively. Examining the second line of Eq.~(\ref{T1calc1}, \ref{T1vartrans}), we see that sending $x_1\rightarrow 1$ amounts to sending $\alpha_1\rightarrow 0$, which in turn increases the UV divergence of the integral. Thus, the divergence as $x_1\rightarrow 0$ corresponds to the IR pole of our original integral, such that we must keep the second term on the right-hand side of Eq.~(\ref{x1int2}). Applying similar reasoning to the second integral appearing in Eq.~(\ref{T1calc6}), our final result for the soft triangle integral is then
\begin{align}
    \label{T1res}
    T_1(p,p_2;k_q)&=\frac{i\,\Gamma(1+\epsilon)\Gamma(-\epsilon)}{(4\pi)^{2-\epsilon}} \frac{\left(\Gamma(-\epsilon)\Gamma(1+2\epsilon)(-u)^{-\epsilon} + \Gamma(\epsilon)(-t)^{-\epsilon}\right)}{u} + \mathcal{O}(k_q)\,.
\end{align}
Up to $\mathcal{O}(1/\epsilon)$ the result above can also be written as
\begin{align}
    \label{T1resprime}
    T_1(p,p_2;k_q)&=\frac{i\,r_{\Gamma}}{(4\pi)^{2-\epsilon}} \frac{1}{\epsilon^2} \frac{(-u)^{-\epsilon} + (-t)^{-\epsilon}}{u} + \mathcal{O}(k_q)\,,
\end{align}
with 
\begin{align}
    r_{\Gamma } = \frac{\Gamma^2(1-\epsilon)\Gamma(1+\epsilon)}{\Gamma(1-2\epsilon)}\,.
\end{align}
Using the parameterisation for $t, u$ in Eq.~\eqref{mandiesparam} and taking $x\to 1$, $\epsilon \to 0$ we find 
\begin{align}
    \label{T1exp}
    T_1(p,p_2;k_q)&=\frac{i\,\Gamma(1+\epsilon)\Gamma(-\epsilon)}{(4\pi)^{2-\epsilon}} \frac{\left(\Gamma(-\epsilon)\Gamma(1+2\epsilon)(-u)^{-\epsilon} + \Gamma(\epsilon)(-t)^{-\epsilon}\right)}{u} + \mathcal{O}(k_q)\,.
\end{align}

All other soft integrals we encounter can be evaluated using similar methods. In particular, we quote the results for the triangle integral
\begin{align}
    \label{T2res}
T_2(p,p_2,k_q)&=\int\frac{{\rm d}^d k}{(2\pi)^d}\frac{1}{k^2\,2p\cdot k\,
(k+k_q)^2}\\
&=-\frac{i}{(4\pi)^{d/2}}
\frac{\Gamma^2(1+\epsilon)\Gamma^2(-\epsilon)}{\Gamma(1-\epsilon)}
\frac{(-t)^{-\epsilon}}{-t}\,.\notag
\end{align}
We also need the box integrals
\begin{align}
    B_1(p_1,p_2;k_2)&=\int\frac{{\rm d}^d k_1}{(2\pi)^d}
    \frac{1}{k_1^2\,2p_1\cdot k_1\, 2p_2\cdot k\,2p_1\cdot(k_1-k_2)}\notag\\
    &=-\frac{i}{(4\pi)^{d/2}}\Gamma(3-d/2)\Gamma(2-d/2)\Gamma(d/2-2)
    \frac{(2p_1\cdot k_2)^{d/2-3}}{(2p_1\cdot p_2)}+{\cal O}(k_2)
    \label{B1res}
\end{align}
and
\begin{align}
B_2(p_1,p_2;k_2)&=\int\frac{{\rm d}^d k_1}{(2\pi)^d}
    \frac{1}{k_1^2\,(k_1-k_2)^2\,2p_1\cdot (k_1-k_2)\,2p_2\cdot k_1};
    \notag\\
    &=\frac{i}{(4\pi)^{d/2}}\frac{\Gamma(4-d/2)\Gamma^3(d/2-2)
    \Gamma(3-d/2)}{(d/2-3)\Gamma(d-4)}\notag\\
    &\quad\times\frac{1}{(2p_1\cdot k_2)
    (2p_2\cdot k_2)}\left(-\frac{(2p_1\cdot p_2)}{(2p_1\cdot k_2)
    (2p_2\cdot k_2)}\right)^{2-d/2}.
    \label{B2res}
\end{align}
Note that the latter integral, owing to the presence of additional quadratic denominators, is UV-finite and so requires no additional regularisation in the soft limit. 

As well as the above scalar integrals, we also need certain vector integrals. First, there is the triangle integral:
\begin{equation}
    T_2^\mu(p_1,p_2;k_2)=\int\frac{{\rm d}^d k_1}{(2\pi)^d}\frac{k^{\mu}}{k_1^2\,2p_1\cdot k_1\,(k_1+k_2)^2},
    \label{T2mudef}
\end{equation}
which is related to the scalar integral of Eq.~(\ref{T2res}). Introducing Feynman parameters, one finds that this assumes the form
\begin{align}
T_2^\mu(p_1,p_2;k_2)
&=2\int\frac{{\rm d}^d\tilde{k}}{(2\pi)^d}\left(\prod_i\int {\rm d}\alpha_i
\right)\frac{\delta(1-\alpha_1-\alpha_2-\alpha_3)}{(\alpha_1+\alpha_2)^3
[\tilde{k}^2-M^2]^3}\left(
\tilde{k}-\frac{\alpha_2 k_q+\alpha_3 p}{\alpha_1+\alpha_2}
\right)^\mu,
    \label{T2mucalc1}
\end{align}
where 
\begin{equation}
M^2=\frac{\alpha_2\alpha_3(2p_1\cdot k_2)}{(\alpha_1+\alpha_2)^2},\quad
\tilde{k}^\mu=k^\mu+\frac{(\alpha_2 k_q+\alpha_3 p)^\mu}
{\alpha_1+\alpha_2}.
    \label{T2mucalc2}
\end{equation}
We may ignore the first term in the final brackets of Eq.~(\ref{T2mucalc1}), as this will vanish by the symmetry of the integrand. After performing the momentum integral using Eq.~(\ref{momint}), the remaining integrals can be carried out by setting
\begin{equation}
(\alpha_1,\alpha_2,\alpha_3)=(xy,x(1-y),(1-x)),
    \label{Tmucalc3}
\end{equation}
such that one finds
\begin{align}
    T_2^\mu(p_1,p_2;k_2)
&=-\frac{i(2p_1\cdot k_2)^{d/2-3}}{(4\pi)^{d/2}}
\Gamma(3-d/2)\notag\\
&\times \int_0^1 {\rm d}x\int_0^1 {\rm d}y \,x^{-d/2}(1-x)^{d/2-3}
(1-y)^{d/2-3}[x(1-y)k_q^\mu+(1-x)p^\mu].
\label{Tmucalc4}
\end{align}
In the first term, the $y$ integral is not singular in $d=4$, and thus there is no leading (double) pole in $\epsilon$. The second term does not contribute once the $p^\mu$ factor is contracted into the relevant Feynman diagrams. Thus $T_2^\mu(p_1,p_2;k_2)$ may be ignored for our purposes. 

Finally, we require the vector box integral
\begin{equation}
    B_2^\mu(p_1,p_2;k_2)=\int \frac{{\rm d}^d k}{(2\pi)^d}\frac{k^\mu}
    {k^2\,(k-k_q)^2\,[2p\cdot(k-k_q)](2p_2\cdot k)}.
    \label{B2mudef}
\end{equation}
This can be carried out using similar methods to above, and the result is
\begin{align}
    B_2^\mu(p_1,p_2;k_2)&=\frac{i}{(4\pi)^{d/2}}
    \frac{\Gamma(4-d/2)\Gamma(d/2-1)\Gamma^2(d/2-2)\Gamma(3-d/2)}
    {(d-6)\Gamma(d-3)}
    \left(-\frac{(2p_1\cdot p_2)}{(2p_q\cdot k_2)(2p_2\cdot k_2)}
    \right)^{2-d/2}\notag\\
    &\times \left[
    \frac{2p_2^\mu}{(2p_2\cdot k_2)(2p_1\cdot p_2)}+\frac{2k_2^\mu}{(2p_2\cdot k_q)(2p_1\cdot p_2)}
    -\frac{p^\mu}{(2p_1\cdot p_2)(2p_1\cdot k_2)}
    \right].
    \label{B2mures}
\end{align}

\bibliographystyle{JHEP}
\bibliography{refs}

\end{document}